\documentstyle[preprint,prd,aps]{revtex}

\def\beq{\begin{equation}}
\def\eeq{\end{equation}}
\def\beqa{\begin{eqnarray}}
\def\eeqa{\end{eqnarray}}

\def\za{\alpha}
\def\zb{\beta}
\def\lsim{\mathrel{\raise.3ex\hbox{$<$\kern-.75em\lower1ex\hbox{$\sim$}}} }
\def\gsim{\mathrel{\raise.3ex\hbox{$>$\kern-.75em\lower1ex\hbox{$\sim$}}} }

\begin{document}
\draft
\preprint{\vbox{\hbox{IPAS-HEP-k006}\hbox{KEK-TH-714}
\hbox{Dec 2000} \hbox{ed. Mar 2001} }}

\title{One-loop Neutron Electric 
Dipole Moment from Supersymmetry without R parity}
\author{\bf Y.-Y. Keum$\!^1$ 
$\!$\footnote{E-mail: keum@phys.sinica.edu.tw} \ 
and Otto C. W. Kong$\!^{1,2}$
$\!$\footnote{E-mail: kongcw@phys.sinica.edu.tw}
}
\address{$^1\!$Institute of Physics, Academia Sinica, 
Nankang, Taipei, Taiwan 11529\\
$^2\!$Theory Group, KEK, Tsukuba, Ibaraki, 305-0801, Japan}
\maketitle

\begin{abstract}
We present a detailed analysis together with exact numerical calculations
on one-loop contributions to the neutron electric dipole moment from 
supersymmetry without R parity, focusing on the gluino, chargino, and
neutralino contributions. Apart from the neglected family mixing among
quarks, complete formulae are given for the various contributions 
through the quark dipole operators, to which the present study is restricted.
We discuss the structure and main features of the R-parity violating
contributions and the interplay between the R-parity conserving and 
violating parameters. In particular, the parameter combination
$\mu_i^*\lambda^{\!\prime}_{i\scriptscriptstyle 1\!1}$, under
the optimal parametrization adopted, is shown to be solely responsible
for the R-parity violating contributions in the supersymmetric loop
diagrams. While $\mu_i^*\lambda^{\!\prime}_{i\scriptscriptstyle 1\!1}$
could bear a complex phase, the latter is not necessary to have a
R-parity violating contribution.
\end{abstract}
\vspace{10mm}
\pacs{PACS index: 13.40.Em,11.30.Er,12.60.Jv,14.80.Ly }


\section{introduction}
Neutron and electron electric dipole moments (EDMs) are important topics 
for new CP violating physics. They are known to be extremely small
in the Standard Model (SM); in fact, way below the present experimental limit. 
With supersymmetry (SUSY) comes many plausible extra EDM contributions. That has 
led to the so-called SUSY CP problem\cite{susycp} for the  minimal 
supersymmetric standard model (MSSM). If one simply takes the minimal 
supersymmetric spectrum of the SM and imposes nothing more than
the gauge symmetries while still admitting soft SUSY breaking, the generic 
supersymmetric standard model would result. When the large number
of baryon and/or lepton number violating terms in such a generic 
supersymmetric SM are removed by hand, through imposing
an {\it ad hoc} discrete symmetry called R parity, one obtains the
MSSM Lagrangian. In the case of R-parity violation, two recent papers 
focus on the contributions from the extra trilinear terms in the
superpotential and conclude that there is no new EDM contribution at the 
1-loop level\cite{two}. Perhaps it has not been emphasized enough in the 
two papers that they are {\it not} studying the complete theory of SUSY 
without R parity, which is nothing other than the generic 
supersymmetric SM; in particular, they have nelgected admissible RPV 
parameters other than the trilinear ones in the superpotential.
It is interesting to see  that in the  generic case
there are in fact contributions at the 1-loop level, as pointed out
in Refs.\cite{as4,cch}. In particular, Ref.\cite{as4} gives a clear 
illustration of the much overlooked existence of a R-parity 
violating (RPV) contribution to $LR$ squark mixings and the resulting
contribution to neutron EDM through the simple 1-loop gluino diagram.
We would like to emphasize again that the new contribution involves 
both bilinear and trilinear (RPV) couplings in the superpotential. Since  
RPV scenarios studied in the literature typically admit only one of the two 
types of couplings, the contribution has not been previously identified. A 
simple estimate of the bound obtained on the RPV parameters 
(the $\mu_i^* \, \lambda^{\!\prime}_{i{\scriptscriptstyle  1\!1}}$ combination)
given in Ref.\cite{as4} has already illustrated that the bound from the
neutron EDM as one of the most important, being competitive even when compared
with sub-eV neutrino mass bounds and substantially more stringent than most
collider bounds. The present article aims at giving a detailed analysis and
numerical study of the RPV extension of SUSY contributions to neutron EDM. 
Similar new RPV contributions to electron EDM have been noted in Ref.\cite{as4}.
In fact, the complete result for RPV contributions to the masses of
the sleptons and other (color-singlet) scalars has been given in
Ref.\cite{as5}, which focuses mainly on their implications to neutrino
masses.

The complete theory of SUSY without
R parity admits all kinds of RPV terms without bias. It is obviously
better motivated than {\it ad hoc} versions of RPV theories. The large 
number of new parameters involved, however, makes the theory difficult to 
analyze. The question of the specification of flavor bases to define
the parameters in the Lagrangian of the theory unambiguously becomes
more important. In fact, thinking about the theory as the generic
supersymmetric SM instead of as ``MSSM + RPV terms" helps to clarify
many of the issues involved\cite{as8}. From such a perspective,
it has been illustrated\cite{k} that an optimal parametrization, 
called the single-VEV parametrization (SVP), provides a very nice 
formulation which helps to simplify much of the analysis. In particular, 
the SVP gives the complete results for the tree-level mass matrices
of all state, fermions as well as scalars, in the simplest form\cite{as5}.
The formulation has been used to study leptonic phenomenology\cite{k}
and various aspects of neutrino masses\cite{as5,ok,as1,AL,GH}. The present
EDM study (also Refs.\cite{as4,cch}) and  parallel works on
$\mu \to e\,\gamma$\cite{as7} (see also Ref.\cite{cch2}), electron EDM, and 
$b \to s \, \gamma$\cite{as11} 
will further illustrate the advantage of adopting the SVP.

We focus here only on such contributions to
the neutron EDM, based on the valence quark model\cite{HKP}. Hence, we study
only the 1-loop quark EDMs. We will give complete 1-loop formulae for EDMs
of the up- and down-sector quarks, of which the $u$ and $d$ results are used to calculate the neutron EDM through the
\beq \label{vqm}
d_n = {1\over 3} \left (4\, d_d -d_u \right) \, \eta
\eeq
formula, where $\eta \simeq 1.53$ is a QCD correction factor from
renormalization group evolution\cite{IN1,QCD}. This is to be matched with 
the experimental bound\cite{exp} 
\[
d_n < 6.3 \cdot 10^{-26}\,e \cdot \mbox{cm} \; .
\]
In the MSSM case, one has the SUSY loop contributions and the 
charged Higgs contributions. The latter are very negligible. We focus here
in this article on the analogue of the 1-loop SUSY contributions. The latter 
include the gluino loop, the charginolike loop, and the neutralinolike loop.
By the last two, we mean generalization of the chargino and neutralino
loops under the generic picture. The (RPV) mixings of the leptons with the
gauginos and Higgsinos give five (color-singlet) charged leptons and seven
neutral fermions, including the charginos and neutralinos as well as $e$,
$\mu$, $\tau$, and three physical neutrinos. They come from the same set
of electroweak states and should not be separated from one another in the 
analysis. It is no surprise that the physical chargino and neutralino states
dominate the EDM contributions. We use explicit exact mass eigenstate 
expressions in our analysis to illustrate that as well as other interesting
features, starting from the generic electroweak states couplings under the SVP.
An exact numerical calculation is performed. We would like to mention that
the generalization of the charged Higgs loop contribution involves
other different RPV parameters. Moreover, there are many new and 
potentially important contributions including a $t$-quark loop, as also pointed 
out in Ref.\cite{cch}. We will give also the formulae of such
quark-scalar loop contributions, though a detailed study is postponed to a future 
publication. Note that Ref.\cite{cch}, which first appeared around 
the same time as Ref.\cite{as4}, is the only other study of the same topic 
available. In our opinion, our study here is more systematic and complete. 
Ref.\cite{cch} does not include, for example, the RPV $LR$ scalar mixing and 
the resulted gaugino loop contribution to EDM. Moreover, to the best of our knowlegde,
the present study includes the first exact numerical calculation performed. 
Ref.\cite{cch} also quotes a $\cos\!\zb$ dependence of the major charginolike
contribution, hence a weakening of the bound in the large $\tan\!\zb$ regime 
--- a result with which
we disagree. Our careful numerical study illustrates many more interesting features,
as the discussion below will speak for itself.

This paper is organized as follows:
We first summarize the formulation and notation used in Sec.~II, where
we also elaborate in some detail on the electroweak fermion field couplings 
needed to study the quark EDMs. Next, Sec.~III contains results presented 
recently in Ref.\cite{as5} on all the scalar masses, listed here so that 
the present paper will be self-contained.
Of most importance here is the RPV contributions to $LR$ mixings,
which play a central part in the EDM contributions. The quark EDMs are
analyzed in Sec.~IV.  Some results
from our numerical calculations are presented in Sec.~V, after which we 
conclude in Sec.~VI. An appendix gives some background formulae on
the color-singlet fermion masses. Note that our formulae and calculations
here naturally include the R-parity conserving MSSM part, though we will
concentrate on the role of the RPV parameters and their unique contributions.
Corresponding studies of the MSSM case can be found, for example, in
Refs.\cite{IN1,KiOs,IN2,NGKO}, to which we refer the readers for comparison.

\section{Formulation and Notation} 
We summarize our formulation and notation below.
The most general renormalizable superpotential for the generic
supersymmetric SM  can be written then as
\begin{equation}
W = \varepsilon_{ab}\left[ \mu_{\alpha}  \hat{H}_u^a \hat{L}_{\alpha}^b 
+ y_{ik}^u \hat{Q}_i^a   \hat{H}_{u}^b \hat{U}_k^{\scriptscriptstyle C}
+ \lambda_{\alpha jk}^{\!\prime}  \hat{L}_{\alpha}^a \hat{Q}_j^b
\hat{D}_k^{\scriptscriptstyle C} + 
\frac{1}{2}\, \lambda_{\alpha \beta k}  \hat{L}_{\alpha}^a  
 \hat{L}_{\beta}^b \hat{E}_k^{\scriptscriptstyle C} \right] + 
\frac{1}{2}\, \lambda_{ijk}^{\!\prime\prime}  
\hat{U}_i^{\scriptscriptstyle C} \hat{D}_j^{\scriptscriptstyle C}  
\hat{D}_k^{\scriptscriptstyle C}  \; ,
\end{equation}
where  $(a,b)$ are $SU(2)$ indices, $(i,j,k)$ are the usual family (flavor) 
indices, and $(\za, \zb)$ are the extended flavor indices going from $0$ to $3$.
At the limit where $\lambda_{ijk}, \lambda^{\!\prime}_{ijk},  
\lambda^{\!\prime\prime}_{ijk}$, and $\mu_{i}$  all vanish, 
one recovers the expression for the R-parity preserving case ({\it i.e.,} MSSM), 
with $\hat{L}_{0}$ identified as $\hat{H}_d$. Without R parity imposed,
the latter is not {\it a priori} distinguishable from the $\hat{L}_{i}$'s.
Note that $\lambda$ is antisymmetric in the first two indices, as
required by  the $SU(2)$  product rules, as shown explicitly here with 
$\varepsilon_{\scriptscriptstyle 12} =-\varepsilon_{\scriptscriptstyle 21}=1$.
Similarly, $\lambda^{\!\prime\prime}$ is antisymmetric in the last two indices
from $SU(3)_{\scriptscriptstyle C}$.

R parity is exactly an {\it ad hoc} symmetry put in to make $\hat{L}_{0}$
stand out from the other $\hat{L}_i$'s as the candidate for  $\hat{H}_d$.
It is defined in terms of baryon number, lepton number, and spin as, 
explicitly, ${\mathcal R} = (-1)^{3B+L+2S}$. The consequence is that 
the accidental symmetries of baryon number and lepton number in the SM 
are preserved, at the expense of making particles and superparticles having 
a categorically different quantum number, R parity. The latter is actually 
not the most effective discrete symmetry to control superparticle 
mediated proton decay\cite{pd}, but is most restrictive in terms
of what is admitted in the Lagrangian, or the superpotential alone.

A naive look at the scenario suggests that the large number of new 
couplings makes the task formidable. However, it becomes quite
manageable with an optimal choice of flavor bases, the SVP\cite{k}. 
In fact, doing phenomenological studies without specifying a choice 
of flavor bases is ambiguous. It is like doing SM quark physics with 18
complex Yukawa couplings instead of the 10 real physical parameters.
For SUSY without R parity, the choice of an optimal
parametrization mainly concerns the 4 $\hat{L}_\alpha$ flavors.
Under the SVP
\footnote{Note that our notation here is a bit different from that
in Ref.\cite{k}; we follow mostly the notation in Refs.\cite{as4} and 
\cite{as5} while improving and elaborating further whenever appropriate.
We will clarify all notation used as our discussion goes along\cite{as8}. }
, flavor bases are chosen such that 
1/ among the $\hat{L}_\alpha$'s, only  $\hat{L}_0$, bears a VEV
({\it i.e.,} $\langle \hat{L}_i \rangle \equiv 0$);
2/  $y^{e}_{jk} (\equiv \lambda_{0jk} =-\lambda_{j0k}) 
=\frac{\sqrt{2}}{v_{\scriptscriptstyle 0}}\,{\rm diag}
\{m_{\scriptscriptstyle 1},
m_{\scriptscriptstyle 2},m_{\scriptscriptstyle 3}\}$;
3/ $y^{d}_{jk} (\equiv \lambda^{\!\prime}_{0jk}) 
= \frac{\sqrt{2}}{v_{\scriptscriptstyle 0}}\,{\rm diag}\{m_d,m_s,m_b\}$; 
4/ $y^{u}_{ik}=\frac{\sqrt{2}}{v_{\scriptscriptstyle u}}\,
V_{\!\mbox{\tiny CKM}}^{\!\scriptscriptstyle T}\; {\rm diag}\{m_u,m_c,m_t\}$, where 
$v_{\scriptscriptstyle 0}\equiv \sqrt{2} \, \langle \hat{L}_0 \rangle$
and $v_{\scriptscriptstyle u}\equiv \sqrt{2} \,
\langle \hat{H}_{u} \rangle$.
The big advantage of here is that the (tree-level) mass 
matrices for all the fermions  do not involve any of
the trilinear RPV couplings, though the approach makes  no assumption 
on any RPV coupling including even those from soft SUSY breaking,
and all the parameters used are uniquely defined, with the exception of
some removable phases. 

We are interested in scalar-fermion-fermion couplings similar to
those of the charginos and neutralinos in the MSSM. The gaugino couplings
are, of course, standard. Coming from the gauge interaction parts, they
have nothing to do with R parity. The ``Higgsino-like" part is, however,
different.  Without R parity and in a generic flavor basis of the four $\hat{L}_\za$'s,
the $\hat{H}_d$ of the MSSM is hidden among the latter. The SVP, however, identifies
$\hat{L}_0$ as the one having ``Higgs" properties of $\hat{H}_d$, though it
still maintains (RPV) couplings similar to those of the   $\hat{L}_i$'s. 
We write the components of a $\hat{L}_\za$ fermion 
doublet as ${l}_{\!\scriptscriptstyle \za}^{\scriptscriptstyle 0}$
and ${l}_{\!\scriptscriptstyle \za}^{\!\!\mbox{ -}}$, and their scalar
partners as $\tilde{l}_{\!\scriptscriptstyle \za}^{\scriptscriptstyle 0}$
and $\tilde{l}_{\!\scriptscriptstyle \za}^{\!\!\mbox{ -}}$. Apart from being 
better motivated theoretically, the common notation helps to trace the 
flavor structure. However, we will also use notation of the form
${h}_{\!\scriptscriptstyle d}^{\star}$ and
$\tilde{h}_{\!\scriptscriptstyle d}^{\star}$ as alternative notation for
$\tilde{l}_{\!\scriptscriptstyle 0}^{\star}$ and
${l}_{\!\scriptscriptstyle 0}^{\star}$ in some places below. This is 
unambiguous under our formulation. We will also referred to 
${h}_{\!\scriptscriptstyle d}^{\star}$ and
$\tilde{h}_{\!\scriptscriptstyle d}^{\star}$ as the Higgs boson and Higgsino,
respectively, while they are generally also included in the terms slepton
and lepton.
 
Note that in the left-handed lepton and slepton field notation
introduced above, we have dropped the commonly used $L$ subscript, for
simplicity. For the components of the three right-handed leptonic 
superfields, we use ${l}_i^{\scriptscriptstyle +}$ and
$\tilde{l}_i^{\scriptscriptstyle +}$, with again the $R$ subscript
dropped. The notation for the quark and squark fields will be standard,
with the $L$ and $R$ subscripts. A normal quark state, such as 
$d_{\!\scriptscriptstyle L_k}$, denotes a mass eigenstate, while a squark
state the supersymmetric partner of one. A quark or squark state with
a $\prime$ denotes one with the quark state being the $SU(2)$ partner 
of a mass eigenstate. For instance,  
$\tilde{u}^{\prime}_{\!\scriptscriptstyle L_3}$ is the up-type squark state
from $\hat{Q}_{\scriptscriptstyle 3}$ which contains the exact
left-handed $b$ quark according to our parametrization of the Lagrangian.

The up-sector Higgs boson is unaffected by R-parity violation. The scalar and
fermion states of the doublet are denoted by 
${h}_{\!\scriptscriptstyle u}^{\!\scriptscriptstyle +}$, 
${h}_{\!\scriptscriptstyle u}^{\!\scriptscriptstyle 0}$ and
$\tilde{h}_{\!\scriptscriptstyle u}^{\!\scriptscriptstyle +}$, 
$\tilde{h}_{\!\scriptscriptstyle u}^{\!\scriptscriptstyle 0}$,
respectively. 

The identity of the charginos and neutralinos is unambiguous in the MSSM.
Without R parity, they mix with the charged leptons and neutrinos. In fact,
the true charged leptons and neutrinos are the light mass eigenstates of
the full $5\times 5$ and $7\times 7$ mass matrices, respectively. The
mass eigenstates deviate from the ${l}_i^{\!\!\mbox{ -}}$'s and
${l}_i^{\scriptscriptstyle 0}$'s. Though the latter deviations are
practically negligible in the limit of small $\mu_i$'s, it still helps 
to distinguish the electroweak states from the mass eigenstates. Moreover,
large R-parity violation, especially in terms of a large
${\mu}_{\scriptscriptstyle 3}$, is not definitely ruled out\cite{k}.
Here, in this paper, we make the explicit distinction and reserve the
terms chargino and neutralino for the heavy states beyond the physical charged
leptons $e$, $\mu$, and $\tau$ and neutrinos. The $m_i$'s introduced 
above are the Yukawa contributions to the physical charged lepton masses, hence
approximately equal to the latter. The readers are referred to the appendix  
for more  details about the fermion mass terms. 

We are now ready to spell out the couplings concerning the (color-singlet) 
charged and neutral fermions from the superpotential. We have 
\beqa
{\cal{L}}_{\chi}
&=&  y_{\!\scriptscriptstyle u_i} \,
V_{\!\mbox{\tiny CKM}}^{ij} \;
\tilde{h}_{\!\scriptscriptstyle u}^{\!\scriptscriptstyle +} \,
\left[ \;
\tilde{u}_{\!\scriptscriptstyle R_i}^c \,
d_{\!\scriptscriptstyle L_j} +
u_{\!\scriptscriptstyle R_i}^c \,
\tilde{d}_{\!\scriptscriptstyle L_j}  \,
\right] 
+ y_{\!\scriptscriptstyle d_i} \,
{l}_{\scriptscriptstyle 0}^{\!\!\mbox{ -}} \,
\left[ \;
\tilde{d}_{\!\scriptscriptstyle R_i}^c \,
u_{\!\scriptscriptstyle L_i}^{\prime}  +
d_{\!\scriptscriptstyle R_i}^c \,
\tilde{u}_{\!\scriptscriptstyle L_i}^{\prime}  \,
\right]  
 +  \lambda^{\!\prime}_{ijk} \;
{l}_i^{\!\!\mbox{ -}} \,
\left[ \;
\tilde{d}_{\!\scriptscriptstyle R_k}^c \,
u_{\!\scriptscriptstyle L_j}^{\prime}  +
d_{\!\scriptscriptstyle R_k}^c \,
\tilde{u}_{\!\scriptscriptstyle L_j}^{\prime}  \,
\right] 
\nonumber \\
&& - \;
y_{\!\scriptscriptstyle u_i} \,
\tilde{h}_{\!\scriptscriptstyle u}^{\!\scriptscriptstyle 0} \,
\left[ \;
\tilde{u}_{\!\scriptscriptstyle R_i}^c \,
u_{\!\scriptscriptstyle L_i} +
u_{\!\scriptscriptstyle R_i}^c \,
\tilde{u}_{\!\scriptscriptstyle L_i} \,
\right] 
- y_{\!\scriptscriptstyle d_i} \,
{l}_{\scriptscriptstyle 0}^{\scriptscriptstyle 0} \,
\left[ \;
\tilde{d}_{\!\scriptscriptstyle R_i}^c \,
d_{\!\scriptscriptstyle L_i}  +
d_{\!\scriptscriptstyle R_i}^c \,
\tilde{d}_{\!\scriptscriptstyle L_i} \,
\right]
- \lambda^{\!\prime}_{ijk} \;
{l}_i^{\scriptscriptstyle 0} \,
\left[ \;
\tilde{d}_{\!\scriptscriptstyle R_k}^c \,
d_{\!\scriptscriptstyle L_j}  +
d_{\!\scriptscriptstyle R_k}^c \,
\tilde{d}_{\!\scriptscriptstyle L_j}  \,
\right]
\nonumber \\ 
&& + \;
 y_{\!\scriptscriptstyle e_i} \,
\left[ \;
{l}_{\scriptscriptstyle 0}^{\!\!\mbox{ -}} \,
{l}_i^{\scriptscriptstyle +} \,
\tilde{l}_i^{\scriptscriptstyle 0}  \,
-
{l}_i^{\!\!\mbox{ -}}  \,
{l}_i^{\scriptscriptstyle +} \,
\tilde{l}_{\scriptscriptstyle 0}^{\scriptscriptstyle 0} \,
\right]
+ y_{\!\scriptscriptstyle e_i} \,
\left[ \;
{l}_{\scriptscriptstyle 0}^{\!\!\mbox{ -}} \,
{l}_i^{\scriptscriptstyle 0}  \,
\tilde{l}_i^{\scriptscriptstyle +} \,
-
{l}_i^{\!\!\mbox{ -}} \,
{l}_{\scriptscriptstyle 0}^{\scriptscriptstyle 0} \,
\tilde{l}_i^{\scriptscriptstyle +} \,
\right]
 + y_{\!\scriptscriptstyle e_i} \,
\left[ \;
{l}_i^{\scriptscriptstyle 0}  \,
{l}_i^{\scriptscriptstyle +} \,
\tilde{l}_{\scriptscriptstyle 0}^{\!\!\mbox{ -}} \,
-
{l}_{\scriptscriptstyle 0}^{\scriptscriptstyle 0} \,
{l}_i^{\scriptscriptstyle +} \,
\tilde{l}_i^{\!\!\mbox{ -}}  \,
\right] 
\nonumber \\
&& + \; 
 \lambda_{ijk} \;
{l}_i^{\!\!\mbox{ -}} \,
{l}_k^{\scriptscriptstyle +} \,
\tilde{l}_j^{\scriptscriptstyle 0}  \, 
+  \lambda_{ijk} \;
{l}_i^{\!\!\mbox{ -}} \,
{l}_j^{\scriptscriptstyle 0} \,
\tilde{l}_k^{\scriptscriptstyle +} 
- \lambda_{ijk} \;
{l}_i^{\scriptscriptstyle 0} \,
{l}_k^{\scriptscriptstyle +} \,
\tilde{l}_j^{\!\!\mbox{ -}}  \,
\quad + \quad \mbox{h.c.} \;\; ,
\label{Lc}\eeqa
where
\beq \label{yy}
y_{\!\scriptscriptstyle u_i} = 
\frac {g_{\scriptscriptstyle 2} \, m_{u_i} }
{\sqrt{2}\, M_{\!\scriptscriptstyle W} \,\sin\!\beta}
\;\; , \hspace{10mm}
y_{\!\scriptscriptstyle d_i} = 
\frac {g_{\scriptscriptstyle 2} \, m_{d_i} }
{\sqrt{2}\, M_{\!\scriptscriptstyle W} \,\cos\!\beta}
\;\; , \hspace{10mm}
y_{\!\scriptscriptstyle e_i} = 
\frac {g_{\scriptscriptstyle 2} \, m_{i} }
{\sqrt{2}\, M_{\!\scriptscriptstyle W} \,\cos\!\beta}
\;\; 
\eeq
are the (diagonal) quark and charged lepton Yukawa couplings, and 
$\tan\!\beta =\frac{v_{\scriptscriptstyle u}}{v_{\scriptscriptstyle 0}}$\cite{app}. 
Recall that $\lambda^{\!\prime}_{{\scriptscriptstyle 0}jk}$ 
corresponds to the down-quark Yukawa coupling matrix,  and 
$\lambda_{{\scriptscriptstyle 0}jk}$ corresponds to the charged 
lepton Yukawa coupling matrix, both of which are diagonal under the SVP;
in addition, we have 
$u_{\!\scriptscriptstyle L_i}^{\prime} = V_{\!\mbox{\tiny CKM}}^{\dag\, ij}\, 
u_{\!\scriptscriptstyle L_j}$ being the $SU(2)$ partner of the mass eigenstate
$d_{\!\scriptscriptstyle L_i}$, and $\tilde{u}_{\!\scriptscriptstyle L_i}^{\prime}$
its scalar partner. We also use below 
$\tilde{d}_{\!\scriptscriptstyle L_i}^{\,\prime}$, which is, explicitly,
$V_{\!\mbox{\tiny CKM}}^{ij}\, \tilde{d}_{\!\scriptscriptstyle L_j}$.
There are some more scalar-fermion-fermion terms 
besides those given in ${\cal{L}}_{\chi}$. These extra terms
are slepton-quark-quark terms. We will see
that the latter are actually also involved in 1-loop EDM diagrams,
though not the major focus of this paper. With the above explicit 
listed terms, however, it is straightforward to see what the extra 
terms are like.  

In both of the above expressions for ${\cal{L}}_{\chi}$, 
there is a clear distinction between the MSSM terms and the RPV terms. The 
nice feature is a consequence of the SVP. The simple structure of
the trilinear coupling contributions to the $d$-quark and charged lepton masses,
which is equivalent to that of the R-parity conserving limit, is what makes
the analysis simple and easy to handle. We want to emphasize that the
above expressions are exact tree-level results without hidden assumptions
behind their validity. They are good even when there is large R-parity violation.

\section{Squark and slepton masses} 
The soft SUSY breaking part of the Lagrangian, in terms of scalar parts
of the superfield multiplets, can be written as follows:
\beqa
V_{\rm soft} &=& \epsilon_{\!\scriptscriptstyle ab} 
  B_{\za} \,  H_{u}^a \tilde{L}_\za^b +
\epsilon_{\!\scriptscriptstyle ab} \left[ \,
A^{\!\scriptscriptstyle U}_{ij} \, 
\tilde{Q}^a_i H_{u}^b \tilde{U}^{\scriptscriptstyle C}_j 
+ A^{\!\scriptscriptstyle D}_{ij} 
H_{d}^a \tilde{Q}^b_i \tilde{D}^{\scriptscriptstyle C}_j  
+ A^{\!\scriptscriptstyle E}_{ij} 
H_{d}^a \tilde{L}^b_i \tilde{E}^{\scriptscriptstyle C}_j   \,
\right] + {\rm h.c.}\nonumber \\
&+&
\epsilon_{\!\scriptscriptstyle ab} 
\left[ \,  A^{\!\scriptscriptstyle \lambda^\prime}_{ijk} 
\tilde{L}_i^a \tilde{Q}^b_j \tilde{D}^{\scriptscriptstyle C}_k  
+ \frac{1}{2}\, A^{\!\scriptscriptstyle \lambda}_{ijk} 
\tilde{L}_i^a \tilde{L}^b_j \tilde{E}^{\scriptscriptstyle C}_k  
\right] 
+ \frac{1}{2}\, A^{\!\scriptscriptstyle \lambda^{\prime\prime}}_{ijk} 
\tilde{U}^{\scriptscriptstyle C}_i  \tilde{D}^{\scriptscriptstyle C}_j  
\tilde{D}^{\scriptscriptstyle C}_k  + {\rm h.c.}
\nonumber \\
&+&
 \tilde{Q}^\dagger \tilde{m}_{\!\scriptscriptstyle {Q}}^2 \,\tilde{Q} 
+\tilde{U}^{\dagger} 
\tilde{m}_{\!\scriptscriptstyle {U}}^2 \, \tilde{U} 
+\tilde{D}^{\dagger} \tilde{m}_{\!\scriptscriptstyle {D}}^2 
\, \tilde{D} 
+ \tilde{L}^\dagger \tilde{m}_{\!\scriptscriptstyle {L}}^2  \tilde{L}  
  +\tilde{E}^{\dagger} \tilde{m}_{\!\scriptscriptstyle {E}}^2 
\, \tilde{E}
+ \tilde{m}_{\!\scriptscriptstyle H_{\!\scriptscriptstyle u}}^2 \,
|H_{u}|^2 
\nonumber \\
&& + \frac{M_{\!\scriptscriptstyle 1}}{2} \tilde{B}\tilde{B}
   + \frac{M_{\!\scriptscriptstyle 2}}{2} \tilde{W}\tilde{W}
   + \frac{M_{\!\scriptscriptstyle 3}}{2} \tilde{g}\tilde{g}
+ {\rm h.c.}\; ,
\label{soft}
\eeqa
where we have separated the R-parity conserving ones from the 
RPV ones ($H_{d} \equiv \tilde{L}_0$) for the $A$ terms. Note that 
$\tilde{L}^\dagger \tilde{m}_{\!\scriptscriptstyle \tilde{L}}^2  \tilde{L}$,
unlike the other soft mass terms, is given by a 
$4\times 4$ matrix. Explicitly, 
$\tilde{m}_{\!\scriptscriptstyle {L}_{00}}^2$ is
$\tilde{m}_{\!\scriptscriptstyle H_{\!\scriptscriptstyle d}}^2$ 
of the MSSM case while 
$\tilde{m}_{\!\scriptscriptstyle {L}_{0k}}^2$'s give RPV mass mixings.
The other notation is obvious.

The SVP also simplifies much the otherwise extremely complicated expressions
for the mass-squared matrices of the scalar sectors. First, we will look 
at the squarks sectors. The masses of up squarks obviously have no RPV 
contribution. The down-squark sector, however, has an interesting result. 
We have the mass-squared matrix as follows: 
\beq \label{MD}
{\cal M}_{\!\scriptscriptstyle {D}}^2 =
\left( \begin{array}{cc}
{\cal M}_{\!\scriptscriptstyle LL}^2 & {\cal M}_{\!\scriptscriptstyle RL}^{2\dag} \\
{\cal M}_{\!\scriptscriptstyle RL}^{2} & {\cal M}_{\!\scriptscriptstyle RR}^2
 \end{array} \right) \; ,
\eeq
where
\beqa
{\cal M}_{\!\scriptscriptstyle LL}^2 &=&
\tilde{m}_{\!\scriptscriptstyle {Q}}^2 +
m_{\!\scriptscriptstyle D}^\dag m_{\!\scriptscriptstyle D}
+ M_{\!\scriptscriptstyle Z}^2\, \cos\!2 \beta 
\left[ -\frac{1}{2} + \frac{1}{3} \sin\!^2 \theta_{\!\scriptscriptstyle W}\right] \; ,
\nonumber \\
{\cal M}_{\!\scriptscriptstyle RR}^2 &=&
\tilde{m}_{\!\scriptscriptstyle {D}}^2 +
m_{\!\scriptscriptstyle D} m_{\!\scriptscriptstyle D}^\dag
+ M_{\!\scriptscriptstyle Z}^2\, \cos\!2\beta 
\left[  - \frac{1}{3} \sin\!^2 \theta_{\!\scriptscriptstyle W}\right] \; ,
\eeqa 
and
\beqa 
({\cal M}_{\!\scriptscriptstyle RL}^{2})^{\scriptscriptstyle T} 
&=& 
A^{\!{\scriptscriptstyle D}} \frac{v_{\scriptscriptstyle 0}}{\sqrt{2}}
- (\, \mu_{\scriptscriptstyle \za}^*\lambda^{\!\prime}_{{\scriptscriptstyle \za}jk}\,)
\; \frac{v_{\scriptscriptstyle u}}{\sqrt{2}} \; 
\nonumber \\ &=& 
\left[ A_d -  \mu_{\scriptscriptstyle 0}^* \, \tan\!\beta \right]
\,m_{\!\scriptscriptstyle D}\;
+ \frac{\sqrt{2}\, M_{\!\scriptscriptstyle W} \cos\!\beta}
{g_{\scriptscriptstyle 2} } \,
\delta\! A^{\!{\scriptscriptstyle D}}
- \frac{\sqrt{2}\, M_{\!\scriptscriptstyle W} \sin\!\beta}
{g_{\scriptscriptstyle 2} } \,
(\, \mu_i^*\lambda^{\!\prime}_{ijk}\, ) \; .
\label{RL}
\eeqa
Here, $m_{\!\scriptscriptstyle D}$ is the down-quark mass matrix, 
which is diagonal under the parametrization adopted; 
$A_d$ is a constant (mass) parameter representing the 
``proportional" part of the $A$ term and the matrix 
$\delta\! A^{\!{\scriptscriptstyle D}}$ is the ``proportionality" violating 
part; $(\, \mu_i^*\lambda^{\!\prime}_{ijk}\, )$, and similarly 
$(\, \mu_{\scriptscriptstyle \za}^*\lambda^{\!\prime}_{{\scriptscriptstyle \za}jk}\,)$, 
denotes the $3\times 3$ matrix $(\;)_{jk}$ with elements listed
\footnote{
Note that we use this kind of bracket notation for matrices extensively 
here. In this case, the repeated index $i$ is to be summed 
over as usual and, hence, is dummy.}
. The   
$(\, \mu_{\scriptscriptstyle \za}^*\lambda^{\!\prime}_{{\scriptscriptstyle \za}jk}\,)$
term is the full $F$-term contribution, while the 
$(\, \mu_i^*\lambda^{\!\prime}_{ijk}\, )$ part separated out in the last expression
gives the RPV contributions.  

Next we go on to the slepton sector. From Eq.(\ref{soft}) above, we 
can see that the charged Higgs bosons should be
considered on the same footing together with the sleptons. We have hence an
$8\times 8$ mass-squared matrix of the following $1+4+3$ form:
\beq \label{ME}
{\cal M}_{\!\scriptscriptstyle {E}}^2 =
\left( \begin{array}{ccc}
\widetilde{\cal M}_{\!\scriptscriptstyle H\!u}^2 &
\widetilde{\cal M}_{\!\scriptscriptstyle LH}^{2\dag}  & 
\widetilde{\cal M}_{\!\scriptscriptstyle RH}^{2\dag}
\\
\widetilde{\cal M}_{\!\scriptscriptstyle LH}^2 & 
\widetilde{\cal M}_{\!\scriptscriptstyle LL}^{2} & 
\widetilde{\cal M}_{\!\scriptscriptstyle RL}^{2\dag} 
\\
\widetilde{\cal M}_{\!\scriptscriptstyle RH}^2 &
\widetilde{\cal M}_{\!\scriptscriptstyle RL}^{2} & 
\widetilde{\cal M}_{\!\scriptscriptstyle RR}^2  
\end{array} \right) \; ,
\eeq
where
\beqa
\widetilde{\cal M}_{\!\scriptscriptstyle H\!u}^2 &=&
\tilde{m}_{\!\scriptscriptstyle H_{\!\scriptscriptstyle u}}^2
+ \mu_{\!\scriptscriptstyle \za}^* \mu_{\scriptscriptstyle \za}
+ M_{\!\scriptscriptstyle Z}^2\, \cos\!2 \beta 
\left[ \,\frac{1}{2} - \sin\!^2\theta_{\!\scriptscriptstyle W}\right]
\nonumber \\
&+&  M_{\!\scriptscriptstyle Z}^2\,  \sin\!^2 \beta \;
[1 - \sin\!^2 \theta_{\!\scriptscriptstyle W}]  \; ,
\nonumber \\
\widetilde{\cal M}_{\!\scriptscriptstyle LL}^2 &=&
\tilde{m}_{\!\scriptscriptstyle {L}}^2 +
m_{\!\scriptscriptstyle L}^\dag m_{\!\scriptscriptstyle L}
+ (\mu_{\!\scriptscriptstyle \za}^* \mu_{\scriptscriptstyle \zb})
+ M_{\!\scriptscriptstyle Z}^2\, \cos\!2 \beta 
\left[ -\frac{1}{2} +  \sin\!^2 \theta_{\!\scriptscriptstyle W}\right] \; 
\nonumber \\
&+& \left( \begin{array}{cc}
 M_{\!\scriptscriptstyle Z}^2\,  \cos\!^2 \beta \;
[1 - \sin\!^2 \theta_{\!\scriptscriptstyle W}] 
& \quad 0_{\scriptscriptstyle 1 \times 3} \quad \\
0_{\scriptscriptstyle 3 \times 1} & 0_{\scriptscriptstyle 3 \times 3}  
\end{array} \right) \; ,
\nonumber \\
\widetilde{\cal M}_{\!\scriptscriptstyle RR}^2 &=&
\tilde{m}_{\!\scriptscriptstyle {E}}^2 +
m_{\!\scriptscriptstyle E} m_{\!\scriptscriptstyle E}^\dag
+ M_{\!\scriptscriptstyle Z}^2\, \cos\!2 \beta 
\left[  - \sin\!^2 \theta_{\!\scriptscriptstyle W}\right] \; ,
\eeqa
and
\beqa 
\widetilde{\cal M}_{\!\scriptscriptstyle LH}^2
&=& (B_{\za}^*)  
+ \left( \begin{array}{c} 
{1 \over 2} \,
M_{\!\scriptscriptstyle Z}^2\,  \sin\!2 \beta \;
[1 - \sin\!^2 \theta_{\!\scriptscriptstyle W}]  \\
0_{\scriptscriptstyle 3 \times 1} 
\end{array} \right)
\; ,
\nonumber \\
\widetilde{\cal M}_{\!\scriptscriptstyle RH}^2
&=&  -\,(\, \mu_i^*\lambda_{i{\scriptscriptstyle 0}k}\, ) \; 
\frac{v_{\scriptscriptstyle 0}}{\sqrt{2}} \; 
= (\, \mu_k^* \, m_k \, ) \hspace*{1in} \mbox{ (no sum over $k$)} \quad \; ,
\nonumber \\
(\widetilde{\cal M}_{\!\scriptscriptstyle RL}^{2})^{\scriptscriptstyle T} 
&=& \left(\begin{array}{c} 
0  \\   A^{\!{\scriptscriptstyle E}} 
\end{array}\right)
 \frac{v_{\scriptscriptstyle 0}}{\sqrt{2}}
 - (\, \mu_{\scriptscriptstyle \za}^*\lambda_{{\scriptscriptstyle \za\zb}k}\, ) \; 
\frac{v_{\scriptscriptstyle u}}{\sqrt{2}} \; 
\nonumber \\
&=& [A_e - \mu_{\scriptscriptstyle 0}^* \, \tan\!\beta ] 
\left(\begin{array}{c}  
0  \\   m_{\!{\scriptscriptstyle E}} 
\end{array}\right) \,
+ \frac{\sqrt{2}\, M_{\!\scriptscriptstyle W} \cos\!\beta}
{g_{\scriptscriptstyle 2} } \,
\left(\begin{array}{c} 
0  \\ \delta\! A^{\!{\scriptscriptstyle E}}
\end{array}\right)
-  \left(\begin{array}{c}  
- \mu_{k}^* \, m_k\, \tan\!\beta \\ 
\frac{\sqrt{2}\, M_{\!\scriptscriptstyle W} \sin\!\beta}
{g_{\scriptscriptstyle 2} } \,(\, \mu_i^*\lambda_{ijk}\, ) 
\end{array}\right) \; .
\label{ERL}
\eeqa
The notation and results here are similar to the squark case above, with some 
difference. We have $A_e$ and $\delta\! A^{\!{\scriptscriptstyle E}}$,
or the extended matrices {\tiny $\left(\begin{array}{c} 
0  \\   \star
\end{array}\right)$} incorporating them, denoting the splitting of the $A$ term,
with proportionality defined with respect to 
$m_{\!\scriptscriptstyle E}$; $m_{\!\scriptscriptstyle L}=
\mbox{diag}\{0,m_{\!\scriptscriptstyle E}\}= \mbox{diag}\{0,m_{\!\scriptscriptstyle 1},
m_{\!\scriptscriptstyle 2},m_{\!\scriptscriptstyle 3}\}$. Recall that the $m_i$'s
are approximately the charged lepton masses. 
A $4\times 3$ matrix $(\, \mu_i^*\lambda_{i{\scriptscriptstyle \zb}k}\, )$
gives the RPV contributions to 
$(\widetilde{\cal M}_{\!\scriptscriptstyle RL}^{2})^{\scriptscriptstyle T}$.
In the above expression, we separate explicitly the first row of the former,
which corresponds to mass-squared terms of the type
$\tilde{l}^{\scriptscriptstyle +} h_{\scriptscriptstyle d}^{\!\!\mbox{ -}}$ 
type ($h_{\scriptscriptstyle d}^{\!\!\mbox{ -}} \equiv 
\tilde{l}_{\scriptscriptstyle 0}^{\!\!\mbox{ -}}$). The nonzero 
$\widetilde{\cal M}_{\!\scriptscriptstyle RH}^2$ and the $B_i^*$'s in 
$\widetilde{\cal M}_{\!\scriptscriptstyle LH}^2$ are also RPV contributions.
The former is a $\tilde{l}^{\scriptscriptstyle +} 
(h_{\scriptscriptstyle u}^{\scriptscriptstyle +})^{\dag} $
type, while the latter a $\tilde{l}^{\!\!\mbox{ -}} 
h_{\scriptscriptstyle u}^{\scriptscriptstyle +} $  term.
Note that the parts with the 
$[1 - \sin\!^2 \theta_{\!\scriptscriptstyle W}]$ factor are singled out 
as they are extra contributions to the masses of the charged Higgs bosons
({\it i.e.,} $l_{\scriptscriptstyle 0}^{\!\!\mbox{ -}}
\equiv h_{\!\scriptscriptstyle d}^{\!\!\mbox{ -}}$ and
$h_{\!\scriptscriptstyle u}^{\!\scriptscriptstyle +}$).
The latter are the result of quartic terms in the scalar potential and
the fact that the Higgs doublets bear VEVs.

The neutral scalar mass terms, in terms of the
$(1+4)$ complex scalar fields  $\phi_n$'s, can be written in two parts
--- a simple $({\cal M}_{\!\scriptscriptstyle {\phi}{\phi}\dag}^2)_{mn} \,
\phi_m^\dag \phi_n$ part and a Majorana-like part in the form 
${1\over 2} \,  ({\cal M}_{\!\scriptscriptstyle {\phi\phi}}^2)_{mn} \,
\phi_m \phi_n + \mbox{h.c.}$ As the neutral scalars originate
from chiral doublet superfields, the existence of the Majorana-like
part is a direct consequence of the electroweak symmetry
breaking VEVs, hence restricted to the scalars playing the Higgs boson
role only. They come from the quartic terms of the Higgs fields in
the scalar potential. We have, explicitly,
\beqa \label{Mpp}
{\cal M}_{\!\scriptscriptstyle {\phi\phi}}^2 =
{1\over 2} \, M_{\!\scriptscriptstyle Z}^2\,
\left( \begin{array}{ccc}
 \sin\!^2\! \beta  &  - \cos\!\beta \, \sin\! \beta
& \quad 0_{\scriptscriptstyle 1 \times 3} \\
 - \cos\!\beta \, \sin\! \beta & \cos\!^2\! \beta 
& \quad 0_{\scriptscriptstyle 1 \times 3} \\
0_{\scriptscriptstyle 3 \times 1} & 0_{\scriptscriptstyle 3 \times 1} 
& \quad 0_{\scriptscriptstyle 3 \times 3} 
\end{array} \right) \; ,
\eeqa
and
\beqa 
{\cal M}_{\!\scriptscriptstyle {\phi}{\phi}\dag}^2 &=&
 {\cal M}_{\!\scriptscriptstyle {\phi\phi}}^2 +
\left( \begin{array}{cc}
\tilde{m}_{\!\scriptscriptstyle H_{\!\scriptscriptstyle u}}^2
+ \mu_{\!\scriptscriptstyle \za}^* \mu_{\scriptscriptstyle \za}
+ M_{\!\scriptscriptstyle Z}^2\, \cos\!2 \beta 
\left[-\frac{1}{2}\right]   
& - (B_\za) \\
- (B_\za^*) &
\tilde{m}_{\!\scriptscriptstyle {L}}^2 
+ (\mu_{\!\scriptscriptstyle \za}^* \mu_{\scriptscriptstyle \zb})
+ M_{\!\scriptscriptstyle Z}^2\, \cos\!2 \beta 
\left[ \frac{1}{2}\right]\end{array} \right)  \; .
\label{Mp}
\eeqa
Note that ${\cal M}_{\!\scriptscriptstyle {\phi\phi}}^2$ here is 
real (see the Appendix), while 
${\cal M}_{\!\scriptscriptstyle {\phi}{\phi}\dag}^2$ does have complex entries.
The full $10\times 10$ (real and symmetric) mass-squared matrix for 
the real scalars is then given by
\beq \label{MSN}
{\cal M}_{\!\scriptscriptstyle S}^2 =
\left( \begin{array}{cc}
{\cal M}_{\!\scriptscriptstyle SS}^2 &
{\cal M}_{\!\scriptscriptstyle SP}^2 \\
({\cal M}_{\!\scriptscriptstyle SP}^{2})^{\!\scriptscriptstyle T} &
{\cal M}_{\!\scriptscriptstyle PP}^2
\end{array} \right) \; ,
\eeq
where the scalar, pseudoscalar, and mixing parts are
\beqa
{\cal M}_{\!\scriptscriptstyle SS}^2 &=&
\mbox{Re}({\cal M}_{\!\scriptscriptstyle {\phi}{\phi}\dag}^2)
+ {\cal M}_{\!\scriptscriptstyle {\phi\phi}}^2 \; ,
\nonumber \\
{\cal M}_{\!\scriptscriptstyle PP}^2 &=&
\mbox{Re}({\cal M}_{\!\scriptscriptstyle {\phi}{\phi}\dag}^2)
- {\cal M}_{\!\scriptscriptstyle {\phi\phi}}^2 \; ,
\nonumber \\
{\cal M}_{\!\scriptscriptstyle SP}^2 &=& -
 \mbox{Im}({\cal M}_{\!\scriptscriptstyle {\phi}{\phi}\dag}^2) \; ,
\label{lastsc}
\eeqa
respectively. If $\mbox{Im}({\cal M}_{\!\scriptscriptstyle {\phi}{\phi}\dag}^2)$
vanishes, the scalars and pseudo-scalars decouple from one another and 
the unphysical Goldstone mode would be found among the latter. Finally, we
note that the $B_\za$ entries may also be considered as a kind of $LR$ 
mixing.

We would like to emphasize that the above scalar mass results are complete 
--- all RPV contributions, SUSY breaking or otherwise, are included. The 
simplicity of the result is a consequence of the SVP.
Explicitly, there are no RPV $A$-term contributions due to the vanishing
of VEVs $v_i\equiv \sqrt{2}\langle\hat{L}_i\rangle$. The Higgs-boson-slepton
results given as in Eqs.(\ref{ME}) and (\ref{MSN}) contain a redundancy of 
parameters and hide the unphysical Goldstone state. However, the results
as they are given here are good enough for some purposes
including the present EDM discussion. 

Note that in the following discussion, we will not consider flavor 
changing scalar mass terms from soft SUSY breaking, which could be suppressed
through a flavor-blind SUSY breaking mediating mechanism such as 
gauge mediation. A major concentration of our study, however, will be on
the effects of the flavor changing scalar mass terms from RPV superpotential
parameters, which give interesting new results.

\section{Contributions to neutron EDM at 1-loop}
In this section, we will discuss the RPV 1-loop contributions to neutron
EDM systematically, drawing comparison with the R-parity conserving (MSSM) case
wherever it would be useful. We will give complete 1-loop formulae for quark
EDM's for our generic supersymmetric SM. We will not, however, go into 
the chromoelectric dipole operator or Weinberg gluonic operator contributions. 
Following the common practice, family mixing will largely be 
neglected, though we will comment on some particularly interesting aspects 
of the issue. We will also compare the results with Feynman diagrams
given more or less in an electroweak basis. Naively, such diagrams, with a minimal 
number of mass insertions admitted, should represent a first order result, 
at least where mass mixings are small. Note that the latter is true for the 
small-$\mu_i$ scenario, which is our major focus. Afterall, a mass eigenstate 
result may be considered as corresponding to taking an infinite summation over 
electroweak-state diagrams with all possible mass insertions. The comparison
helps to illustrate better the physics hidden behind the complicated formulae.
In addition to a short Letter by the present authors\cite{as4}, some parts of 
the results here have also been discussed in Ref.\cite{cch}.
  
\subsection{Gluino Contributions}
The most direct contributions come from a gaugino loop, as shown in 
Fig.~1. The diagram looks the same as the MSSM gluino
and neutralino diagrams with two gauge coupling vertices. As pointed out
in our previous short Letter\cite{as4}, the new RPV contributions here are a 
simple result of the RPV $LR$ squark mixings [{\it cf.} Eq.(\ref{RL})].
In Ref.\cite{as4}, we focused on the dominant gluino loop contribution. We 
first give some more details of that analysis before going into the 
other contributions. Notice that though both the
$u$ and $d$ quarks get EDMs from gaugino loops in the the MSSM, only the $d$
quark has the RPV contribution. The $u$-squark sector simply has no RPV 
$LR$ mixings. In Fig.~1, as well as the subsequent diagrams,
we use a family index $k$ for the external quark lines though only the
$k=1$ case would be the $d$- (or $u$-) quark EDM we are mainly concerned
with. With the only possible exception of $d_{\scriptscriptstyle 2}$,
which corresponds to the $s$ quark\cite{HKP}, $k\ne 1$ results are not 
relevant for neutron EDM. 

For illustrative purposes, we first neglect interfamily mixings among the 
squarks. The $\tilde{d}$ mass-squared matrix, of the form given in 
Eq.(\ref{MD}) but reduced to one family, is Hermitian and can be 
diagonalized by the unitary transformation
\beq \label{UD}
{\cal D}^{\dag}_d \,{\cal M}_{\!\scriptscriptstyle D}^2 \,{\cal D}_d = 
\mbox{diag}\{M^2_{\!\scriptscriptstyle \tilde{d}-}, 
M^2_{\!\scriptscriptstyle \tilde{d}+}\} \; ,
\eeq 
with
\beq \label{mass}
M^2_{\!\scriptscriptstyle \tilde{d}\mp} =
{1 \over 2}\,\,\left[ ({\cal M}_{\!\scriptscriptstyle LL}^2 
+ {\cal M}_{\scriptscriptstyle RR}^2) \mp
\sqrt{({\cal M}_{\!\scriptscriptstyle LL}^2 
- {\cal M}_{\!\scriptscriptstyle RR}^2)^2 
+ 4 \,|{\cal M}_{\!\scriptscriptstyle RL}|^2}
\right] \; 
\eeq 
and
\beq
{\cal D}_d = 
\left( \begin{array}{cc}
\cos{\theta \over 2} & -\sin{\theta \over 2}\,e^{-i\phi} \\
\sin{\theta \over 2}\,e^{-i\phi} & \cos{\theta \over 2}
 \end{array} \right) \; .
\eeq
Here, ${\cal M}_{\!\scriptscriptstyle RL}^2 = 
|{\cal M}_{\!\scriptscriptstyle RL}^2| 
\, e^{i\phi}$ and the range of $\theta$ can be choosen so that 
$-{\pi \over 2} \leq \theta \leq {\pi \over 2}$ with
$\tan\!\theta = {2 \; |{\cal M}_{\!\scriptscriptstyle RL}^2| 
\over {\cal M}_{\!\scriptscriptstyle LL}^2 
- {\cal M}_{\!\scriptscriptstyle RR}^2}$.
In terms of the mass eigenstates $\tilde{d}_-$ and $\tilde{d}_+$,
the gluino contribution to the EDM is then given by\cite{IN1,IN2}
\beq \label{G01}
\left({d_{\scriptscriptstyle d} \over e}\right)_{\!\!\tilde{g}} =
-{2 \alpha_s \over 3 \pi} \left[ \,\mbox{Im}(\Gamma_{\!d}^{11})\;
{M_{\!\scriptscriptstyle \tilde{g}} \over M^2_{\!\scriptscriptstyle \tilde{d-}}} \;
{\cal Q}_{\tilde{d}} \; 
B \!\left( {M_{\!\scriptscriptstyle \tilde{g}}^2 \over M^2_{\!\scriptscriptstyle \tilde{d}-}} \right)
+ \mbox{Im}(\Gamma_{\!d}^{12})\;
{M_{\!\scriptscriptstyle \tilde{g}} \over M^2_{\!\scriptscriptstyle \tilde{d+}} } \; 
{\cal Q}_{\tilde{d}} \; 
B \!\left( {M_{\!\scriptscriptstyle \tilde{g}}^2 \over M^2_{\!\scriptscriptstyle \tilde{d}+}} \right)
\, \right] \; ,
\eeq
where ${\cal Q}_{\tilde{d}}=-{1\over 3}$, 
$\Gamma_{\!d}^{1k} = {\cal D}_{d2k}{\cal D}^{*}_{d1k}$, giving
$\mbox{Im}(\Gamma_{\!d}^{11}) = - \mbox{Im}(\Gamma_{\!d}^{12}) 
= {1 \over 2} \sin\!\theta \, \sin\!\phi$, and
\beq \label{Bx}
B(x) = {1 \over 2\,(x-1)^2} \left[1 + x + {2\,x \ln x \over (1-x) } \right] \; .
\eeq
Introducing $M_{\!\scriptscriptstyle \tilde{d}}^2 = 
(M_{\!\scriptscriptstyle \tilde{d}-}^2 + M_{\!\scriptscriptstyle \tilde{d}+}^2)/2$
and using the identity relation $x \,B(x) = B(1/x)$, the gluino contribution 
becomes the often-quoted
\beq \label{G02}
\left({d_{\scriptscriptstyle d} \over e}\right)_{\!\!\tilde{g}} =
-{2 \alpha_s \over 3 \pi} \;
{M_{\!\scriptscriptstyle \tilde{g}} \over M^2_{\!\scriptscriptstyle \tilde{d}} } \; 
{\cal Q}_{\tilde{d}}\; 
\mbox{Im}(\delta^{\scriptscriptstyle D}_{\!\scriptscriptstyle 1\!1}) \;
F \!\left( {M_{\!\scriptscriptstyle \tilde{g}}^2 \over M^2_{\!\scriptscriptstyle \tilde{d}}} \right) \; ,
\eeq
where $\delta_{\!\scriptscriptstyle 1\!1}^{\scriptscriptstyle D}$ is 
${\cal M}_{\!\scriptscriptstyle RL}^2 / M_{\!\scriptscriptstyle \tilde{d}}^2$ 
(with ${\cal M}_{\!\scriptscriptstyle RL}^2$ restricted to the $\tilde{d}$ family)
and
\beq \label{Fx}
F(x) = {1 \over (1-x)^3} \left[ { 1+ 5 x \over 2} +
{(2+x) x \ln x \over (1-x)} \right] \; .
\eeq

The EDM expression above is, in fact, the same as that of the MSSM case,
except that $\delta_{\!\scriptscriptstyle 1\!1}^{\scriptscriptstyle D}$,
or equivalently ${\cal M}_{\!\scriptscriptstyle RL}^2$, has now an extra 
RPV part. From the general result given in Eq.(\ref{RL}), we have, for 
the $\tilde{d}$ squark,
\beq \label{delta-D}
\delta_{\!\scriptscriptstyle 1\!1}^{\scriptscriptstyle D} 
 M^2_{\!\scriptscriptstyle \tilde{d}} =
\left[ A_d -  \mu_{\scriptscriptstyle 0}^* \, \tan\!\beta \right]\,m_{d}\;
+ \frac{\sqrt{2}\, M_{\!\scriptscriptstyle W} \cos\!\beta}
{g_{\scriptscriptstyle 2} } \,
\delta\! A^{\!{\scriptscriptstyle D}}_{\scriptscriptstyle 11}
- \frac{\sqrt{2}\, M_{\!\scriptscriptstyle W} \sin\!\beta}
{g_{\scriptscriptstyle 2} } \,
(\, \mu_i^*\lambda^{\!\prime}_{i\scriptscriptstyle 11}\, ) \; .
\eeq
Note that the $\mu_i^*\lambda^{\!\prime}_{i\scriptscriptstyle 11}$ term
does contain nontrivial CP violating phases and gives RPV contributions to
neutron EDM. Though the above analysis neglects interfamily mixings, it is 
clear from Fig.~1 that the $LR$ squark mixing 
$\delta_{\!\scriptscriptstyle 1\!1}^{\scriptscriptstyle D}$ is what is directly
responsible for the EDM contribution. Including interfamily mixings would
complicate the mass eigenstate analysis but not modify the EDM result in
any substantial way. 

\subsection{Neutralinolike Contributions}
The contributions from the (electroweak) neutral gaugino loop, as shown in 
Fig.~1, are expected to be small, due to the much smaller gauge couplings.
Apart from the neutral gaugino loops, there may be other neutralinolike 
loop contributions. In the MSSM, one has to consider possible contributions
from the Higgsino parts and the gaugino-Higgsino mixing parts. Part of such contributions involves no $LR$ squark mixing. The latter feature compensates
for the smaller Yukawa-type couplings involved\cite{KiOs}. Without R parity, 
the gauginos and Higgsinos mix with the leptons. We have seven neutral fermions, 
all among which give rise to a neutralinolike loop contribution. 

The neutral fermion loop contribution to $u$ and $d$ quark EDMs is given by
\beq
\left({d_{\scriptscriptstyle f} \over e} \right)_{\!\!\chi^{0}} = 
{\alpha_{\!\mbox{\tiny em}} \over 4 \pi \,\sin\!^2\theta_{\!\scriptscriptstyle W}} \;
{\cal Q}_{\!\tilde{f}} \;  
\sum_{\scriptscriptstyle \tilde{f}\mp} 
\sum_{n=1}^{7} \,\mbox{Im}({\cal N}_{\!fn\mp}) \;
{{M}_{\!\scriptscriptstyle \chi^0_n} \over M_{\!\scriptscriptstyle \tilde{f}\mp}^2} \;
B\!\left({{M}_{\!\scriptscriptstyle \chi^0_{n}}^2 \over M_{\!\scriptscriptstyle \tilde{f}\mp}^2} \right) \; ,
\eeq 
where 
\beqa
{\cal N}_{\! fn-}&=&
\left[-\sqrt{2} \left\{ \tan\!\theta_{\!\scriptscriptstyle W} \,
({\cal Q}_{\!f} -T_{3f})\, \mbox{\boldmath $X$}_{\!\!1n}
+ T_{3f}\, \mbox{\boldmath $X$}_{\!\!2n} \right\}{\cal D}^{*}_{\!f11} - 
{y_{\!\scriptscriptstyle f} \over g_{\scriptscriptstyle 2} } \, 
\mbox{\boldmath $X$}_{\!\!bn} \, {\cal D}^{*}_{\!f21} \;
 - \delta_{b{\scriptscriptstyle 4}}\,
{\lambda_{k{\scriptscriptstyle 1\!1}}^{\!\prime} \over g_{\scriptscriptstyle 2} }\; 
\mbox{\boldmath $X$}_{\!\!(k+4)n} \, {\cal D}^{*}_{d21}  \,
\right] 
\nonumber \\
&& \hspace{10mm} \cdot\;
\left[ \sqrt{2} \, \tan\!\theta_{\!\scriptscriptstyle W} \,
{\cal Q}_{\!f} \, \mbox{\boldmath $X$}_{\!\!1n} \, {\cal D}_{\!f21} 
- {y_{\!\scriptscriptstyle f} \over g_{\scriptscriptstyle 2} } \, 
\mbox{\boldmath $X$}_{\!\!bn} \, {\cal D}_{\!f11} \,
- \delta_{b{\scriptscriptstyle 4}}\,
{\lambda_{h{\scriptscriptstyle 1\!1}}^{\!\prime} \over g_{\scriptscriptstyle 2} }\,\, 
\mbox{\boldmath $X$}_{\!\!(h+4)n} \, {\cal D}_{d11}  \,
\right] \; ,
\nonumber \\
{\cal N}_{\! fn+}&=&
\left[-\sqrt{2} \left\{ \tan\!\theta_{\!\scriptscriptstyle W} \,
({\cal Q}_{\!f} -T_{3f})\, \mbox{\boldmath $X$}_{\!\!1n}
+ T_{3f}\, \mbox{\boldmath $X$}_{\!\!2n} \right\}{\cal D}^{*}_{\!f12} - 
{y_{\!\scriptscriptstyle f} \over g_{\scriptscriptstyle 2} }  \, 
\mbox{\boldmath $X$}_{\!\!bn} \, {\cal D}^{*}_{\!f22} \;
 - \delta_{b{\scriptscriptstyle 4}}\,
{\lambda_{k{\scriptscriptstyle 1\!1}}^{\!\prime} \over g_{\scriptscriptstyle 2} }\; 
\mbox{\boldmath $X$}_{\!\!(k+4)n} \, {\cal D}^{*}_{d22}  \,
\right] \;
\nonumber \\
&& \hspace{10mm} \cdot\;
\left[ \sqrt{2} \, \tan\!\theta_{\!\scriptscriptstyle W} \,
{\cal Q}_{\!f} \, \mbox{\boldmath $X$}_{\!\!1n} \, {\cal D}_{\!f22} 
- {y_{\!\scriptscriptstyle f} \over g_{\scriptscriptstyle 2} } \, 
\mbox{\boldmath $X$}_{\!\!bn} \, {\cal D}_{\!f12} \,
- \delta_{b{\scriptscriptstyle 4}}\,
{\lambda_{h{\scriptscriptstyle 1\!1}}^{\!\prime} \over g_{\scriptscriptstyle 2} }\,\, 
\mbox{\boldmath $X$}_{\!\!(h+4)n} \, {\cal D}_{d12}  \,
\right] \; ,
\eeqa
with $b=3\,(4)$ for $T_{3f}={1\over 2}\,(-{1 \over 2})$ [{\it i.e.,} for $f$ 
being the $u$ ($d$) quark], and $y_{\!\scriptscriptstyle f}$ the corresponding 
Yukawa coupling [{\it cf.} Eq.(\ref{yy})].
Recall, from Eq.(\ref{UD}), that ${\tilde{d}_\mp}$ 
denotes the two $\tilde{d}$ mass eigenstates and ${\cal D}_{d}$ 
the diagonalizing matrix; likewise,  ${\tilde{u}_\mp}$ and ${\cal D}_{u}$ 
are the corresponding notations for the $u$-quark case. Finally, the
$\mbox{\boldmath $X$}_{\!\!ij}$'s are elements of the matrix that diagonalizes 
${\cal M_{\scriptscriptstyle N}}$, as defined explicitly in the Appendix.
Note that the last term in each set of brackets of the ${\cal N}_{\! fn\mp}$ 
expressions is nonvanishing only for $f=d$, as indicated by the 
$\delta_{b{\scriptscriptstyle 4}}$ symbol. Similar to the case of the
gaugino contributions, RPV contributions here exist only for the $d$-quark 
EDM. We would like to emphasize that the formulae here represent
the full 1-loop result for our generic supersymmetric SM. 

Each expression for ${\cal N}_{\! fn\mp}$ above is a product of two
terms, from the two loop vertices involving the $L$- and $R$-handed quark
fields, respectively, as a chirality flip has to be imposed within the loop.
For each vertex, the three terms (in each set of brackets) correspond to the
gauge, quark Yukawa, and RPV $\lambda^{\!\prime}$ couplings, with the last
existing only for the $d$-quark case. With two gauge coupling vertices, we have
the gaugino diagram. Diagrams with a gauge and a 
$y_{\!\scriptscriptstyle f}$ coupling are shown explicitly in Fig.~2
for the $d$ quark, in which $y_{\!\scriptscriptstyle d}$ is indicated
by $\lambda^{\!\prime}_{{\scriptscriptstyle 0}kk}$, {\it i.e.,} with the 
notation used in our superpotential and a general, unspecified, 
quark family index (~$y_{\!\scriptscriptstyle d} \equiv
 \lambda^{\!\prime}_{{\scriptscriptstyle 01\!1}}$ in our notation~). Note
that the diagrams require no $LR$ mixing on the squark line. This
is the important MSSM contribution, the chargino counterpart of which
typically receives the major attention. The latter is numerically a bit
larger. The type of diagrams has no RPV contribution.

There is, however, a RPV analogue to Fig.~2. With the notation as given, this
is obviously given by replacing $\lambda^{\!\prime}_{{\scriptscriptstyle 0}kk}$
in the figure with a RPV $\lambda^{\!\prime}_{ikk}$. This is shown
explicitly in Fig.~3. From the latter figure, one can see the the first
order result would come from a ${l}_i^{\scriptscriptstyle 0}$-gaugino
mass mixing. However, under our formulation, the latter is vanishing
[see Eq.(\ref{mn}) in the Appendix]. Looking at the type of contributions from 
our EDM formula here, while a  $\lambda^{\!\prime}_{k{\scriptscriptstyle 1\!1}}$
coupling may not be small, it comes into the formula with a
$\mbox{\boldmath $X$}_{\!\!(k+4)n}$. For $n=1$--$4$, corresponding to the
heavy neutralino mass eigenstates, $\mbox{\boldmath $X$}_{\!\!(k+4)n}$
measures a RPV mixing in the neutral fermion masses  
$\cal{M_{\scriptscriptstyle N}}$. For $n=5$--$7$, at least one of the three
$\mbox{\boldmath $X$}_{\!\!(k+4)n}$'s is expected to be of order 1, but the 
corresponding (physical neutrino) mass eigenvalues 
${M}_{\!\scriptscriptstyle \chi^0_n}$'s
give a strong suppression factor in the resulting EDM contributions. There is
also a further suppression from the mixing factor of the gaugino part,
{\it e.g.,} $\mbox{\boldmath $X$}_{\!\!1(k+4)}$. Furthermore,
there is, potentially, a GIM cancellation among the seven mass eigenstates.
Our numerical results, however, do show that the type of 
contribution is generally important. We will discuss the issue more
carefully below, using the explicit example of its charginolike counterpart.

Last, we come to the diagrams with no gauge vertices, which we show in Fig.~4
for the $d$ quark, using our generic 
$\lambda^{\!\prime}_{{\scriptscriptstyle \za}jk}$ notation. Such a diagram
in the MSSM case is totally negligible due to the small Yukawa couplings
involved and the suppression from the $LR$ squark mixing required. In fact,
as shown in the figure, the minimal mass insertion needed for the diagram
corresponds to a Majorana mass among the 
${l}_{\scriptscriptstyle \za}^{\scriptscriptstyle 0}$'s, which is again
vanishing. The situation is similar to that of Fig.~3. The contribution is 
expected to be always dominated by the latter one. 

\subsection{Charginolike Contributions}
Next we come to the charged fermion counterpart. 
With similar notation as used in the neutral fermion loop formula above,
the charged fermion loop contributions to $u$- and $d$-quark EDMs can be written as
\beq \label{edmco}
\left({d_{\scriptscriptstyle f} \over e} \right)_{\!\!\chi^{\mbox{-}}} = 
{\alpha_{\!\mbox{\tiny em}} \over 4 \pi \,\sin\!^2\theta_{\!\scriptscriptstyle W}} \; 
\sum_{\scriptscriptstyle \tilde{f}'\mp} 
\sum_{n=1}^{5} \,\mbox{Im}({\cal C}_{\!fn\mp}) \;
{{M}_{\!\scriptscriptstyle \chi^{\mbox{-}}_n} \over 
M_{\!\scriptscriptstyle \tilde{f}'\mp}^2} \;
\left[ {\cal Q}_{\!\tilde{f}'} \; 
B\!\left({{M}_{\!\scriptscriptstyle \chi^{\mbox{-}}_{n}}^2 \over 
M_{\!\scriptscriptstyle \tilde{f}'\mp}^2} \right) 
+ ( {\cal Q}_{\!{f}} - {\cal Q}_{\!\tilde{f}'} ) \;
A\!\left({{M}_{\!\scriptscriptstyle \chi^{\mbox{-}}_{n}}^2 \over 
M_{\!\scriptscriptstyle \tilde{f}'\mp}^2} \right) 
\right] \; ,
\eeq 
for $f$ being $u$ ($d$) quark and $f'$ being $d$ ($u$), where
\beqa
{\cal C}_{un-} &=&  
{y_{\!\scriptscriptstyle u} \over g_{\scriptscriptstyle 2} } \,\, 
\mbox{\boldmath $V$}^{\!*}_{\!\!2n} \, {\cal D}_{d11} \;
\left( - \mbox{\boldmath $U$}_{\!1n} \,{\cal D}^{*}_{d11} 
+ {y_{\!\scriptscriptstyle d} \over g_{\scriptscriptstyle 2} }\,\, 
\mbox{\boldmath $U$}_{\!2n}\,  {\cal D}^{*}_{d21}
+ {\lambda^{\!\prime}_{k\scriptscriptstyle 1\!1} \over g_{\scriptscriptstyle 2} }\,\, 
\mbox{\boldmath $U$}_{\!(k+2)n}\,  {\cal D}^{*}_{d21} \right) \; ,
\nonumber \\
{\cal C}_{un+} &=&
 {y_{\!\scriptscriptstyle u} \over g_{\scriptscriptstyle 2} } \,\, 
\mbox{\boldmath $V$}^{\!*}_{\!\!2n} \, {\cal D}_{d12} \;
\left( - \mbox{\boldmath $U$}_{\!1n} \, {\cal D}^{*}_{d12} 
+ {y_{\!\scriptscriptstyle d} \over g_{\scriptscriptstyle 2} }\,\, 
\mbox{\boldmath $U$}_{\!2n}\,  {\cal D}^{*}_{d22}
+ {\lambda^{\!\prime}_{k\scriptscriptstyle 1\!1} \over g_{\scriptscriptstyle 2} }\,\, 
\mbox{\boldmath $U$}_{\!(k+2)n}\,  {\cal D}^{*}_{d22} \right) \; ,
\nonumber \\
{\cal C}_{dn-} &=& 
\left( {y_{\!\scriptscriptstyle d} \over g_{\scriptscriptstyle 2} }\,\, 
\mbox{\boldmath $U$}_{\!2n} 
+ {\lambda^{\!\prime}_{k\scriptscriptstyle 1\!1} \over g_{\scriptscriptstyle 2} }\,\, 
\mbox{\boldmath $U$}_{\!(k+2)n} \right)\! {\cal D}_{u11} \;
\left( - \mbox{\boldmath $V$}^{\!*}_{\!\!1n} \,{\cal D}^{*}_{u11} 
+ {y_{\!\scriptscriptstyle u} \over g_{\scriptscriptstyle 2} } \,
\mbox{\boldmath $V$}^{\!*}_{\!\!2n} \, {\cal D}^{*}_{u21} \right) \; ,
\nonumber \\
{\cal C}_{dn+} &=& 
\left( {y_{\!\scriptscriptstyle d} \over g_{\scriptscriptstyle 2} }\,\, 
\mbox{\boldmath $U$}_{\!2n} 
+ {\lambda^{\!\prime}_{k\scriptscriptstyle 1\!1} \over g_{\scriptscriptstyle 2} }\,\, 
\mbox{\boldmath $U$}_{\!(k+2)n} \right)\! {\cal D}_{u12} \;
\left( - \mbox{\boldmath $V$}^{\!*}_{\!\!1n} \, {\cal D}^{*}_{u12} 
+ {y_{\!\scriptscriptstyle u} \over g_{\scriptscriptstyle 2} } \,
 \mbox{\boldmath $V$}^{\!*}_{\!\!2n} \, {\cal D}^{*}_{u22} \right) \; ,
\nonumber \\
&& \mbox{\hspace*{2.5in}\small(only repeated index $i$ is to be summed)} \; ;
\label{Cnmp}
\eeqa
{\boldmath $V$} and {\boldmath $U$} are diagonalizing matrices of the charged 
fermion mass matrix $\cal M_{\scriptscriptstyle C}$ as defined by 
Eq.(\ref{umcv}) in the Appendix; and the function $A(x)$ is given by
\beq
A(x) = {1 \over 2 \, (1-x)^2} \left(3 - x + {2\ln x \over 1-x} \right) \; .
\eeq
The first and second parts of Eq.(\ref{edmco}) come from the cases 
in which the photon is emitted from the squark and fermion lines inside the loop,
respectively. 

The basic feature of the ${\cal C}_{\!fn\mp}$'s terms is similar to the 
previous case of ${\cal N}_{\!fn\mp}$. They do not give rise to a charged 
gaugino loop contribution, though, as $R$-handed quarks do not couple to 
$\tilde{W}^{\pm}$. Taking the available gauge coupling term within each
${\cal C}_{\!fn\mp}$ and a $y_{\!\scriptscriptstyle f}$ to form an 
admissible product, we have the contribution corresponding to a diagram in 
Fig.~5. This is the dominating MSSM contribution, besides the equally
competitive gluino one. Note again that no $LR$ squark mixing is required
the diagrams.

The diagram given in Fig.~5b has the RPV analogue, which requires a 
${l}_k^{\!\!\mbox{ -}}$--$\tilde{W}^{\scriptscriptstyle +}$
mass insertion for the first order result, as shown in in Fig.~6. 
This is the $SU(2)$ partner of Fig.~3, something we promise to discuss in 
more detail. Looking at the contributions from our EDM formula here, 
we have again a $\lambda^{\!\prime}_{k{\scriptscriptstyle 1\!1}}$ coming
in with a $\mbox{\boldmath $U$}_{\!(k+2)n}$, as versus the
$\mbox{\boldmath $X$}_{\!\!(k+4)n}$ in the neutral case above. 
For $n=1$--$2$, corresponding to the heavy chargino mass eigenstates, 
$\mbox{\boldmath $U$}_{\!(k+2)n}$ measures a RPV mass mixing; it is of order 1
for $n=k+2$, but then ${M}_{\!\scriptscriptstyle \chi^{\mbox{-}}_n}$ 
is the small $m_k$, roughly the charged lepton mass. When one sums over
for $n=1$--$5$, it is easy to see that the result is proportional to
the imaginary part of
\beq
\sum_{n=1}^{5} \, \mbox{\boldmath $V$}^{\!*}_{\!\!1n}\,
{M}_{\!\scriptscriptstyle \chi^{\mbox{-}}_n} 
 \mbox{\boldmath $U$}_{\!(k+2)n} \; 
F_{\!\!\scriptscriptstyle B\!A}\!\!
\left({M}_{\!\scriptscriptstyle \chi^{\mbox{-}}_{n}}^2\right) \;
\lambda^{\!\prime}_{k\scriptscriptstyle 1\!1} \; ,
\eeq
where $F_{\!\!\scriptscriptstyle B\!A}\!\!
\left({M}_{\!\scriptscriptstyle \chi^{\mbox{-}}_{n}}^2\right)$
denotes a function on ${M}_{\!\scriptscriptstyle \chi^{\mbox{-}}_{n}}^2$ 
corresponding to the large brackets in Eq.(\ref{edmco}) with functions $B$ and $A$.
If the $F_{\!\!\scriptscriptstyle B\!A}$ could be factored out, together with
$\lambda^{\!\prime}_{k\scriptscriptstyle 1\!1}$, 
what is left inside the summation is nothing other than the 
${l}_k^{\!\!\mbox{ -}}$--$\tilde{W}^{\scriptscriptstyle +}$ 
mass term, which is zero. This is a GIM-like cancellation mechanism, violated only
to the extent that the loop integrals involved as given by the $B$ and $A$ 
functions are not universal. Our numerical calculations, however, show that
the cancellation is very substantially violated, for generic values of 
chargino masses. Let us then look at the 
contribution from an individual mass eigenstate more closely. To get an analytical
approximation, we used the perturbative results for $\mbox{\boldmath $U$}_{\!(k+2)n}$
given in the Appendix. For the $n=1$ and $2$ parts in the above sum, we have 
\beq \label{exp} \nonumber
\mbox{\boldmath $V$}^{\!*}_{\!\!1a}\,
{\mu_k^*} \, R_{\!\scriptscriptstyle R_{2a}} F_{\!\!\scriptscriptstyle B\!A}\!\!
\left( M_{c{\scriptscriptstyle a}}^2 \right) \,
\lambda^{\!\prime}_{k\scriptscriptstyle 1\!1} \; ,
\eeq
with basically the same source of RPV complex phase as the gluino case, namely,
from $\mbox{Im}(\mu_k^* \lambda^{\!\prime}_{k\scriptscriptstyle 1\!1})$, 
except that it is one value of $k$ for each diagram here.  
Note that the explicit $M_{c{\scriptscriptstyle a}}$ factors are canceled.
The $n=k+2$ part is largely suppressed due to a small
mass eigenvalue and the very small RPV $R$-handed mixing given by 
$\mbox{\boldmath $V$}^{\!*}_{\!\!1(k+2)}$ as shown in the Appendix. Note
that $\mbox{\boldmath $V$}^{\!*}_{\!\!1a} \simeq R_{\!\scriptscriptstyle R_{1a}}^*$,
and  $R_{\!\scriptscriptstyle R}$ is a $2\times 2$ unitary matrix of order 1
elements. The expressions give an idea about
the strength of the RPV contributions. In the limit of degenerate
charginos, {\it i.e.,} $M_{c{\scriptscriptstyle 1}}=M_{c{\scriptscriptstyle 2}}$,
the $F_{\!\!\scriptscriptstyle B\!A}$ function factors out of the sum of the
$n=1$ and $2$ parts and the GIM-like cancellation is clearly illustrated, simply
from the unitarity of $R_{\!\scriptscriptstyle R}$. In fact, our
numerical calculations give a cancellation up to 1 part in $10^4$ in
such a situation if we enforce the condition. However, that requires a every
substantial complex phase for $\mu_{\scriptscriptstyle 0}$, hence of not the most
phenomenological interest. Another interesting point on the parameter space appears
when the $R_{\!\scriptscriptstyle R}$ matrix is diagonal, at which the contribution
also goes away, as is obvious from the above expression. This happens at the
point where the condition 
$M_{\scriptscriptstyle 2}^*\, \sin\!\zb = \mu_{\scriptscriptstyle 0} \,\cos\!\zb$
is satisfied. This feature will be confirmed by our numerical results presented
below.

The rest of the charginolike contributions each has at least one 
$y_{\!\scriptscriptstyle f}$ coupling and no gauge coupling vertex. 
Again, a $LR$ squark mixing is required. However, of the first order 
electroweak-state diagrams, as depicted in Fig.~7, each has an admissible 
$\mu_{\scriptscriptstyle \za}^*$ mass insertion on the fermion line. This
looks a bit different from the neutral case above. Apart from an admissible 
RPV down-squark mixing for the case of the $u$ quark, a RPV diagram again has the 
RPV parameter combination  $\mu_k^* \lambda^{\!\prime}_{k\scriptscriptstyle 1\!1}$ 
(no sum). Except for large-$\mu_i$ case, there is no 
reason to expect the contribution to be of any significance compared to 
the other major contributions.

Our EDM formulae neglect interfamily mixings. However, general, unspecified, 
family indices are use in our figures to make the flavor structure transparent.
We note that in the case of the contributions depicted by Figs.~4 and 7, off-diagonal 
$\mu_i^*\lambda^{\!\prime}_{ijk}$-type RPV mixings could play a role to let 
the higher family squarks mediating  more important contributions to $u$- and
$d$-quark EDMs.

\subsection{The Quark-Scalar Loop Contributions in Brief}
There are superpartners to the diagrams in
Fig.~7, {\it i.e.,} with quarks and charged scalars running inside the
loop and a $B_\za$ coupling on the scalar line. These diagrams are shown
explicitly in Fig.~8. Here, the $\za=0$
case gives the MSSM charged Higgs boson contribution. With the $B_i$'s, we have
the RPV contributions. Analytically, we have the formula
\beq
\left({d_{\scriptscriptstyle f} \over e} \right)_{\!\!\phi^{\mbox{-}}} =
 {\alpha_{\!\mbox{\tiny em}} \over 4 \pi \,\sin\!^2\theta_{\!\scriptscriptstyle W}} \;
\sum_{m}^{\prime}  
\sum_{n=1}^{3} \,
\mbox{Im}({\cal C}^{\!\scriptscriptstyle L}_{\scriptscriptstyle inm} \,
{\cal C}^{\!\scriptscriptstyle R^*}_{\scriptscriptstyle inm})_{\!f} \;
{{M}_{\!\scriptscriptstyle f'_n} \over M_{\!\scriptscriptstyle \tilde{\ell}_m}^2} \;
\left[  ({\cal Q}_{\!f} -{\cal Q}_{\!f'}) \;
B\!\left({{M}_{\!\scriptscriptstyle f'_n}^2 \over M_{\!\scriptscriptstyle \tilde{\ell}_m}^2} \right) 
+ {\cal Q}_{\!f'} \;
A\!\left({{M}_{\!\scriptscriptstyle f'_n}^2 \over M_{\!\scriptscriptstyle \tilde{\ell}_m}^2} \right)  \right] \; ,
\eeq 
where, for $f=u$,
\beqa
{\cal C}^{\!\scriptscriptstyle R}_{\scriptscriptstyle inm}
&=& \frac{y_{\!\scriptscriptstyle d}}{g_{\scriptscriptstyle 2}} \,
  {\cal D}^{l^*}_{1m} 
 +  \frac{\lambda_{i1k}^{\!\prime *}}{g_{\scriptscriptstyle 2}}
 \, {\cal D}^{l^*}_{(i+2)m} \;,
\nonumber \\
{\cal C}^{\!\scriptscriptstyle L}_{\scriptscriptstyle inm} 
&=&  \frac{y_{\!\scriptscriptstyle u}}{g_{\scriptscriptstyle 2}} \,
 {\cal D}^{l^*}_{2m}  \; ,
\eeqa
and, for $f=d$,
\beqa
{\cal C}^{\!\scriptscriptstyle R}_{\scriptscriptstyle inm}
&=&  \frac{y_{\!\scriptscriptstyle u}}{g_{\scriptscriptstyle 2}} \,
  {\cal D}^{l}_{1m}  \;,
\nonumber \\
{\cal C}^{\!\scriptscriptstyle L}_{\scriptscriptstyle inm} 
&=&  \frac{y_{\!\scriptscriptstyle d}}{g_{\scriptscriptstyle 2}} \,
 {\cal D}^{l}_{2m} 
 +  \frac{\lambda_{ij{\scriptscriptstyle 1}}^{\!\prime}}{g_{\scriptscriptstyle 2}}
 \, {\cal D}^{l}_{(i+2)m}  \; ,
\eeqa
with the $\sum_{m}^{\prime}$ denoting a sum over (seven) nonzero mass eigenstates 
of the charged scalar; {\it i.e.,} the unphysical Goldstone mode is dropped 
from the sum, ${\cal D}^{l}$ being the diagonalization matrix, {\it i.e.,}
${\cal D}^{l\dag}  {\cal M}_{\!\scriptscriptstyle E}^2 \, {\cal D}^{l}
= \mbox{diag}\{\,M_{\!\scriptscriptstyle \tilde{\ell}_m}^2, m=1\,\mbox{--}\,8\,\}$.
 
With similar notation as used in the charged scalar loop formula above,
the  neutral scalar loop contributions to the EDMs can be written as
\beq 
\left({d_{\scriptscriptstyle f} \over e} \right)_{\!\!\phi^{0}} = 
{\alpha_{\!\mbox{\tiny em}} \over 4 \pi \,\sin\!^2\theta_{\!\scriptscriptstyle W}} \; 
\sum_{m}^{\prime} 
\sum_{n=1}^{3} \,
\mbox{Im}({\cal N}_{\!\scriptscriptstyle inm}^{\!\scriptscriptstyle L}\,
 {\cal N}_{\!\scriptscriptstyle inm}^{\!\scriptscriptstyle R^*}) \;
{{M}_{\!\scriptscriptstyle f_n} \over 
M_{\!\scriptscriptstyle S_m}^2} \; {\cal Q}_{\!{f}}  \;
A\!\left({ M_{\!\scriptscriptstyle f_{n}}^2 \over 
M_{\!\scriptscriptstyle S_m}^2} \right)  \; ,
\eeq 
where, for $f=u$,
\begin{eqnarray}
 {\cal N}^{\!\scriptscriptstyle R}_{\scriptscriptstyle inm}
& =&  - \frac{y_{\!\scriptscriptstyle u}}{g_{\scriptscriptstyle 2}} \,
{1 \over \sqrt{2}} \,
 [ {\cal D}^{s}_{\!1m} - i \, {\cal D}^{s}_{\!6m} ]\; ,
\nonumber \\
 {\cal N}^{\!\scriptscriptstyle L}_{\scriptscriptstyle inm}
& = &  
- \frac{y_{\!\scriptscriptstyle u}}{g_{\scriptscriptstyle 2}} \,
{1 \over \sqrt{2}} \,
[ {\cal D}^{s}_{\!1m} + i \, {\cal D}^{s}_{\!6m} ]\; ,
\end{eqnarray}
and, for $f=d$,
\begin{eqnarray}
 {\cal N}^{\!\scriptscriptstyle R}_{\scriptscriptstyle inm}
& =&  - \frac{y_{\!\scriptscriptstyle d}}{g_{\scriptscriptstyle 2}} \,
{1 \over \sqrt{2}} \,
[ {\cal D}^s_{\!2m} - i \, {\cal D}^s_{\!7m} ]
- \frac{\lambda_{h1k}^{\!\prime*}}{g_{\scriptscriptstyle 2}} \,
{1 \over \sqrt{2}} \,
[ {\cal D}^s_{\!(h+2)m} - i \, {\cal D}^s_{\!(h+7)m} ] \; ,
\nonumber \\
 {\cal N}^{\!\scriptscriptstyle L}_{\scriptscriptstyle inm}
& = &  
- \frac{y_{\!\scriptscriptstyle d}}{g_{\scriptscriptstyle 2}} \,
{1 \over \sqrt{2}} \,
[ {\cal D}^{s}_{\!(i+2)m} + i \, {\cal D}^{s}_{\!(i+7)m} ]
- {\lambda_{hk1}^{\!\prime} \over g_{\scriptscriptstyle 2} } \, 
{1 \over \sqrt{2}} \,
[ {\cal D}^{s}_{\!(h+2)m} + i \, {\cal D}^{s}_{\!(h+7)m} ] \; ,
\end{eqnarray}
with again the unphysical Goldstone mode to be dropped from the sum over
scalar mass eigenstates (nine nonzero), and the obvious notation
$({\cal D}^{s})^{\!\scriptscriptstyle T}  {\cal M}_{\!\scriptscriptstyle S}^2 \, 
{\cal D}^{s} = \mbox{diag}\{\, M_{\!\scriptscriptstyle S_m}^2, 
m=1\,\mbox{--}\,10\,\}$, ${\cal D}^{s}$ being an orthogonal matrix. 

Note that the formulae given in this section, like those in the rest of the paper,
have neglected CKM mixings among the quarks. However, unlike the previous cases,
the mixings play an important role in the quark-scalar loop contributions. 
This is a result of the fact that the EDM contributions have a fermion, a
quark in this case, mass dependence, and the mass hierarchy among the quarks.
For instance, the top quark loop is expected to give a dominating contribution.
Generalizing the formulae to include the mixings is straightforward. The formulae
as they are given in this subsection do give a basic illustration of the
major analytical structure of the type of contributions. We will only discuss the
features briefly here.

Diagrams depicting the neutral scalar contributions are like superpartners of 
the type of diagram given in Fig.~3. This is shown in Fig.~8 for both the $u$- and 
$d$-quark cases. Such a diagram could be interpreted as requiring a 
Majorana-like scalar mass insertion. The latter is first
considered in Ref.\cite{GH}, under the name of Majorana masses from the 
``sneutrinos". However, looking at it from the present framework, these diagrams
should be interpreted as having Majorana-like mass insertions among the
${l}_{\!\scriptscriptstyle \za}^{\scriptscriptstyle 0}$'s (and 
${h}_{\!\scriptscriptstyle u}^{\!\scriptscriptstyle 0}$ for the case of the $u$
quark).

The quark-scalar loops given by the above formulae are not quite considered in
the case of the MSSM, as they would be suppressed by the very small Yukawa couplings.
The RPV analogue, however, admits much more general flavor structure, as 
illustrated in the diagrams in Figs.~8 and 9. For instance, a top quark with
order-1 Yukawa coupling could be contributing to the $d$-quark EDM through
a charged scalar loop. Contributions discussed in this subsection depend
on a much larger number of parameters, through the scalar mass matrices
[{\it cf.} Eqs.(\ref{ME})--(\ref{lastsc})]. The extra parameters are related 
directly to Higgs physics. The important RPV $B_i$ parameters here have a strong 
connection with the $\mu_i$'s\cite{app}. Furthermore, the more general flavor 
structure involved means that the EDM contributions involve 
$\lambda_{ijk}^{\!\prime}$ couplings that would also have important roles to 
play in the related processes of $b \to s\, \gamma$ and $b \to d\, \gamma$. 
We will leave all these issues for later studies.

\section{Result from Numerical Calculations}
We now discuss the results we obtained by a careful numerical implementation of our
EDM formulae discussed above, with explicit numerical diagonalization of all
the mass matrices involved. Note that the quark-scalar loop contribution is not 
included in the numerical study. Hence, we are concentrating on the SUSY contributions in the presence of RPV couplings, to be compared directly with
the MSSM results.

As discussed in the previous section, the imaginary part of the combination 
$\mu_i^*\lambda^{\!\prime}_{i\scriptscriptstyle 1\!1}$ is what is central to the
RPV 1-loop EDM contributions. To simplify the discussion, we single out 
$\mu_{\scriptscriptstyle 3}$ and $\lambda^{\!\prime}_{\scriptscriptstyle 31\!1}$
and put the corresponding parameters for $i=1$ and $2$ to be essentially zero
\footnote{In the actual calculation, tiny but nonzero values are used to avoid
possible problems of numerical manipulations.}. 
This applies to all the numerical results discussed here. 

We give in Table~1 some detailed numerical results for four illustrative sample 
cases. To first focus on the RPV contributions which we are particularly interested 
in, cases A and B in the table have all complex phases in the R-parity conserving 
part switched off. For the RPV parameters $\mu_{\scriptscriptstyle 3}$ and
$\lambda^{\!\prime}_{\scriptscriptstyle 31\!1}$, we take the former being real 
and put a $\pi/4$ complex phase into the latter. However, we want to emphasize
again that only the phase of the combination
$\mu_{\scriptscriptstyle 3}^*\lambda^{\!\prime}_{\scriptscriptstyle 31\!1}$
matters. For instance, we have checked explicitly that shifting the overall phase,
or a part of it into $\mu_{\scriptscriptstyle 3}$ instead, gives exactly the 
same results. The difference bewteen case A and case B is only in the value of 
$\mu_{\scriptscriptstyle 3}$ chosen. Case A has small $\mu_{\scriptscriptstyle 3}$,
at $10^{-3}\,\mbox{GeV}$. This is the small-$\mu_i$ scenario, corresponding to
a sub-eV mass for $\nu_\tau$ as suggested, but far from mandated, by the result 
from the Super-Kamiokande (super-K) experiment\cite{sK} 
\footnote{Note that 
$\mu_{\scriptscriptstyle 3} \,\cos\!\beta \sim 10^{-4}\,\mbox{GeV}$
is actually expected, allowing a larger $\mu_{\scriptscriptstyle 3}$
in the case of large $\tan\!\beta$ (see, for example, Ref.\cite{ok}).
Our choice of $\mu_{\scriptscriptstyle 3}$ value here is, admittedly, quite
pushing the limit.  It is mainly for illustrative purposes. 
}. 
In contrast, case B has $\mu_{\scriptscriptstyle 3}=1\,\mbox{GeV}$. 
Note that imposing the $18.2\,\mbox{MeV}$ experimental 
bound\cite{aleph} for the mass of $\nu_\tau$ still admits a relatively
large $\mu_{\scriptscriptstyle 3}$, especially for a large $\tan\!\zb$.
Reading from the results in Ref.\cite{k}, the bound is $\sim 7\,\mbox{GeV}$ 
at $\tan\!\zb=2$ and $\sim 300\,\mbox{GeV}$ at $\tan\!\zb=45$. 
As for $\lambda^{\!\prime}_{\scriptscriptstyle 31\!1}$, the best bound on the 
(from $\tau \to \pi \nu$) is around $0.05\mbox{--}0.1$\cite{lambda}.  Hence, case B 
is still within the admissible region of RPV parameter
space, though beyond the more popular small-$\mu_i$ scenario. Moreover, we 
have illustrated in Ref.\cite{as4} that the 
$\lambda^{\!\prime}_{\scriptscriptstyle 31\!1}$
bound is not strengthened even when the above-mentioned stringently interpreted 
neutrino mass bound from super-K is imposed on the 1-loop neutrino mass contribution
obtainable from the parameter. We therefore use the same 
$\lambda^{\!\prime}_{\scriptscriptstyle 31\!1}$ magnitude of $0.05$ for both, 
and actually all, cases in the table.

Apart from the overall EDM results, we are interested in understanding the relative 
values of the different pieces of the contributions discussed in the previous 
section. Hence, we show, in the table, the contribution from individual pieces,
identified by the couplings of the two loop-vertices given inside the first column. 
They can be matched easily with terms in the formulae. The corresponding Feynman diagrams are marked whenever they are available inside the next column. 
Column 3 of the table indicates whether a $LR$ squark mixing is involved 
in the contribution, and if the RPV part of the latter is available and needed 
for a RPV contribution to come in. The individual contribution that is vanishing 
under cases A and B corresponds to a diagram which does not admit a RPV contribution. 
This is the case for most entries for the $u$-quark EDM and the entries for the
charginolike loop contributions to the $d$-quark EDM not involving a 
$\lambda^{\!\prime}$ coupling loop vertex. Note that a charginolike loop 
for the $d$ quark involves $LR$ mixing of $\tilde{u}$ which has no RPV part.
As emphasized in Ref.\cite{as4}, the RPV contributions are far more 
prominent for the $d$ quark part. Again, 
$\lambda^{\!\prime}_{\scriptscriptstyle 31\!1}$ is the only nonzero 
$\lambda^{\!\prime}$ coupling introduced, though we use a more general $\lambda^{\!\prime}$ notation within the table to remind the readers of
the more general flavor structure admissible, as illustrated in the previous
section. We also recall from the discussions there that a
$\lambda^{\!\prime}_{ijk}$ coupling always comes into the EDM picture
accompanied by a RPV fermion mixing matrix element proportional to $\mu_i^*$.
The EDM results are basically in direct proportion to 
$\mu_i^* \, \lambda^{\!\prime}_{i\scriptscriptstyle 1\!1}$
over the whole region of parameter space studied.
Hence, comparing cases A and B, we see an {\it exact} factor of $10^{-3}$ 
suppression in case A for contributions with one $\lambda^{\!\prime}$ loop vertex 
or involving a RPV $LR$ mixing, except the very small
$y_{\!\scriptscriptstyle d}^2$ term. 

The overall EDM numbers of case A are still
below the present experimental bound. This is smaller than
a naive estimated result, as given for the gluino contribution by Eq.(\ref{G02}), 
for example, due to some unavoidable partial cancellation. For the gluino
result, in particular, the cancellation is between the contributions from
the two squark mass eigenstates, which would in fact be exact when the latter states
are degenerate. This slightly weakens the EDM bound of the previously estimated 
$\mbox{Im}(\mu_i^*\lambda^{\!\prime}_{i\scriptscriptstyle 1\!1})<10^{-6}\,\mbox{GeV}$
result given in Ref.\cite{as4}. However, we see here that with a
unification-type relationship imposed on the gaugino masses, we are likely to have
a chargino contribution larger than the gluino one. Moreover, the values of
the other SUSY parameters chosen here can be pushed to increase the EDM
number, as discussed below. As for case B, the numbers are beyond the experimental
bound, indicating that the RPV parameters, or the involved complex phase,
would have to be further constrained.

Case C of Table~1 gives a different scenario. Here, we allow complex phases,
and hence EDM contributions, of the MSSM part. The RPV parameters are set real.
This ensures no RPV EDM contribution from diagrams involving $LR$ squark mixings.
The interesting point to note here is that there is indeed nonzero RPV
contributions from diagrams with RPV loop vertices. In particular, the RPV
chargino contribution to $d$-quark EDM is comparable in magnitude
to its R-parity conserving counterpart. Hence, the existence of nonzero RPV
parameters, even real, would change the EDM story of the MSSM. In other words,
in the presence of a complex $\mu_{\scriptscriptstyle 0}$, the neutron EDM bound 
actually constrains the magnitude of the the combination of RPV parameters
given by $\mu_i^*\lambda^{\!\prime}_{i\scriptscriptstyle 1\!1}$, instead of
just the phase of it. This is an important
feature that has not been pointed out before. The result of the case serves, 
otherwise, as a check against other neutron EDM studies of the MSSM.

Finally, we illustrate in case D a specific situation with both R-parity 
conserving and violating phases. Here, we pick a case of large $\tan\!\beta$,
for which the larger $\mu_{\scriptscriptstyle 3}$ is still within the limit
admitted by the stringent interpretation of the super-K bound\cite{ok}.
The negative phase chosen for $\lambda^{\!\prime}_{\scriptscriptstyle 31\!1}$ 
gives a cancellation between the two contributions to the imaginary part of
$LR$ $d$ squark mixing. This directly resulted in suppression of the EDM
contributions involving the latter, such as the $d$-quark gluino loop. The
very interesting point to note here is that the same cancellation results
in the chargino and neutralino loop
\footnote{A chargino (neutralino) contribution means exactly the one with a 
physical chargino (neutralino) running in the loop. Hence, it is only a part
of the fermion mass eigenstate sum in the analytical formulae given for
the (total)  charginolike and neutralinolike loop contributions.
The latter, as illustrated in the previus section, are generally dominated 
by the chargino and neutralino mass eigenstates. }
, 
as explicitly illustrated by the entries of the terms with loop vertex 
couplings given by $g\cdot y_{\!\scriptscriptstyle d}$ and 
$g\cdot \lambda^{\!\prime}_{i\scriptscriptstyle 1\!1}$ 
in both cases. Essentially, when we have a small imaginary part for
$\mu_{\scriptscriptstyle \za}^*\lambda^{\!\prime}_{\scriptscriptstyle \za 1\!1}$
(recall that $\za=0$ to $3$, $\lambda^{\!\prime}_{\scriptscriptstyle 0 1\!1}
\equiv  y_{\!\scriptscriptstyle d}$), 
{\it e.g.,} an internal cancellation among the summed terms, the corresponding
$d$ quark EDM contributions, involving the suppressed $LR$ mixing
[~{\it cf.} first line of Eq.(\ref{RL})~] or otherwise, are all suppressed.
The chargino and neutralino contributions actually reflect some
proportionality to the ($F$-term) $LR$ mixings in a way similar to the gluino
contribution. Note, however, that the $A$-term phase contributes to the gluino
diagram while leaving the charginolike and neutralinolike diagrams not quite 
affected, and hence spoils the simultaneous cancellation achieved in the case~D 
results here. Besides, as the $u$-quark EDM results do not have a complete 
matching analogue between the R-parity conserving and violating contributions, the 
simultaneous cancellation mechanism does not exist there. In the current case, 
the $u$-quark parts are suppressed due to large $\tan\!\beta$.

The above sample cases illustrate some of the interesting features about
the RPV contributions to neutron EDM. Next, we discuss how the EDM result
varies with some basic parameters. Fig.~10 shows a logarithmic plot of 
(the magnitude of) the RPV neutron EDM result for $\mu_{\!\scriptscriptstyle 0}$ 
value between $\pm2000\,\mbox{GeV}$, with the other parameters set at the same
values as case~A in Table 1. Note the the small
$|\mu_{\!\scriptscriptstyle 0}|$ ($\lsim 70\,\mbox{GeV}$) region has been ruled out
for having too light a chargino mass\cite{k}. As noted in the previous section,
the gluino and chargino contributions compete with one another with one of them
being dominating in a region of the parameter space. The dip on the $C$~line
for the charginolike contribution corresponds to the case of vanishing
$R$-handed mixing among the charginos, {\it i.e.,} when the condition
$M_{\scriptscriptstyle 2}\,\sin\!\zb + \mu_{\scriptscriptstyle 0}\,\cos\!\zb=0$
is satisfied, as noted above, where the contribution essentially vanishes. Note that 
in the region to the left of the point, the contribution becomes negative. 
The $G$~line is horizontal, as the gluino diagram result has no
dependence on $\mu_{\!\scriptscriptstyle 0}$.

Next, we check the $\tan\!\beta$ dependence and compare with the MSSM result.
In Fig.~11, we give again a plot of the (total) EDM result for the same set of input
parameters as in case~A of Table~1, while varying $\tan\!\beta$. This is the line
marked as ``RPV only". The numerical result confirms our earlier discussion 
that there is not much sensitivity in $\tan\!\beta$ [{\it cf.} expression 
(\ref{exp})], as indicated by the flatness of the line. This is in good contrast
to the MSSM result, also based on the same set of input parameters except with
nonvanishing phases for $A$ and $\mu_{\!\scriptscriptstyle 0}$. The third line,
marked by ``GSSM", gives the total result obtained from our formulae given above
with the same nonvanishing phases for $A$ and $\mu_{\!\scriptscriptstyle 0}$,
as well as $\lambda^{\!\prime}_{\scriptscriptstyle 31\!1}$. Note that the GSSM
result here is more than the sum of the two other lines, due to the presence of
RPV contribution even in the limit of a real 
$\lambda^{\!\prime}_{\scriptscriptstyle 31\!1}$, as discussed for Case~C of 
Table~1 above. The  $\tan\!\beta$ dependence, or the lack of, illustrated in the
figure is quite generic, in a wide region of the parameter space.
 
The dip in the GSSM line of Fig.~11 represents a point where the overall
contribution is small as a result of the cancellation among the different pieces,
a feature that has attracted a lot of attention lately in the case of the MSSM (see, 
for example, Ref.\cite{FO} and Refs.\cite{IN2,NGKO}). The MSSM line in the figure
has a dip only beyond the range of the $\tan\!\beta$ value shown. As also pointed
out in a little bit different setting (with vanishing $A$ phase) for Case~D of 
Table~1 above, the new RPV contribution modifies the overall picture and provides a
possible new cancellation mechanism. The cancellation feature is better illustrated
in Fig.~12, in which the variation against the $A$-term phase is shown, for the small
and large $\tan\!\beta$ cases. Again our GSSM result is compared with that
of the MSSM. One can see that the presence of the RPV contribution shifts the
cancellation points substantially.
Finally, we show variations of the result as a function of the gaugino mass 
parameters here represented by $M_{\scriptscriptstyle 2}$. This is given in Fig.~13
with the four lines marked and correspond to the four cases of Fig.~12, each
with the $A$-term phase set at the position of the dip as given in the latter figure.

\section{conclusion}
We have given explicit formulae and detailed discussions on the full 1-loop 
contributions to quark EDMs for the generic supersymmetric SM (without R parity). The 
extra, RPV, contributions are interesting additions to the R-parity conserving part.
We have given results from an exact numerical study, illustrating various
novel aspects. Our formulation emphasizes the universal structure of the R-parity
conserving and violating parts, which also is illustrated itself well in the results.
The experimental bound on the neutron EDM is hence established as an important
source of constraints on the model parameter space, including the RPV
part. The 1-loop RPV contributions always involve the particular combination
of RPV parameters given by 
$\mu_i^*\lambda^{\!\prime}_{i\scriptscriptstyle 1\!1}$, 
with little sensitivity to the value of $\tan\!\beta$. So far as the RPV
parameters are concerned, this combination is well constrained by the EDM bound.
This applies not only to the complex phase, or imaginary part of, the combination. 
Real $\mu_i^*\lambda^{\!\prime}_{i\scriptscriptstyle 1\!1}$
contribute in the presence of complex phases in the chargino and neutralino
mass entries. Studies with either $\mu_i$- or  $\lambda^{\!\prime}$-
type couplings assumed to be zero miss the class of very interesting
phenomenological features of SUSY without R parity.

\acknowledgements 
O.K. is in great debt to Kingman Cheung, a collaborator on a parallel work
on $\mu \to e\,\gamma$, for discussions on common issues of the two studies.
Y.Y.K. wishes to thank M. Kobayashi and H.Y. Cheng for their hospitality.
His work was in part supported by the National Science Council of R.O.C.
under Grant No. NSC-89-2811-M-001-0053.

\newpage
\appendix
\section{ notes on the fermion masses} 
Under the SVP, the (color-singlet) charged fermion mass 
matrix is given by the simple form 
\beq \label{mc}
{\mathcal{M}_{\scriptscriptstyle C}} =
 \left(
{\begin{array}{ccccc}
{M_{\!\scriptscriptstyle 2}} &  {\sqrt 2} \, M_{\!\scriptscriptstyle W} \cos\!\beta
& 0 & 0 & 0 \\
  {\sqrt 2} \, M_{\!\scriptscriptstyle W} \sin\!\beta
&  {{ \mu}_{\scriptscriptstyle 0}} & {{ \mu}_{\scriptscriptstyle 1}} 
&  {{ \mu}_{\scriptscriptstyle 2}} & {{ \mu}_{\scriptscriptstyle 3}} \\
0 &  0 & \;\; {{m}_{\scriptscriptstyle 1}} \;\;  & 0 & 0 \\
0 & 0 & 0 & \;\; {{m}_{\scriptscriptstyle 2}} \;\; & 0 \\
0 & 0 & 0 & 0 & \;\; {{m}_{\scriptscriptstyle 3}} \;\; 
\end{array}}
\right)  \; ,
\eeq
with explicit bases for right-handed and left-handed states given by
$(-i\tilde{W}^{\scriptscriptstyle +},
\tilde{h}_{\!\scriptscriptstyle u}^{\!\scriptscriptstyle +},
{l}_{\scriptscriptstyle 1}^{\!\scriptscriptstyle +},
{l}_{\scriptscriptstyle 2}^{\!\scriptscriptstyle +},
{l}_{\scriptscriptstyle 3}^{\!\scriptscriptstyle +}\,)$
and
$(-i\tilde{W}^{\!\!\mbox{ -}},
{l}_{\scriptscriptstyle 0}^{\!\!\mbox{ -}},
{l}_{\scriptscriptstyle 1}^{\!\!\mbox{ -}},
{l}_{\scriptscriptstyle 2}^{\!\!\mbox{ -}},
{l}_{\scriptscriptstyle 3}^{\!\!\mbox{ -}}\,)$,
respectively. Here, we allow ${M_{\!\scriptscriptstyle 2}}$ and all four
$\mu_{\scriptscriptstyle \za}$ parameters to be complex, though the $m_i$'s
are restricted to be real, for reasons that will become clear below.
Obviously, each  $\mu_i$ parameter here characterizes 
directly the deviation of the ${l}_i^{\!\!\mbox{ -}}$ from the corresponding 
physical charged lepton  ($\ell_i = e$, $\mu$, and $\tau$), {\it i.e.,} light 
mass eigenstates. For any set of other parameter inputs, the ${m}_i$'s can 
then be determined, through a simple numerical procedure, to guarantee that 
the correct mass eigenvalues of  $m_e$, $m_\mu$, and $m_\tau$  are obtained 
--- an issue first addressed and solved in Ref.\cite{k}. The latter issue
is especially important when $\mu_i$'s not substantially smaller than
${ \mu}_{\scriptscriptstyle 0}$ are considered. Such an odd scenario is
not definitely ruled out\cite{k}. However, for the more popular small-$\mu_i$ scenario,
we have ${l}_i^{\!\!\mbox{ -}} \approx \ell_i^{\!\!\mbox{ -}}$, and deviations
on $m_i$'s from the (real) $\ell_i$ masses are negligible.
Note that the deviation of ${l}_i^{\scriptscriptstyle +}$ from 
$\ell_i^{\scriptscriptstyle +}$ is quite negligible in any case.

We introduce unitary matrices {\boldmath $V$} and {\boldmath $U$} diagonalizing 
the $R$- and $L$-handed states with
\beq
 \label{umcv}
 \mbox{\boldmath $V$}^\dag {\mathcal{M}_{\scriptscriptstyle C}} \,
\mbox{\boldmath $U$} = \mbox{diag} 
\{ {M}_{\!\scriptscriptstyle \chi^{\mbox{-}}_n} \} \equiv 
\mbox{diag} 
\{ {M}_{c {\scriptscriptstyle 1}}, {M}_{c {\scriptscriptstyle 2}},
m_e, m_\mu, m_\tau \}\; .
\eeq			
Here, the mass eigenvalues ${M}_{\!\scriptscriptstyle \chi^{\mbox{-}}_n}$ 
with $n=1$ and $2$, {\it i.e.,}  
${M}_{c {\scriptscriptstyle 1}}$ and ${M}_{c {\scriptscriptstyle 2}}$, 
are the chargino masses. Consider further
\beq
R_{\!\scriptscriptstyle R}^{\dagger} 
  \left( \begin{array}{cc}
  M_{\!\scriptscriptstyle 2}  & {\sqrt 2} \, M_{\!\scriptscriptstyle W} \cos\!\beta \\
 {\sqrt 2} \, M_{\!\scriptscriptstyle W} \sin\!\beta
& {\mu}_{\scriptscriptstyle 0}
  \end{array} \right) R_{\!\scriptscriptstyle L}
  \; = \; {\rm diag} 
  \{ M_{c{\scriptscriptstyle 1}}^{o}, M_{c{\scriptscriptstyle 2}}^{o} \} \;\;\; 
\label{Mcidef}
\eeq
with $M_{c{\scriptscriptstyle 1}}^{o}$ and $M_{c{\scriptscriptstyle 2}}^{o}$ 
being the chargino masses in the ${\mu}_i = 0$ limit. One can then write
the diagonalizing matrices in the block form
\begin{eqnarray} \label{VU}
\mbox{\boldmath $V$} =
  \left( \begin{array}{cc}
  R_{{\!\scriptscriptstyle R}}
  & -R_{{\!\scriptscriptstyle R}} \,V^{\dagger} \\
  V             & I_{\!\scriptscriptstyle 3 \times 3}                   
  \end{array} \right)  
\qquad \qquad \qquad
\mbox{\boldmath $U$} =
  \left( \begin{array}{cc}
  R_{{\!\scriptscriptstyle L}}
  & -R_{{\!\scriptscriptstyle L}}\, U^{\dagger} \\
  U              & I_{\!\scriptscriptstyle 3 \times 3}                   
  \end{array} \right) \; .
\end{eqnarray}
For ${\mu}_i \ll M_{c{\scriptscriptstyle a}}^{o}$ ($a=1$ and $2$),
a block perturbative diagonalization can be performed directly on the matrix 
${\mathcal{M}_{\scriptscriptstyle C}}$ to obtain the following simple result 
\beqa
\mbox{\boldmath $U$}_{\!(i+2)1} & \simeq& 
\frac{  \mu_i^* }{M_{c{\scriptscriptstyle 1}}} \,
R_{{\!\scriptscriptstyle R}_{21}}\; ,
\nonumber \\
\mbox{\boldmath $U$}_{\!(i+2)2} & \simeq& 
\frac{ \mu_i^*  }{M_{c{\scriptscriptstyle 2}}} \,
R_{{\!\scriptscriptstyle R}_{22}}\; ,
\nonumber \\
\mbox{\boldmath $V$}_{\!\!(i+2)a} & \simeq& 
\frac{m_i }{M_{c_{\scriptscriptstyle a}}} \,
\mbox{\boldmath $U$}_{\!(i+2)a}
\qquad\qquad (a=1 \,\mbox{and } 2)\; .
\label{VUele}
\eeqa
Elements in the $R_{{\!\scriptscriptstyle R}}$ and $R_{{\!\scriptscriptstyle L}}$
matrices are all expected to be of order 1. The above expressions illustrate the
magnitudes of given matrix elements involving RPV mixings, as well as their 
dependence on the RPV parameters. 
We note that the $L$-handed mixings are roughly measured
by the ratio of a $\mu_i$ to the chargino mass scale, while $R$-handed mixings are
further suppressed by a charged lepton to chargino mass ratio, hence quite 
negligible under most consideration.

Note that the notation here is different from that given in Ref.\cite{k} and many others
in the literature. More explicitly, we have $R$- and $L$-handed mass eigenstates given by
$(\, \chi_{{\!\scriptscriptstyle +}n} \,)= \mbox{\boldmath $V$}^{\scriptscriptstyle T} \,
[ -i\tilde{W}^{\scriptscriptstyle +},
\tilde{h}_{\!\scriptscriptstyle u}^{\!\scriptscriptstyle +},
{l}_{\scriptscriptstyle 1}^{\!\scriptscriptstyle +},
{l}_{\scriptscriptstyle 2}^{\!\scriptscriptstyle +},
{l}_{\scriptscriptstyle 3}^{\!\scriptscriptstyle +}\,]^{\scriptscriptstyle T}$
and
$(\, \chi_{\!\!\!\mbox{ -}n} \,)= \mbox{\boldmath $U$}^{\dag} \,
[-i\tilde{W}^{\!\!\mbox{ -}},
{l}_{\scriptscriptstyle 0}^{\!\!\mbox{ -}},
{l}_{\scriptscriptstyle 1}^{\!\!\mbox{ -}},
{l}_{\scriptscriptstyle 2}^{\!\!\mbox{ -}},
{l}_{\scriptscriptstyle 3}^{\!\!\mbox{ -}}\,]^{\scriptscriptstyle T}$,
which form the five Dirac fermions 
\[
{\chi}_n^{\!\!\mbox{ -}} = \left( \begin{array}{c}
\chi_{\!\!\!\mbox{ -}n} \\
\chi_{{\!\scriptscriptstyle +}n}^{\dag} 
\end{array} \right)\;.
\]
 The $\mbox{\boldmath $U$}_{\!(i+2)a}$ elements as given
above show no obvious dependence on $\tan\!\beta$, though some nontrivial
dependence is expected through the $R_{{\!\scriptscriptstyle R}_{2a}}$ elements.
Our exact numerical result also indicates a weak sensitivity on the
$\tan\!\beta$ value here. On the other hand, we have, from Eqs.(\ref{VU}) and 
(\ref{VUele}), 
\begin{eqnarray}
\mbox{\boldmath $U$}_{\!a(i+2)} &=&
- \mu_i \cdot \left[ R_{{\!\scriptscriptstyle L}} \, 
\left({\rm diag} 
  \{ M_{c{\scriptscriptstyle 1}}^{o}, M_{c{\scriptscriptstyle 2}}^{o} \}\right)^{-1} \, R_{{\!\scriptscriptstyle R}}^{\dag} \right]_{a2} \; ,
\nonumber \\
\mbox{\boldmath $V$}_{\!a(i+2)} &=&
- \mu_i \cdot \left[ R_{{\!\scriptscriptstyle R}} \, 
\left( {\rm diag} 
  \{ M_{c{\scriptscriptstyle 1}}^{o}, M_{c{\scriptscriptstyle 2}}^{o} \}\right)^{-2} \, R_{{\!\scriptscriptstyle R}}^{\dag} \right]_{a2} \; ,
\eeqa
giving the result
\beqa
\mbox{\boldmath $U$}_{\!1(i+2)} &=&
\frac{ \mu_i \, \sqrt{2}\, M_{\!\scriptscriptstyle W} \cos\!\beta}
{M_{\!\scriptscriptstyle 0}^2} \; ,
\nonumber \\
\mbox{\boldmath $U$}_{\!2(i+2)} &=&
 -\frac{\mu_i \, M_{\!\scriptscriptstyle 2}}{M_{\!\scriptscriptstyle 0}^2} \; ,
\nonumber \\
\mbox{\boldmath $V$}_{\!1(i+2)} &=& \mu_i \, m_i \;
 \frac{\sqrt{2}\, \, M_{\!\scriptscriptstyle W} ( M_{\!\scriptscriptstyle 2}^*\,\sin\!\beta
   + {\mu}_{\scriptscriptstyle 0} \,\cos\!\beta )}{|M_{\!\scriptscriptstyle 0}|^4} \; ,
\nonumber \\
\mbox{\boldmath $V$}_{\!2(i+2)} &=& - \mu_i \, m_i \;
          \frac{( |M_{\!\scriptscriptstyle 2}|^2
  + 2 \, M_{\!\scriptscriptstyle W}^2 \, {\cos}^2\!\!\>{\beta} )}
{|M_{\!\scriptscriptstyle 0}|^4} \; ,
    \end{eqnarray}
where
\[
M_{\!\scriptscriptstyle 0}^2 \; \equiv \;
{\mu}_{\scriptscriptstyle 0} \, M_{\scriptscriptstyle 2} - 
M_{\!\scriptscriptstyle W}^2 \, \sin\! 2\beta 
=  M_{c{\scriptscriptstyle 1}}^{o} \, M_{c{\scriptscriptstyle 2}}^{o} \; .
\]
These RPV elements correspond to those given in Ref.\cite{k} (with  
all complex phases neglected), where the crucial $\cos\!\beta$ dependence of
the nonstandard $Z^0$-boson couplings of the physical charged leptons 
($\ell_i \equiv \chi_{\scriptscriptstyle i+2}$) through
$\mbox{\boldmath $U$}_{\!1(i+2)}$ is emphasized. The difference between
$\mbox{\boldmath $U$}_{\!(i+2)a}$ and $\mbox{\boldmath $U$}_{\!a(i+2)}$ is
hence very important.

The $7\times 7$ Majorana mass matrix for the neutral fermion can be written as
\small\begin{equation}
\label{mn}
\cal{M_{\scriptscriptstyle N}} = 
\left (\begin{array}{ccccccc}
M_{\!\scriptscriptstyle 1} & 0 
& M_{\!\scriptscriptstyle Z} \sin\!\theta_{\!\scriptscriptstyle W}  \sin\!\zb
& - M_{\!\scriptscriptstyle Z} \sin\!\theta_{\!\scriptscriptstyle W}  \cos\!\zb
 & 0 & 0 & 0  \\
0   & M_{\!\scriptscriptstyle 2} 
& -  M_{\!\scriptscriptstyle Z} \cos\!\theta_{\!\scriptscriptstyle W}  \sin\!\zb
&  M_{\!\scriptscriptstyle Z} \cos\!\theta_{\!\scriptscriptstyle W}  \cos\!\zb
 & 0 & 0 & 0  \\
M_{\!\scriptscriptstyle Z} \sin\!\theta_{\!\scriptscriptstyle W}  \sin\!\zb
 & - M_{\!\scriptscriptstyle Z} \cos\!\theta_{\!\scriptscriptstyle W}  \sin\!\zb
& 0  & -\mu_{\scriptscriptstyle 0}  & -\mu_{\scriptscriptstyle 1} 
& -\mu_{\scriptscriptstyle 2} & -\mu_{\scriptscriptstyle 3} \\
- M_{\!\scriptscriptstyle Z} \sin\!\theta_{\!\scriptscriptstyle W}  \cos\!\zb
& M_{\!\scriptscriptstyle Z} \cos\!\theta_{\!\scriptscriptstyle W}  \cos\!\zb
 & -\mu_{\scriptscriptstyle 0} &   0 & 0 & 0 & 0\\
  0  & 0  & -{\mu}_{\scriptscriptstyle 1}   & 0 & 0 & 0 & 0 \\
  0  & 0  & -{\mu}_{\scriptscriptstyle 2}   & 0 & 0 & 0 & 0 \\
  0  & 0  & -{\mu}_{\scriptscriptstyle 3}   & 0 & 0 & 0 & 0 
  \end{array} \right) \; ,
\eeq \normalsize
with explicit basis  $(-i\tilde{B}, -i\tilde{W}, 
\tilde{h}_{\!\scriptscriptstyle u}^{\!\scriptscriptstyle 0}\,, 
\tilde{h}_{\!\scriptscriptstyle d}^{\!\scriptscriptstyle 0}\,, 
{l}_{\scriptscriptstyle 1}^{\scriptscriptstyle 0}\,,
{l}_{\scriptscriptstyle 2}^{\scriptscriptstyle 0}\,,
{l}_{\scriptscriptstyle 3}^{\scriptscriptstyle 0}\,) $. 
Note that $\tilde{h}_{\!\scriptscriptstyle d}^{\!\scriptscriptstyle 0}
\equiv {l}_{\scriptscriptstyle 0}^{\scriptscriptstyle 0}$, and,
from the above discussion of the charged fermions, 
we have, for small $\mu_i$'s,
$(\, {l}_{\scriptscriptstyle 1}^{\scriptscriptstyle 0},
{l}_{\scriptscriptstyle 2}^{\scriptscriptstyle 0},
{l}_{\scriptscriptstyle 3}^{\scriptscriptstyle 0} \, ) 
\approx (\nu_{\scriptscriptstyle e},\nu_{\scriptscriptstyle \mu}, 
\nu_{\scriptscriptstyle \tau})$.
The symmetric, but generally non-Hermitian, matrix can be diagonalized 
by using unitary matrix {\boldmath $X$} such that
\beq
\mbox{\boldmath $X$}^{\!\scriptscriptstyle  T} 
{\cal M_{\scriptscriptstyle N}}\mbox{\boldmath $X$} =
\mbox{diag} \{ {M}_{\!\scriptscriptstyle \chi^0_{n}} \} \; .
\eeq
Again, the first part of the mass eigenvalues, 
${M}_{\!\scriptscriptstyle \chi^0_{n}} $ for $n = 1$--$4$ here, gives
the heavy states, {\it i.e.,} neutralinos. The last part, 
${M}_{\!\scriptscriptstyle \chi^0_{n}} $ for $n = 5$--$7$ are hence
physical neutrino masses at the tree level.

\newpage

\noindent
{\bf Table caption :}\\[.2in]
Table~1 --- Numerical 1-loop neutron EDM results from SUSY without R parity, 
for four illustrative cases.  All EDM numbers are in $e\,\mbox{cm}$. 
Note that the quark EDM numbers are direct output from the numerical program
applying our quark dipole formulae; while the neutron EDM numbers are from
the valence quark model formula, as given in Eq.(\ref{vqm}). 
R-parity violating parameters not given are taken as essentially 
zero. All parameters are taken real except those with complex phases explicitly
listed in each case, where the real number(s) listed then give the magnitude(s).
Parameter $A$ here means a common $A_u$ and $A_d$. Only 
$M_{\scriptscriptstyle 2}$ is shown for the gaugino masses; the others are
fixed by the unification relationship. Explicitly, we use 
$M_{\scriptscriptstyle 1}=0.5\,M_{\scriptscriptstyle 2}$ and
$M_{\scriptscriptstyle 3}=3.5\,M_{\scriptscriptstyle 2}$. The first column under 
``EDM Results" gives the couplings of the loop vertices involved. A $g$ indicates
either one of the electroweak gauge couplings, while a $\lambda^{\!\prime}$ 
coupling means one with the appropriate admissible flavor indices. In
the explicit results of the four cases, the latter is always a
$\lambda^{\!\prime}_{\scriptscriptstyle 31\!1}$. The second column gives
the reference Feynman diagram figures, when available. The third column indicates
whether the particular contribution involves a $LR$ squark mixing. In the case
that the mixing is involved and a R-parity violating (RPV) one is involved in
generating a RPV EDM contribution, it is marked with ``RPV". 

\bigskip
\bigskip

\noindent
{\bf Figure captions :}\\[.2in]
Fig.~1 --- Diagram for $d$-quark EDM from the gaugino loop.\\[.1in]
Fig.~2 --- Diagrams with neutral gaugino-Higgsino mixing  
for $d$-quark EDM.\\[.1in]
Fig.~3 --- R-parity violating neutralinolike loop diagrams for $d$-quark EDM.
Naive electroweak-state analysis suggests that such a diagram is proportional
to a vanishing ${l}_k^{\scriptscriptstyle 0}$-gaugino mass mixing.\\[.1in]
Fig.~4 --- Diagram for $d$-quark EDM suggesting involvement of Majorana 
masses among the ${l}_{\scriptscriptstyle \za}^{\scriptscriptstyle 0}$ or
``neutrinos".\\[.1in]
Fig.~5 --- Diagrams for $u$- and $d$-quark EDMs with charged gaugino-Higgsino mixing. \\[.1in]
Fig.~6 --- R-parity violating charginolike loop diagram for $d$-quark EDM.
Naive electroweak-state analysis suggests that the diagram is proportional
to the vanishing ${l}_k^{\!\!\mbox{ -}}$--$\tilde{W}^{\scriptscriptstyle +}$
mass term.\\[.1in]
Fig.~7 --- Diagrams for $u$- and $d$-quark EDMs with  a $\mu_{\scriptscriptstyle \za}$ mass insertion.\\[.1in]
Fig.~8 --- Diagrams for $u$- and $d$-quark EDMs with a $B_{\za}$ scalar mass insertion.\\[.1in]
Fig.~9 --- Diagrams with a Majorana-like scalar mass insertion 
for $u$- and $d$-quark EDMs. For the $u$-quark case, the direct Majorana-like
$h_{\scriptscriptstyle u}$ mass insertion is explicitly shown. For the $d$-quark 
case, the corresponding direct $h_{\scriptscriptstyle d}$ mass insertion is
obvious, for $\za=\zb=0$; for $\za$ and/or $\zb$ nonzero, the naive direct result
from the diagram would vanish, due to the vanishing VEVs.\\[.1in]
Fig.~10 --- Logarithmic plot of (the magnitude of) the RPV neutron EDM result for 
$\mu_{\!\scriptscriptstyle 0}$ value between $\pm2000\,\mbox{GeV}$,  
with the other parameters set at the same values as case~A in Table 1. 
The lines marked by $G$, $C$, $N$, and ``Total" give the complete
gluino, charginolike, neutralinolike, and total ({\it i.e.,} sum of the three) contributions, respectively. Note that the values of the $N$
contributions and those of the $C$ line for 
$\mu_{\!\scriptscriptstyle 0}<-900\,\mbox{GeV}$ are negative. \\[.1in]
Fig.~11 --- Logarithmic plot of (the magnitude of) the neutron EDM result verses 
$\tan\!\zb$. We show here the MSSM result, our general result with the RPV phase
only, and the generic result  with complex phases of 
both kinds. In particular, the $A$ and $\mu_{\scriptscriptstyle 0}$ 
phases are chosen as $7^o$ and $0.1^o$ respectively, for the MSSM line. They are 
zero for the RPV-only line, with which we have a phase of ${\pi\over 4}$ for 
$\lambda^{\!\prime}_{\scriptscriptstyle 31\!1}$. All the given nonzero values
are used for the three phases for the generic result (from our complete formulae) 
marked by ``GSSM". Again, the other unspecified input parameters are the same as 
for case~A of Table~1.\\[.1in]
Fig.~12 ---  Logarithmic plot of (the magnitude of) the neutron EDM result versus
$\theta_{\!\!\scriptscriptstyle A} $ (the complex phase for the $A$ parameter). 
The four lines shown are characterized by the $\tan\!\zb$ values (~3 or 50~) used 
and whether it is for our GSSM result (again with a phase of ${\pi\over 4}$ for 
$\lambda^{\!\prime}_{\scriptscriptstyle 31\!1}$) --- marked by $G$; or the result
for MSSM --- marked by $M$. Again, the $\lambda^{\!\prime}_{\scriptscriptstyle 31\!1}$
phase is set at ${\pi\over 4}$ for the $G$ lines, and the
$\mu_{\scriptscriptstyle 0}$ phase at $0.1^o$ for all; 
the other unspecified input parameters are the same as for case~A of Table~1. \\[.1in]
Fig.~13 ---  Logarithmic plot of (the magnitude of) the neutron EDM result verses
$M_{\scriptscriptstyle 2}$. The four lines correspond to the four cases of Fig.~12, 
each with $\theta_{\!\!\scriptscriptstyle A} $ set at the dip location, {\it i.e.,} 
$G$-$3$ for GSSM at $\tan\!\zb=3$ with $\theta_{\!\!\scriptscriptstyle A}  = 2^o$, 
$M$-$3$ for MSSM at $\tan\!\zb=3$ with $\theta_{\!\!\scriptscriptstyle A}  = -1^o$, 
$G$-$50$ for GSSM at $\tan\!\zb=50$ with $\theta_{\!\!\scriptscriptstyle A}  = 20^o$, 
$M$-$50$ for MSSM at $\tan\!\zb=50$ with $\theta_{\!\!\scriptscriptstyle A}  = 3^o$. 
All other unspecified input parameters are the same as for Fig.~12. \\[.1in]

\newpage
\thispagestyle{empty}
\vspace*{-.5in}
\begin{tabular}{|c|c|c|c|c|c|c|}\hline
\multicolumn{7}{|c|}{\framebox[3in][c]{\underline{\bf Choice of Parameters}}	} 	\\
\multicolumn{7}{|c|}{$\tilde{m}_{\scriptscriptstyle Q}=300\,\mbox{GeV}$,
$\tilde{m}_{\scriptscriptstyle u}= \tilde{m}_{\scriptscriptstyle d}=200\,\mbox{GeV}$, 
$A=M_{\scriptscriptstyle 2}=300\,\mbox{GeV}$,
 $\mu_{\scriptscriptstyle 0}=-300\,\mbox{GeV}$}	\\
\hline
\multicolumn{3}{|c|}{$\tan\!\zb$} 				
& $3$		& $3$		& $3$		& $50$\\
\multicolumn{3}{|c|}{$\mu_{\scriptscriptstyle 3}$}	
& $1\cdot 10^{-3}\,\mbox{GeV}$	& $1\,\mbox{GeV}$		& $1\,\mbox{GeV}$		& $5\cdot 10^{-3}\,\mbox{GeV}$	\\
\multicolumn{3}{|c|}{$\lambda^{\!\prime}_{\scriptscriptstyle 31\!1}$}	
&  $0.05 $	& $0.05 $	& $0.05 $	& $0.05 $	\\
\multicolumn{3}{|c|}{(complex phases)} 
& $\lambda^{\!\prime}_{\scriptscriptstyle 31\!1}$($\pi/4$) 
& $\lambda^{\!\prime}_{\scriptscriptstyle 31\!1}$($\pi/4$)
& $\mu_{\scriptscriptstyle 0}$($0.5^o$), $A$($10^o$)	
& $\mu_{\scriptscriptstyle 0}$($0.02^o$),  $\mu_{\scriptscriptstyle 3}$($-\pi/4$)
\\
\hline
\multicolumn{7}{|c|}{\framebox[3in][c]{\underline{\bf EDM RESULTS :-}} }	\\
couplings										& Fig.	& $LR$ mixing	
& 			Case A	 & Case B	& Case C	& Case D 	
\\
\hline\hline
\multicolumn{7}{l|}{\framebox[1.3in][c]{\underline{\bf \boldmath $d$ quark EDM}} }	\\
\multicolumn{7}{|l|}{\underline{ Gluino loop} : -}	\\
$\za_{\!\scriptscriptstyle s}$						& 1		& RPV
& $8.8\cdot 10^{-28}$	& $8.8\cdot 10^{-25}$	& -$3.9\cdot 10^{-26}$	& -$6.7\cdot 10^{-29}$ 	\\
\hline
\multicolumn{7}{|l|}{\underline{ Neutralino-like loop} : -}	\\
$g^2$											& 1		& RPV
& -$1.9\cdot 10^{-29}$	& -$1.9\cdot 10^{-26}$	& $8.3\cdot 10^{-28}$	& $2.7\cdot 10^{-30}$ 
\\
$g\cdot y_{\!\scriptscriptstyle d}$						& 2		& no			
	& $\sim 0$	& $\sim 0$		& -$1.6\cdot 10^{-27}$	& -$1.2\cdot 10^{-27}$
\\
$g\cdot \lambda^{\!\prime}_{i{\scriptscriptstyle 1\!1}}$		& 3		& no
& -$1.0\cdot 10^{-28}$	& -$1.0\cdot 10^{-25}$	& $1.1\cdot 10^{-27}$	& $1.1\cdot 10^{-27}$ 
\\
$y_{\!\scriptscriptstyle d}^2$						& 4		& RPV
& $9.7\cdot 10^{-37}$	& $9.7\cdot 10^{-34}$	& -$3.9\cdot 10^{-35}$	& -$2.6\cdot 10^{-33}$
\\
$y_{\!\scriptscriptstyle d}\cdot\lambda^{\!\prime}_{ijk}$
												& 4		& yes
& -$1.7\cdot 10^{-36}$	& -$1.7\cdot 10^{-33}$	& $8.5\cdot 10^{-35}$	& $2.5\cdot 10^{-33}$
\\
two $\lambda^{\!\prime}_{ijk}$						& 4		& yes
& -$2.1\cdot 10^{-39}$	& -$3.4\cdot 10^{-34}$	& $9.0\cdot 10^{-35}$	& -$8.6\cdot 10^{-37}$
\\	
\hline
\multicolumn{7}{|l|}{\underline{ Chargino-like loop} : -}	\\
$g\cdot y_{\!\scriptscriptstyle d}$						& 5		& no			
	& $\sim 0$ 	& $0$		& $2.5\cdot 10^{-26}$	& $1.7\cdot 10^{-26}$ 
\\
$g\cdot \lambda^{\!\prime}_{i{\scriptscriptstyle 1\!1}}$		& 6		& no
& $2.1\cdot 10^{-27}$	& $2.1\cdot 10^{-24}$	& -$1.3\cdot 10^{-26}$	& -$1.7\cdot 10^{-26}$ 
\\
$y_{\!\scriptscriptstyle u}\cdot y_{\!\scriptscriptstyle d}$	& 7		& yes			
	& $\sim 0$ 	& $0$		& -$2.7\cdot 10^{-34}$	& -$8.0\cdot 10^{-36}$
\\
$y_{\!\scriptscriptstyle u}\cdot\lambda^{\!\prime}_{ijk}$
											& 7		& yes
& -$2.1\cdot 10^{-37}$	& -$2.1\cdot 10^{-33}$	& $3.8\cdot 10^{-34}$	& $8.3\cdot 10^{-36}$
\\
\hline\hline
\multicolumn{7}{l|}{\framebox[1.3in][c]{\underline{\bf \boldmath $u$ quark EDM}} }	\\
\multicolumn{7}{|l|}{\underline{ Gluino loop} : -}	\\
$\za_{\!\scriptscriptstyle s}$						& 1		& yes
	& $0$		& $0$ 	& $4.5\cdot 10^{-26}$ & -$1.8\cdot 10^{-30}$
\\
\hline
\multicolumn{7}{|l|}{\underline{ Neutralinolike loop} : -}	\\
$g^2$											& 1		& yes
	& $\sim 0$ 	& $0$		& $2.6\cdot 10^{-27}$	& -$1.4\cdot 10^{-31}$
\\
$g\cdot y_{\!\scriptscriptstyle u}$					& 2		& no	
	& $\sim 0$ 	& $0$		& $2.1\cdot 10^{-28}$	& $5.3\cdot 10^{-31}$
\\
$y_{\!\scriptscriptstyle u}^2$						& 4		& yes
	& $\sim 0$ 	& $0$		& $1.3\cdot 10^{-37}$	& $4.0\cdot 10^{-41}$
\\
\hline
\multicolumn{7}{|l|}{\underline{ Charginolike loop} : -}	\\
$g\cdot y_{\!\scriptscriptstyle u}$						& 5		& no			
	& $\sim 0$ 	& $\sim 0$ 	& -$1.3\cdot 10^{-27}$	& -$3.2\cdot 10^{-30}$
\\
$y_{\!\scriptscriptstyle u}\cdot y_{\!\scriptscriptstyle d}$	& 7		& RPV			
& -$7.6\cdot 10^{-36}$	& -$7.6\cdot 10^{-33}$	& $3.2\cdot 10^{-34}$	& $6.4\cdot 10^{-34}$
\\
$y_{\!\scriptscriptstyle u}\cdot\lambda^{\!\prime}_{ijk}$
											& 7		& RPV
& $9.7\cdot 10^{-36}$	& $9.7\cdot 10^{-33}$	& $-5.1\cdot 10^{-34}$	& -$6.4\cdot 10^{-34}$
\\
\hline
\multicolumn{7}{l|}{\framebox[1.3in][c]{\underline{\bf Neutron EDM}} }	\\
\multicolumn{3}{|l}{\underline{ from Gluino loop} : } 
& $1.8\cdot 10^{-27}$	& $1.8\cdot 10^{-24}$	& -$1.0\cdot 10^{-25}$	& -$1.4\cdot 10^{-28}$
\\
\multicolumn{3}{|l}{\underline{ from Charginolike loop} : } 
& $4.3\cdot 10^{-27}$	& $4.3\cdot 10^{-24}$	& $2.5\cdot 10^{-26}$	& $2.4\cdot 10^{-28}$	
\\
\multicolumn{3}{|l}{\underline{ from Neutralinolike loop} : } 
& -$2.9\cdot 10^{-28}$	& -$2.9\cdot 10^{-25}$	& -$8.6\cdot 10^{-28}$	& -$2.0\cdot 10^{-29}$
\\
\multicolumn{3}{|l}{\underline{TOTAL} : }
& $5.8\cdot 10^{-27}$	& $5.8\cdot 10^{-24}$	& -$7.8\cdot 10^{-26}$	& $8.0\cdot 10^{-29}$
\\
\hline\hline
\end{tabular}

\eject

\begin{figure}
\vspace*{.5in}
\includegraphics{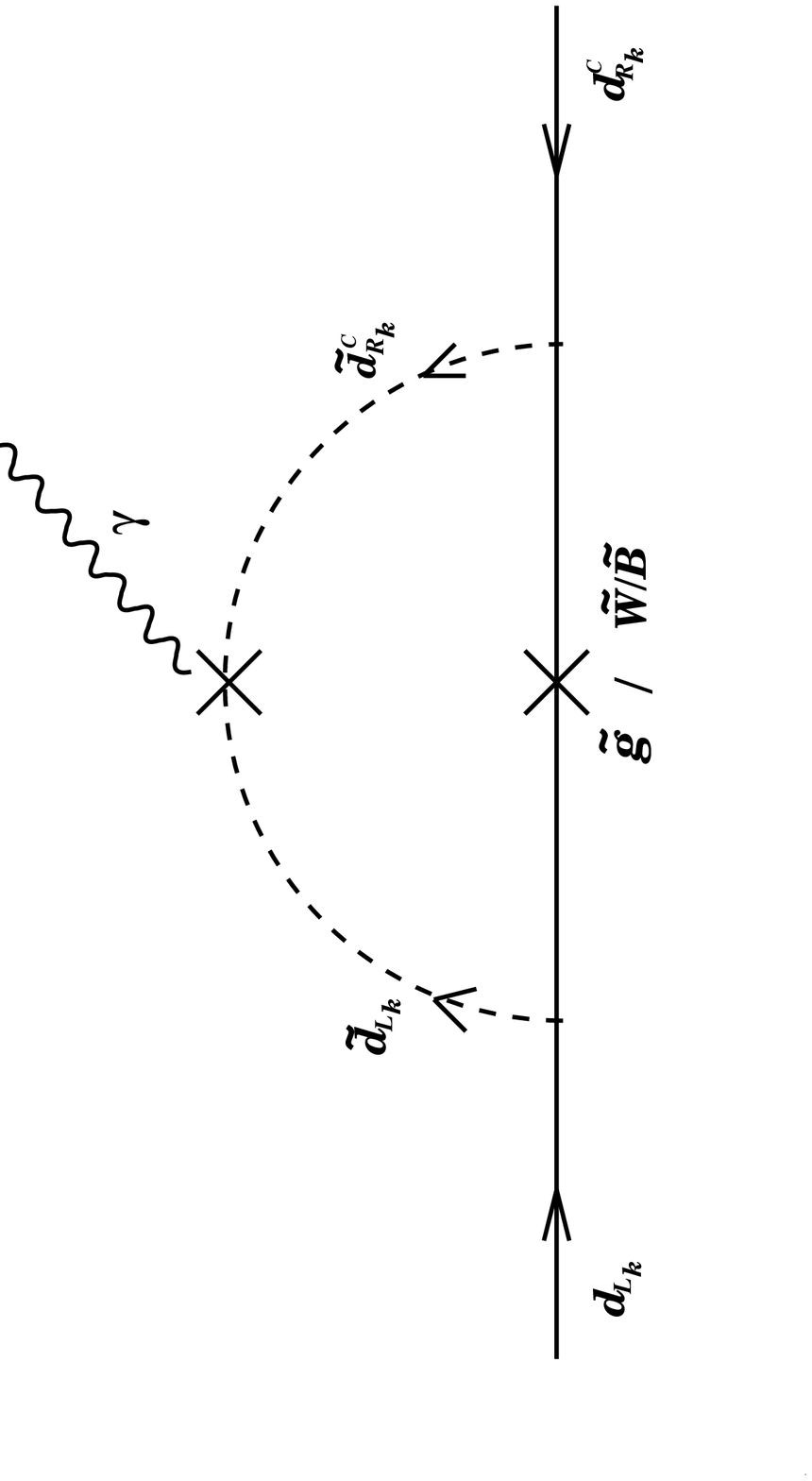}
\vspace*{3in}
\caption{Diagram for $d$-quark EDM from the gaugino loop.}
\end{figure}

\vspace*{1.in}

\eject

\begin{figure}
\vspace*{.5in}
\includegraphics{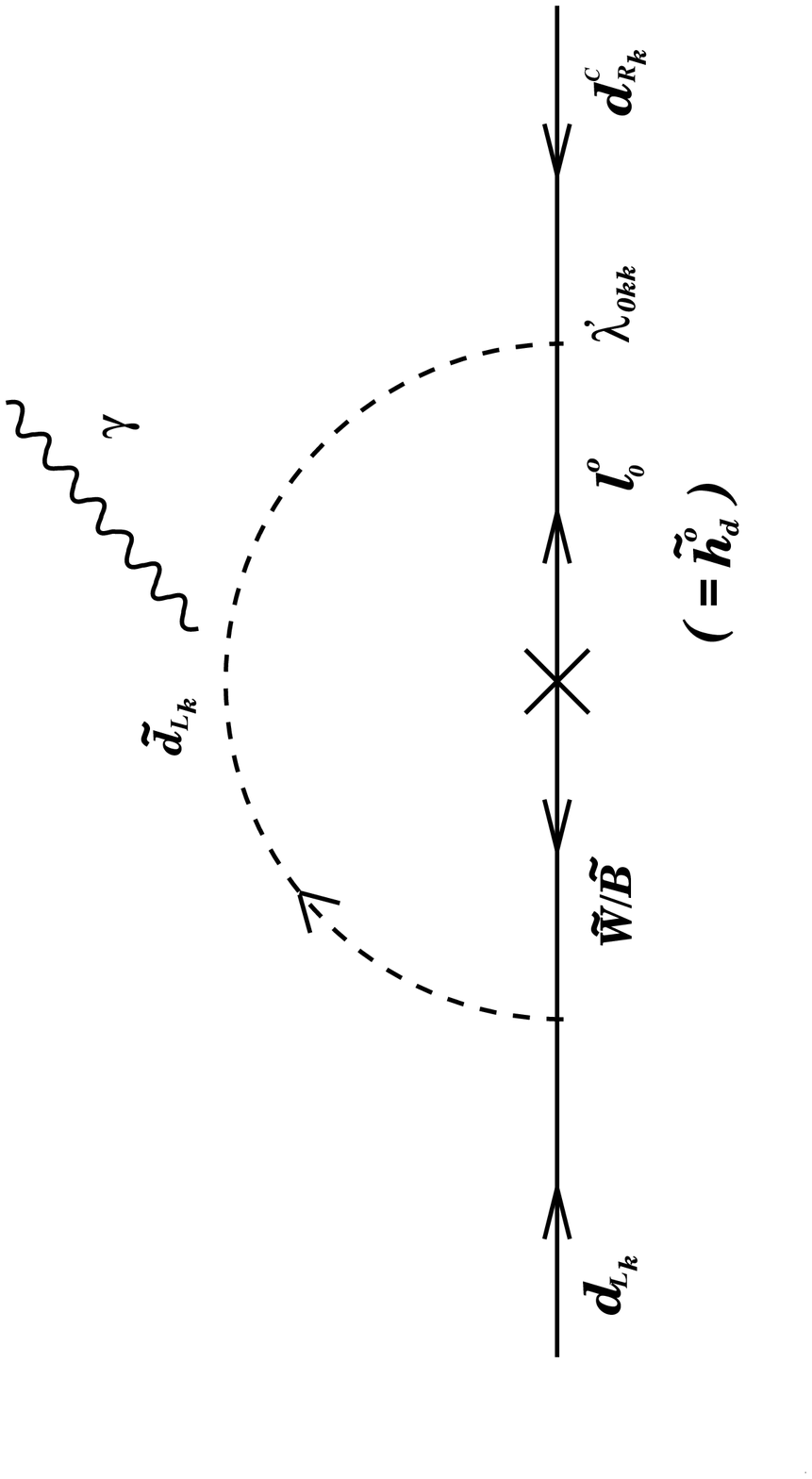}
\vspace*{3in}

\vspace*{.8in}

\includegraphics{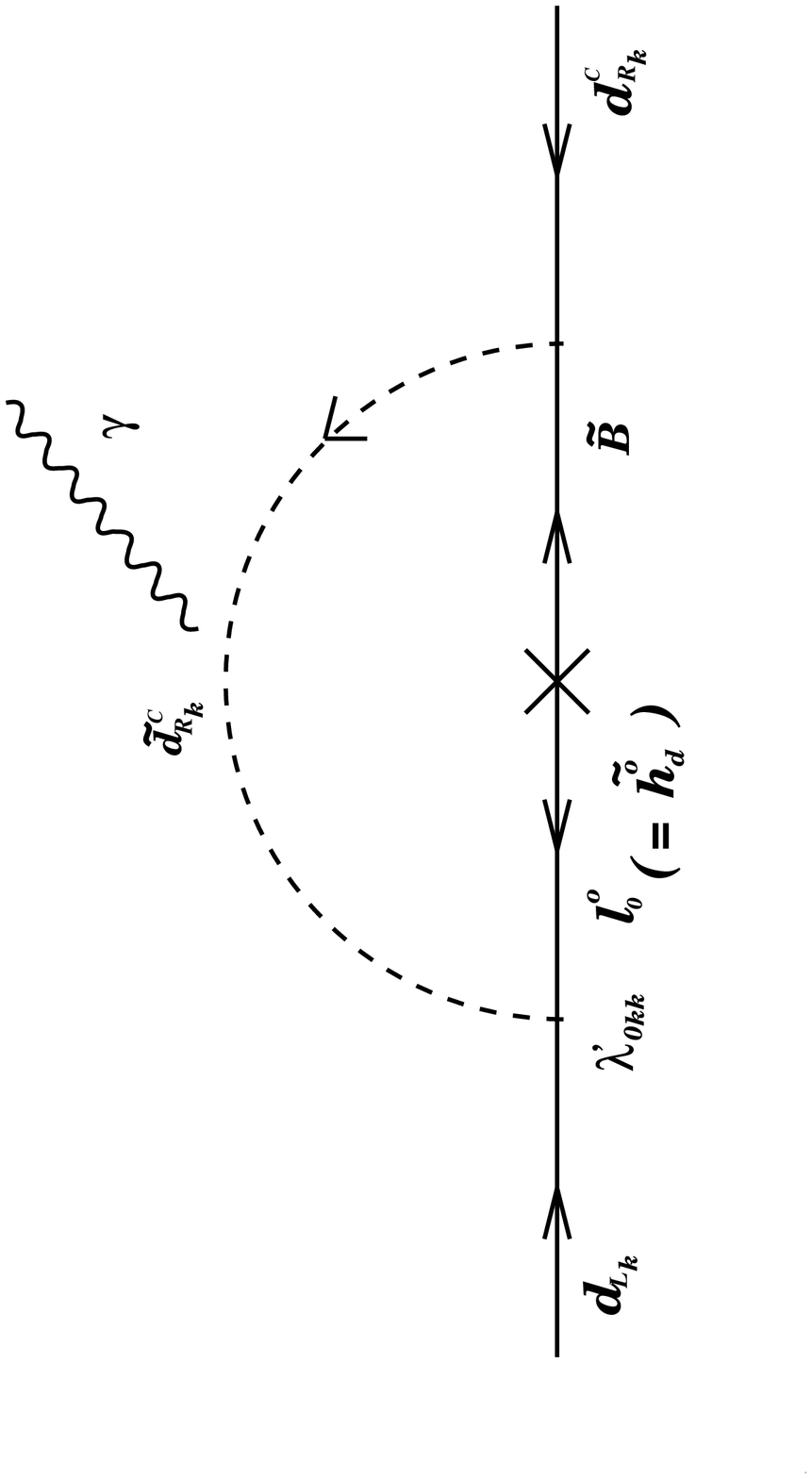}
\vspace*{3.5in}
\caption{Diagrams with neutral gaugino-Higgsino mixing  
for $d$-quark EDM.}
\end{figure}

\eject

\begin{figure}
\vspace*{.5in}
\includegraphics{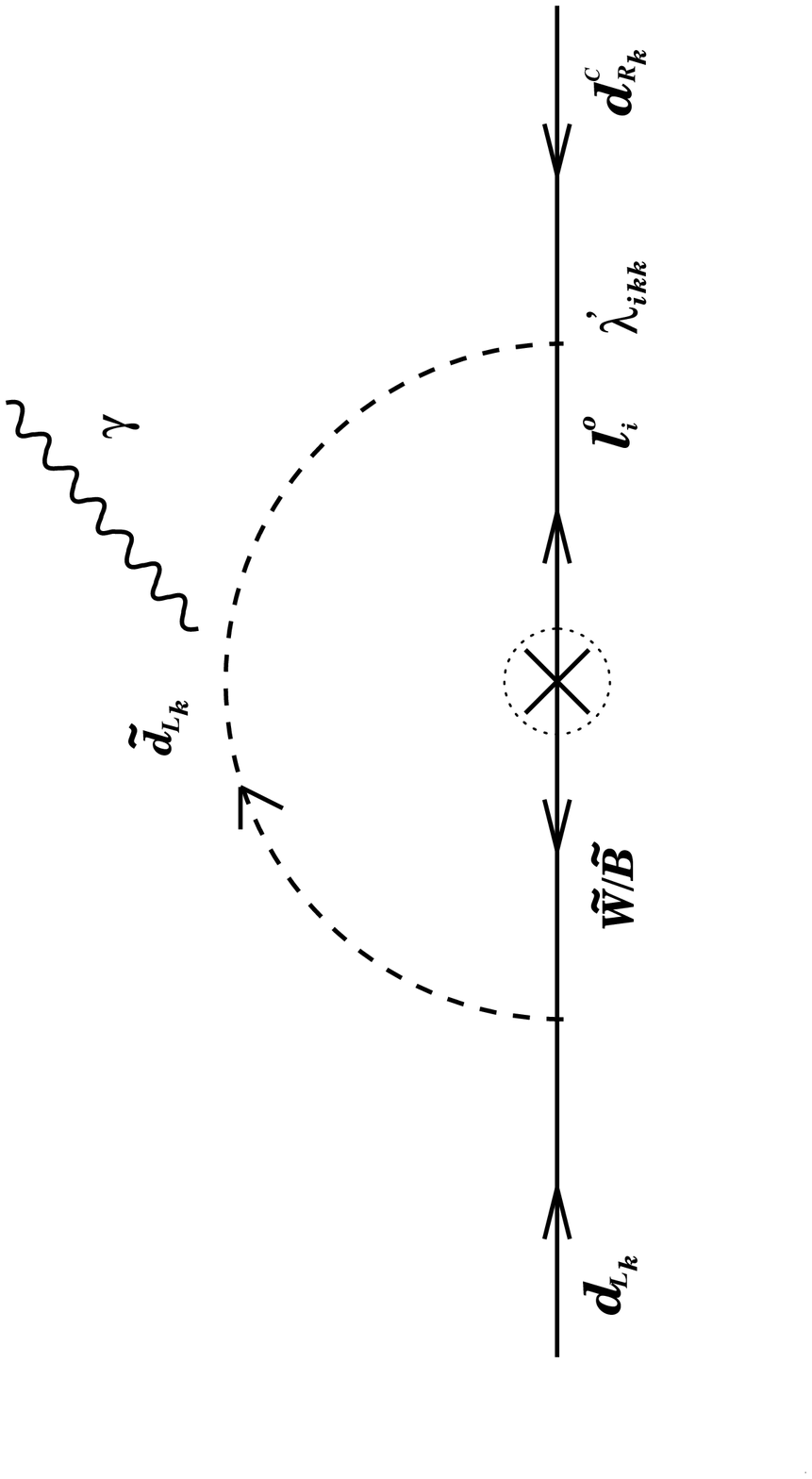}
\vspace*{3in}

\vspace*{.8in}

\vspace*{.5in}
\includegraphics{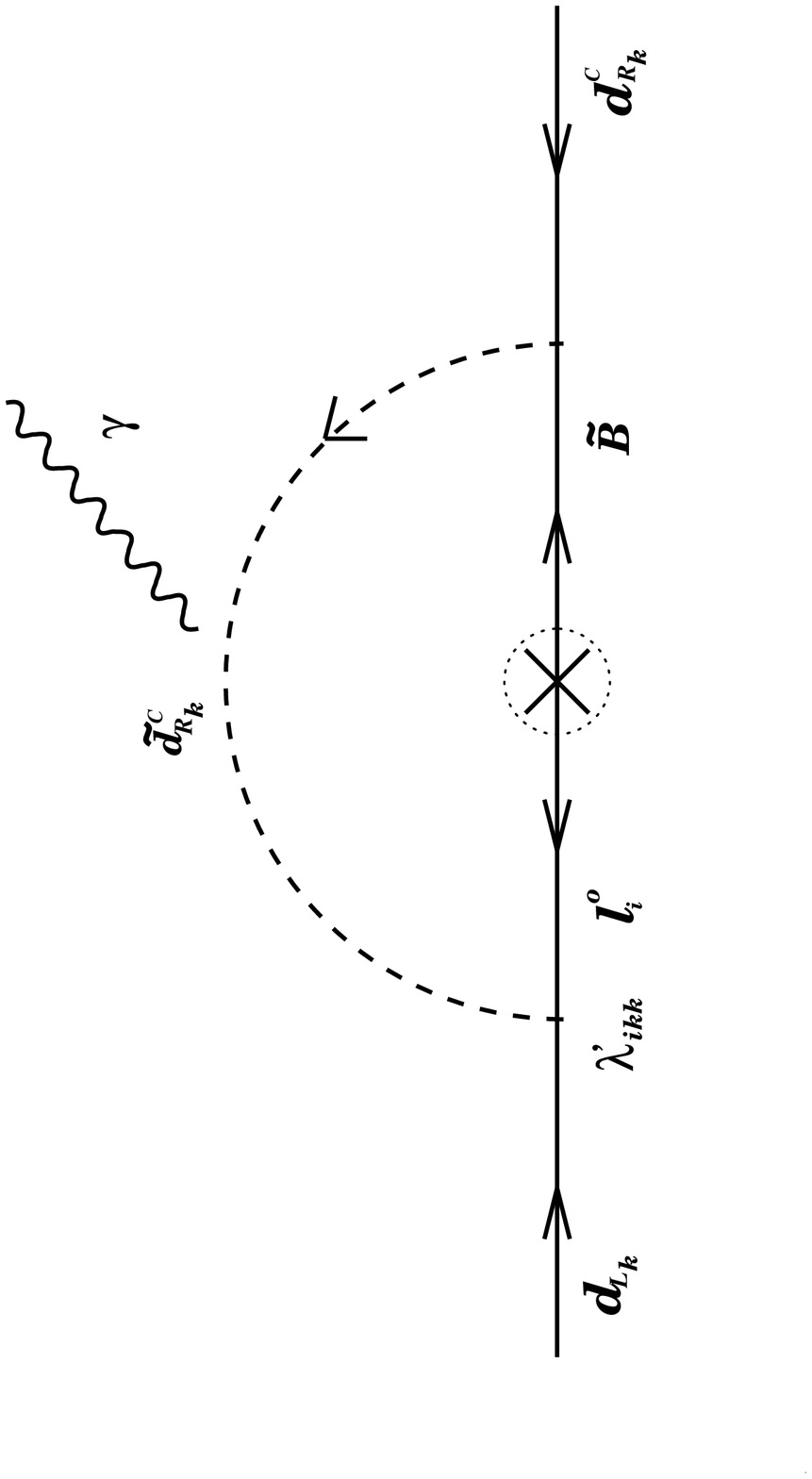}
\vspace*{3in}
\caption{R-parity violating neutralinolike loop diagrams for $d$-quark EDM.
Naive electroweak-state analysis suggests that such a diagram is proportional
to a vanishing ${l}_k^{\scriptscriptstyle 0}$-gaugino mass mixing.}
\end{figure}

\eject

\begin{figure}
\vspace*{.5in}
\includegraphics{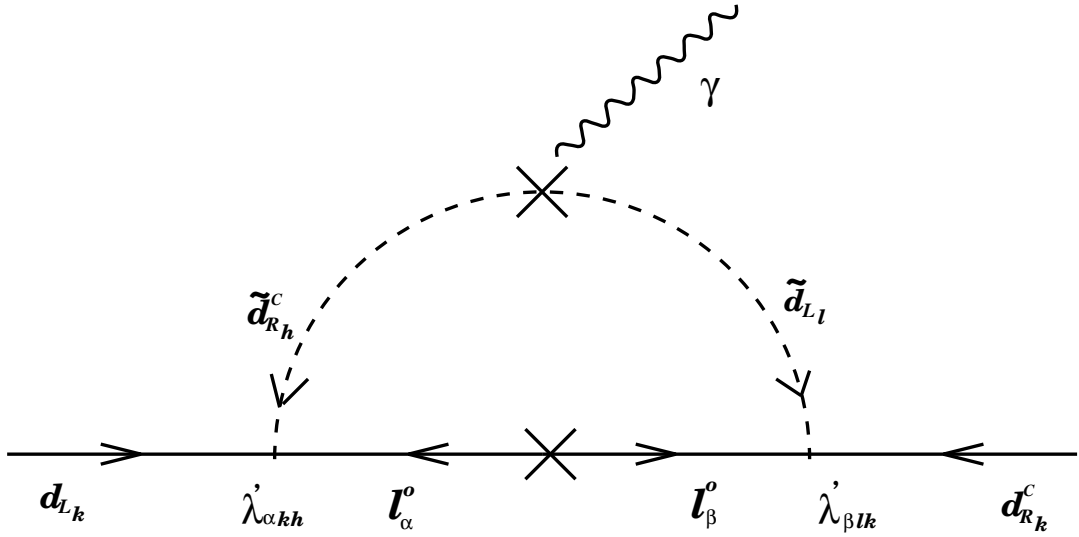}
\vspace*{3in}
\caption{Diagram for $d$-quark EDM suggesting involvement of Majorana 
masses among the ${l}_{\scriptscriptstyle \za}^{\scriptscriptstyle 0}$ or
``neutrinos".}
\end{figure}

\eject

\begin{figure}
\vspace*{.5in}
\includegraphics{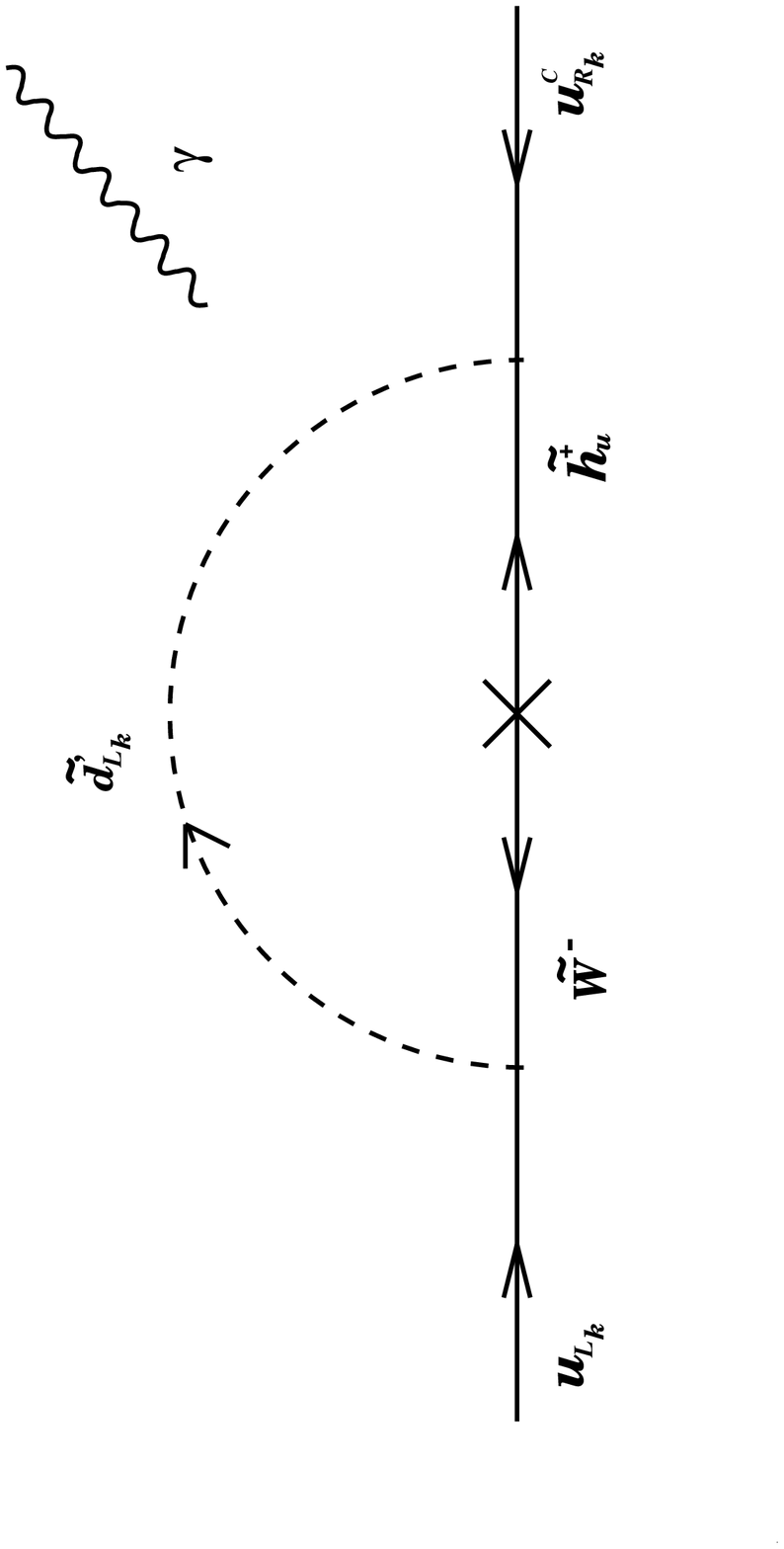}
\vspace*{3in}

\vspace*{.8in}

\includegraphics{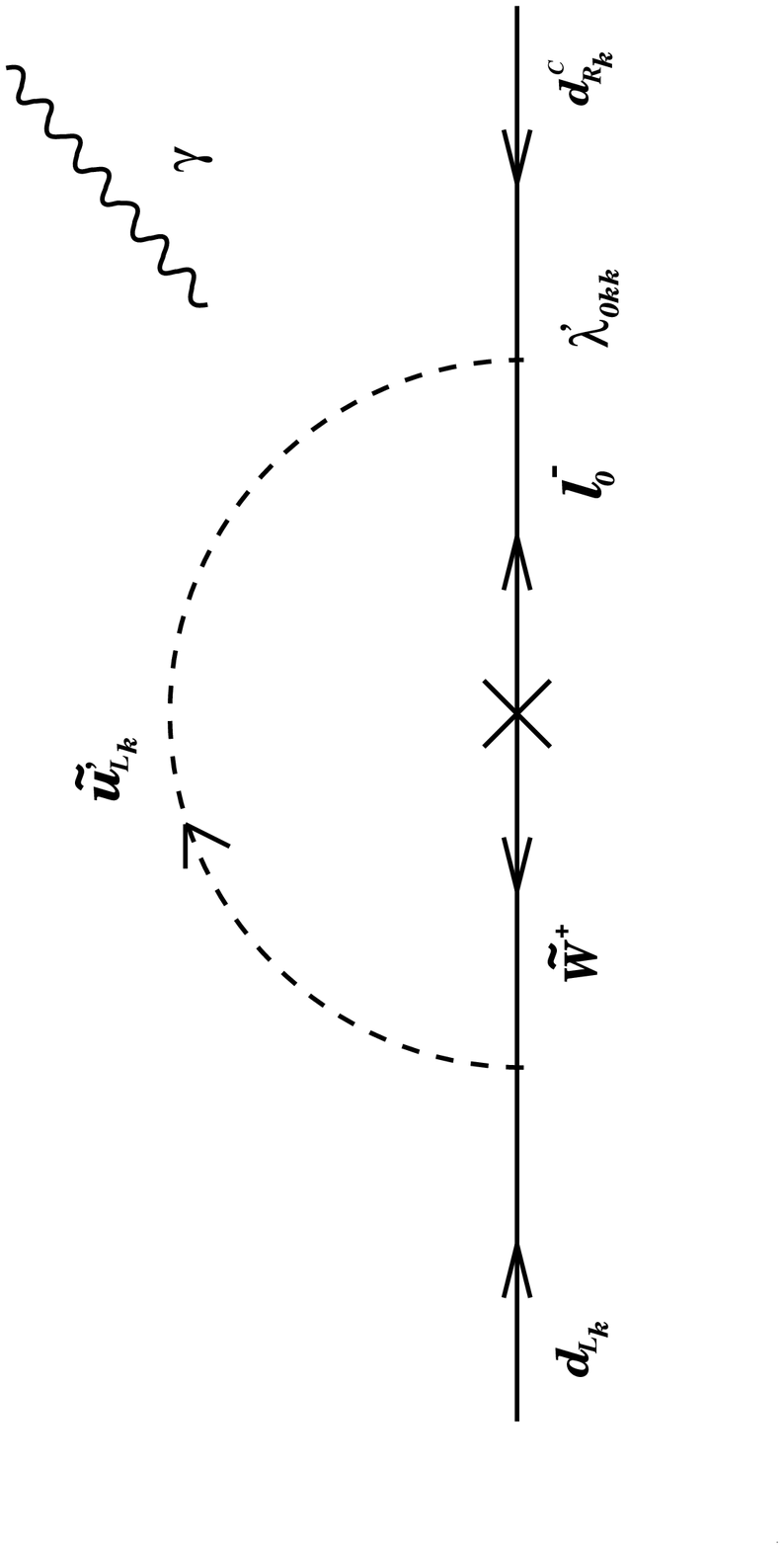}
\vspace*{3.5in}
\caption{Diagrams for $u$- and $d$-quark EDMs with charged gaugino-Higgsino mixing.}
\end{figure}

\eject

\begin{figure}
\vspace*{.5in}
\includegraphics{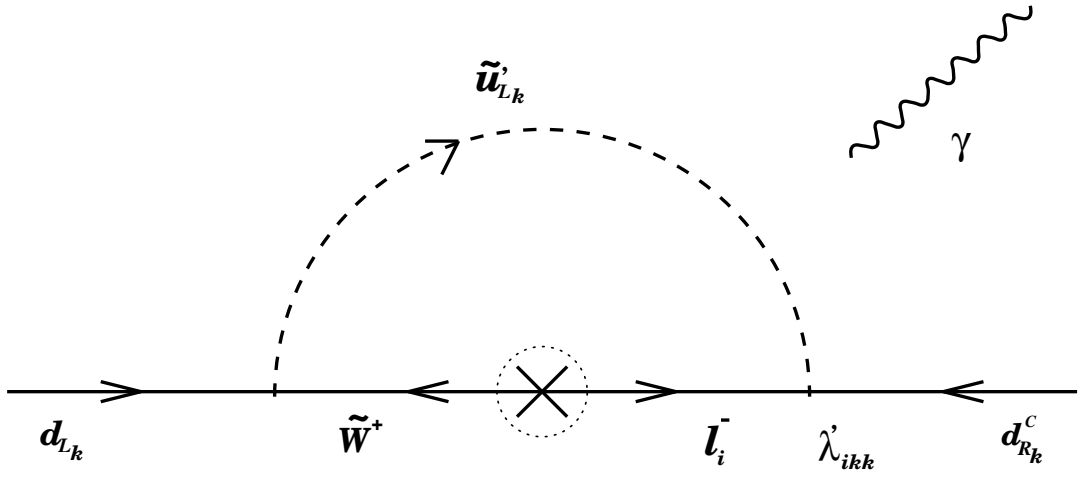}
\vspace*{3in}
\caption{R-parity violating charginolike loop diagram for $d$-quark EDM.
Naive electroweak-state analysis suggests that the diagram is proportional
to the vanishing ${l}_k^{\!\!\mbox{ -}}$--$\tilde{W}^{\scriptscriptstyle +}$
mass term.}
\end{figure}

\eject

\begin{figure}
\includegraphics{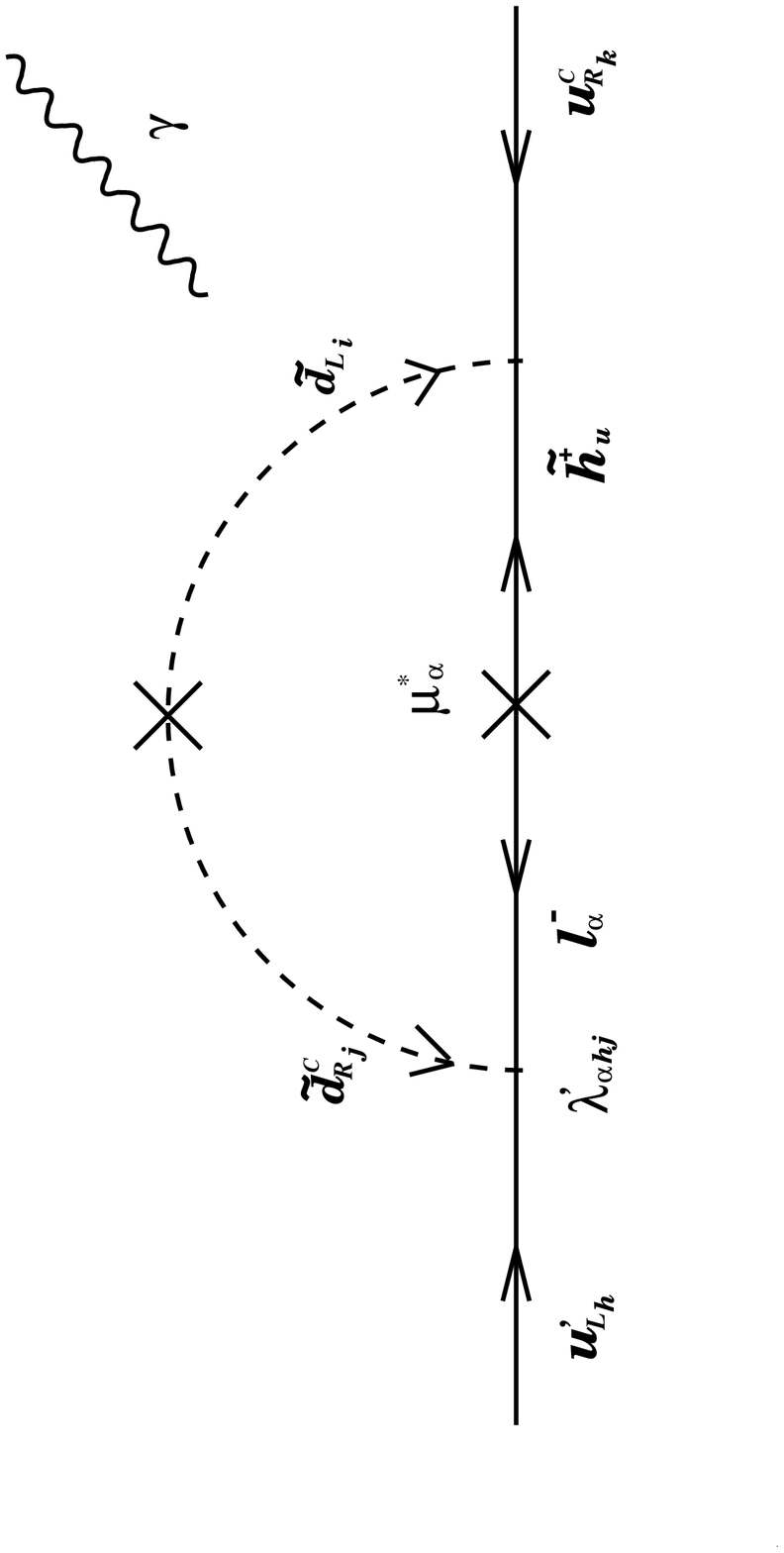}
\vspace*{3in}

\vspace*{.8in}

\vspace*{.5in}
\includegraphics{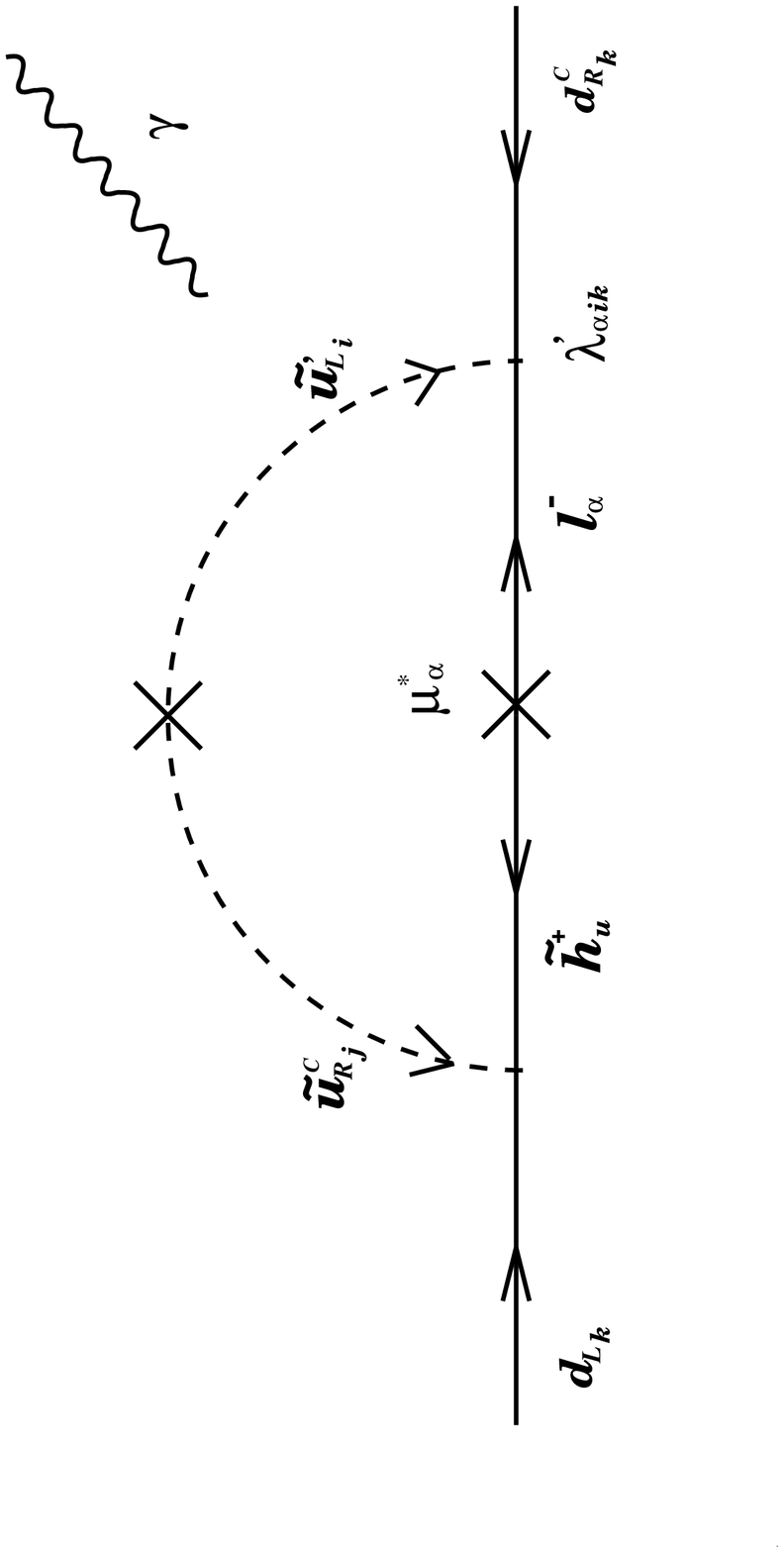}
\vspace*{3in}
\caption{Diagrams for $u$- and $d$-quark EDMs with a $\mu_{\!\scriptscriptstyle \za}$  mass insertion.}
\end{figure}

\eject

\begin{figure}
\includegraphics{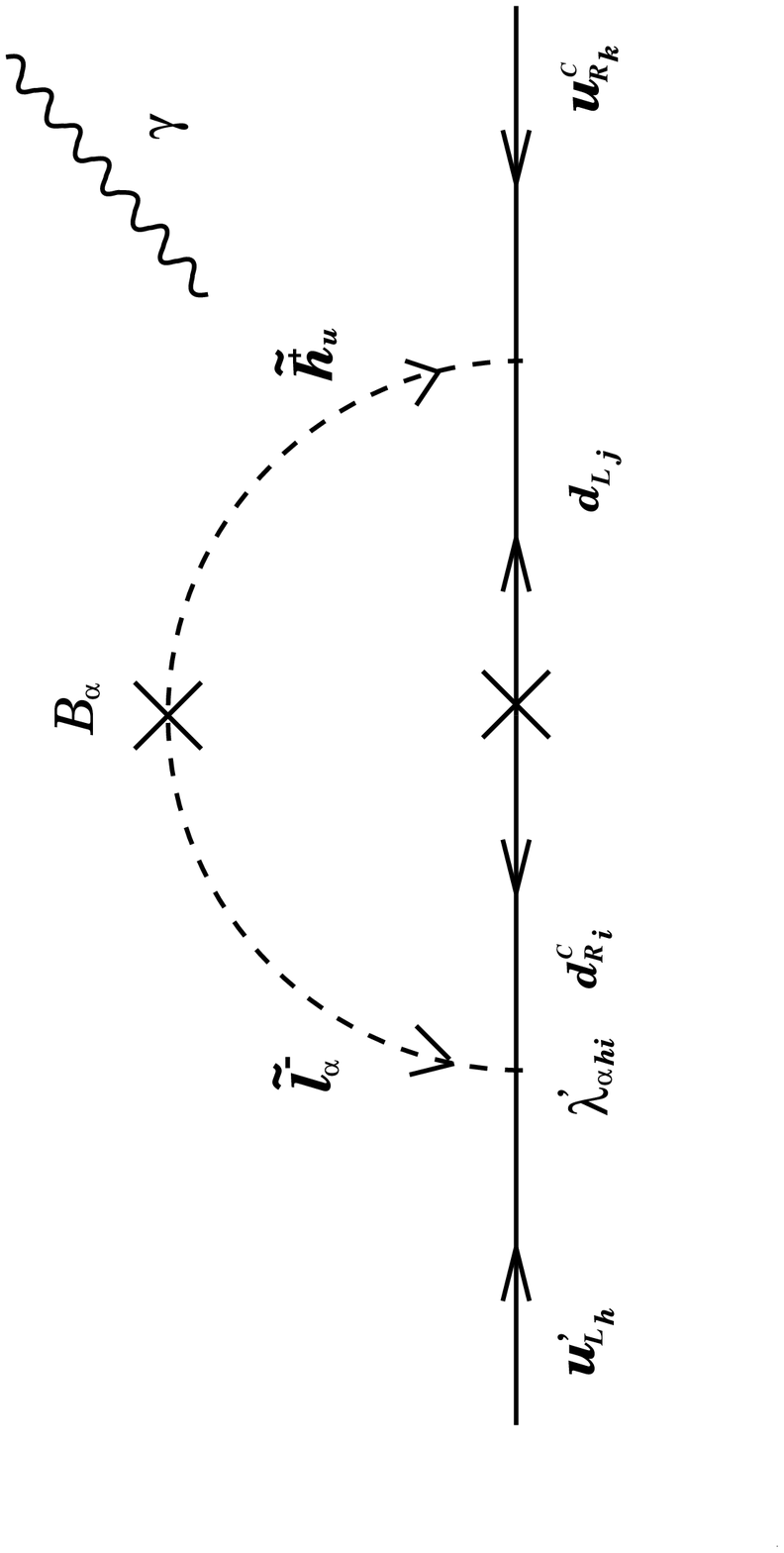}
\vspace*{3in}

\vspace*{.5in}

\includegraphics{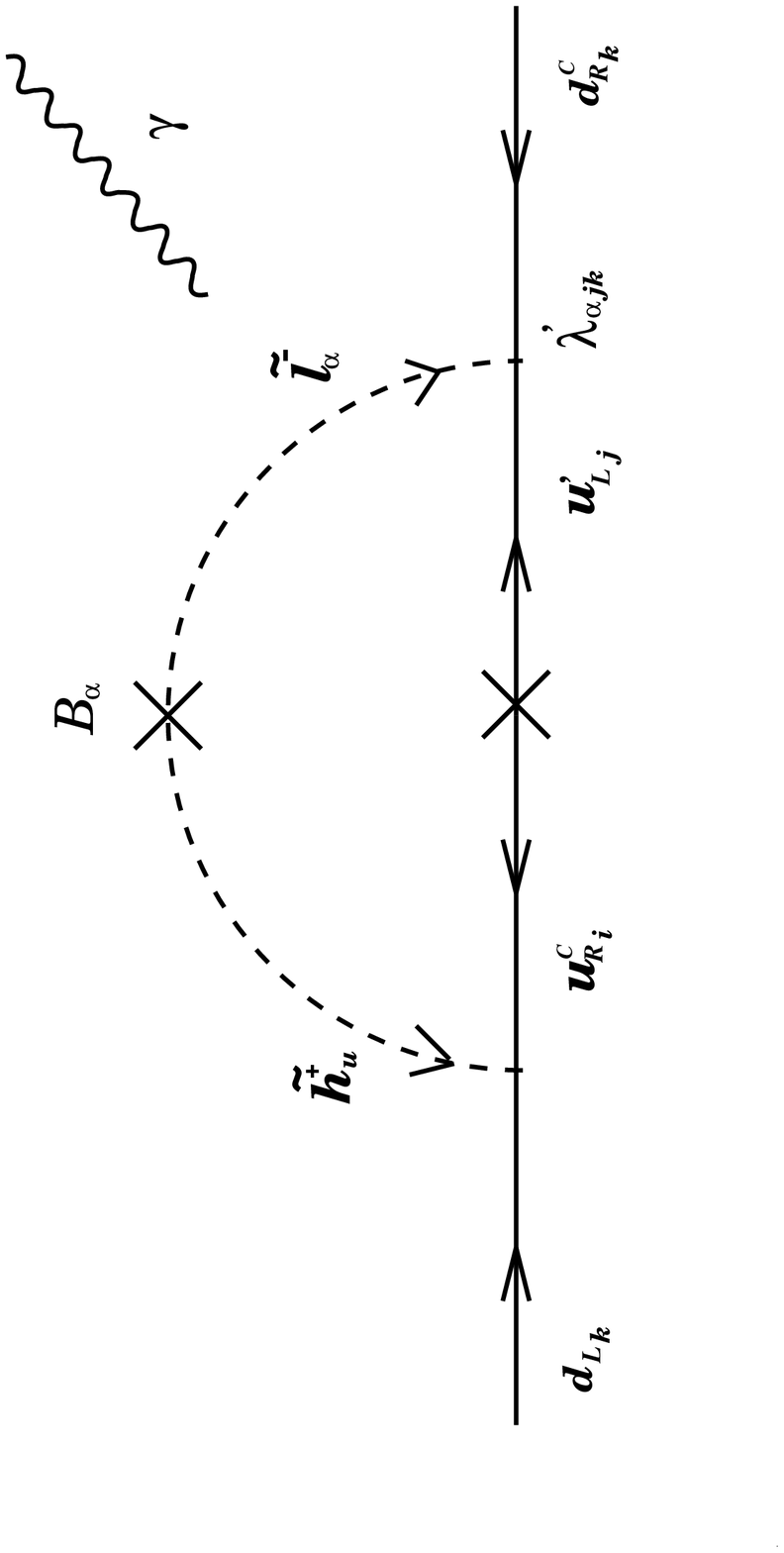}
\vspace*{3in}
\caption{Diagrams for $u$- and $d$-quark EDMs with a $B_{\za}$ scalar mass insertion.}
\end{figure}

\eject

\begin{figure}
\includegraphics{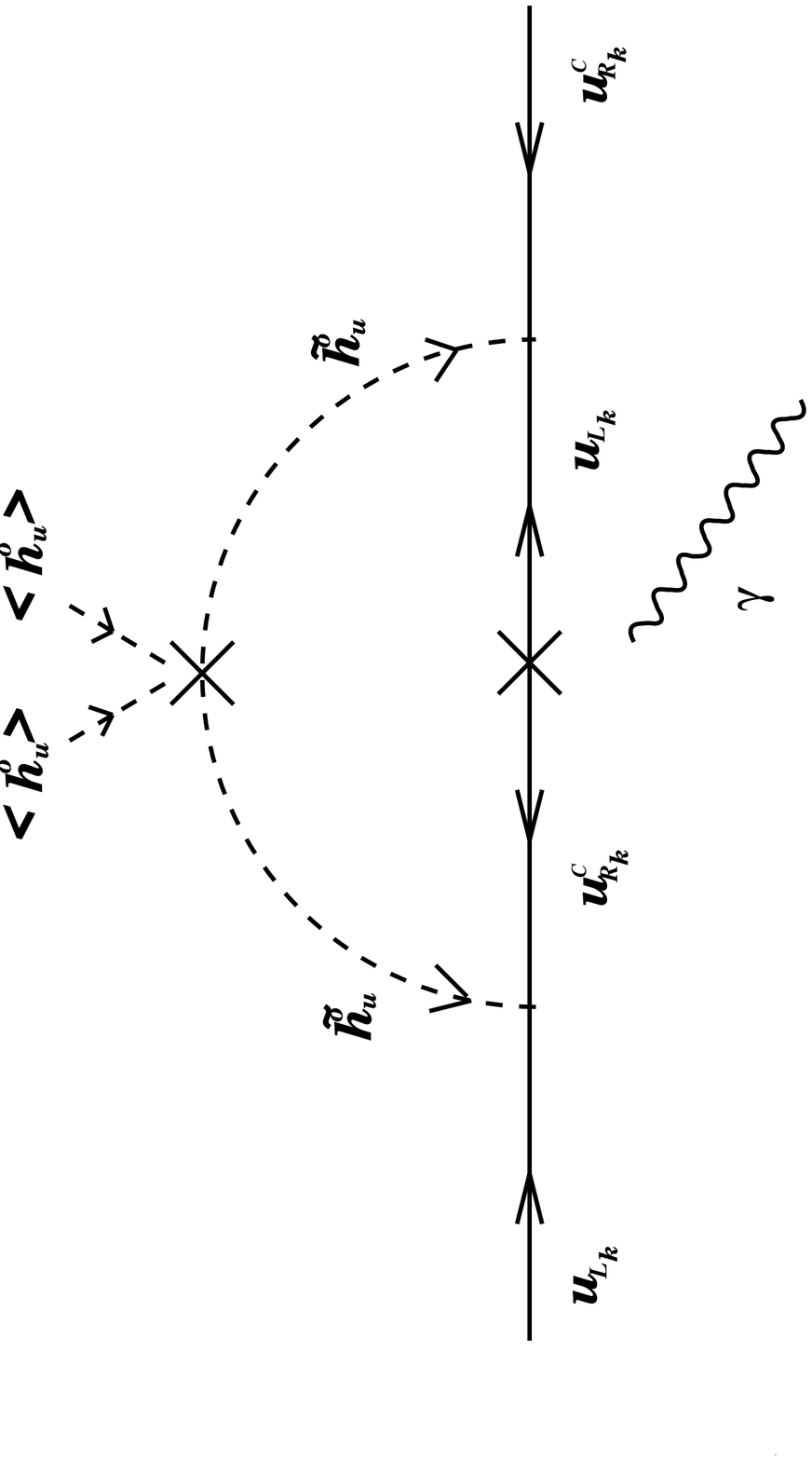}
\vspace*{3in}

\vspace{.7in}

\includegraphics{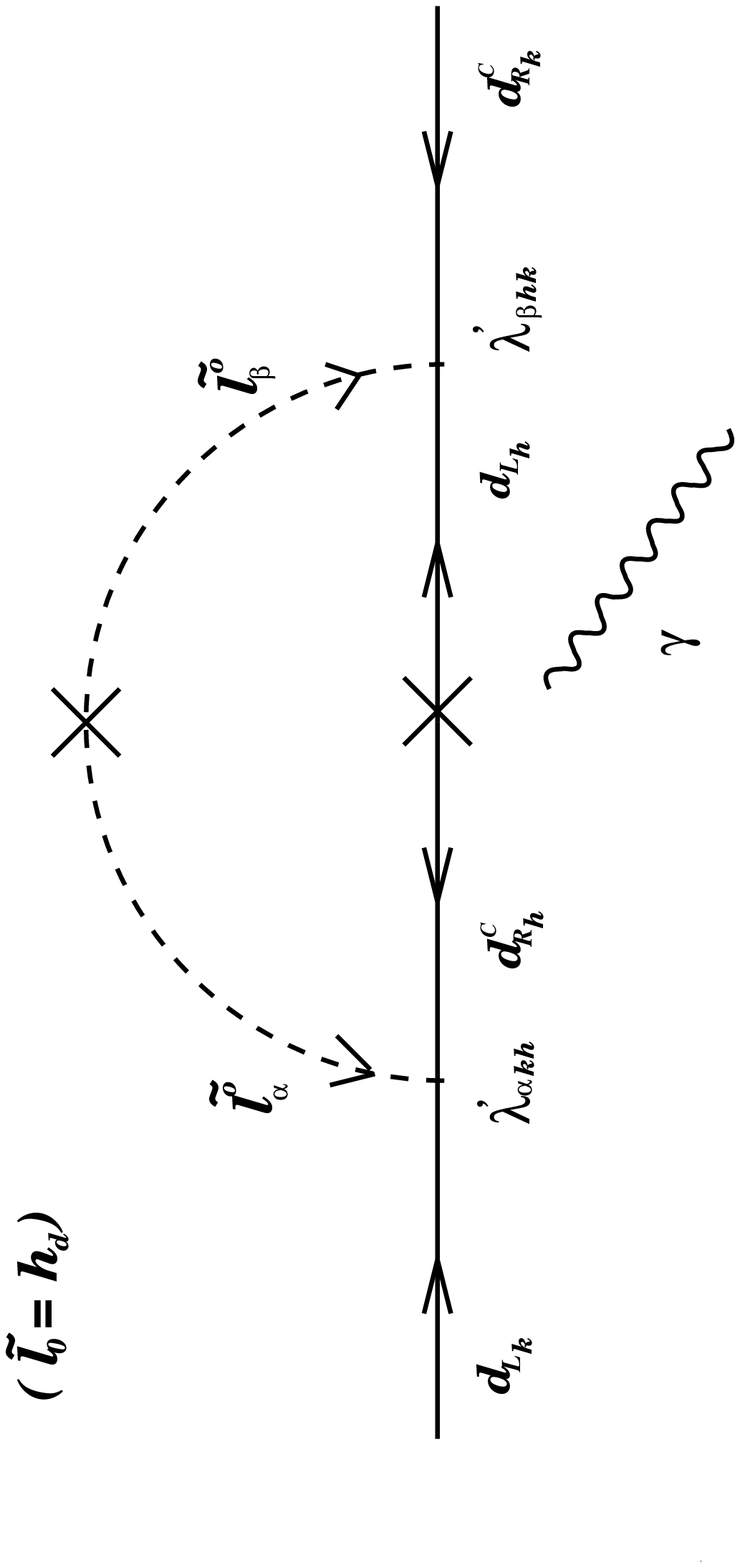}
\vspace*{4in}
\caption{\small Diagrams with a Majorana-like scalar mass insertion 
for $u$- and $d$-quark EDMs. For the $u$-quark case, the direct Majorana-like
$h_{\scriptscriptstyle u}$ mass insertion is explicitly shown. For the $d$-quark 
case, the corresponding direct $h_{\scriptscriptstyle d}$ mass insertion is
obvious, for $\za=\zb=0$; for $\za$ and/or $\zb$ nonzero, the naive direct result
from the diagram would vanish, due to the vanishing VEVs.}
\end{figure}

\eject

\begin{figure}[th]
\vspace*{6.5in}
\includegraphics{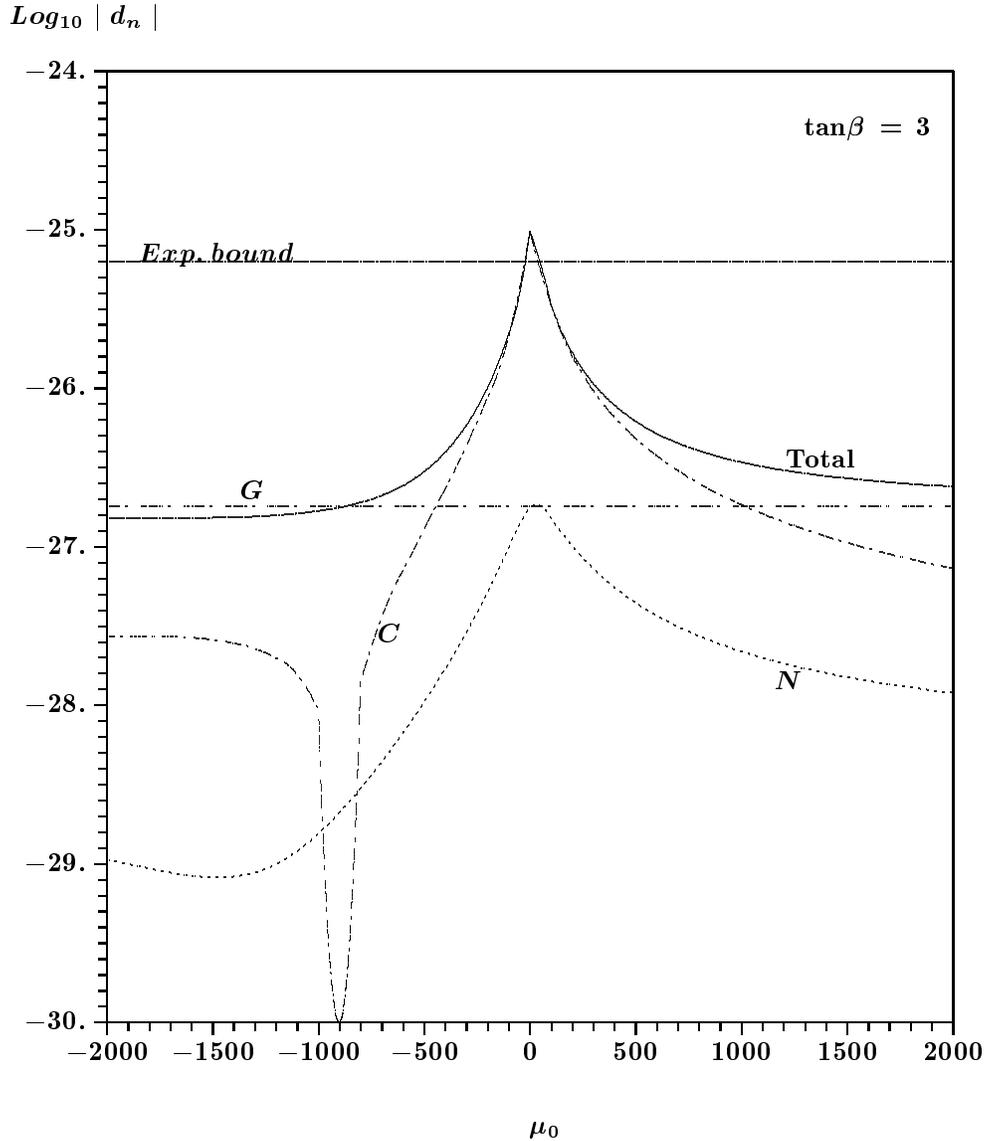}
\vspace*{1in}
\caption{\small Logarithmic plot of (the magnitude of) the RPV neutron EDM result for 
$\mu_{\!\scriptscriptstyle 0}$ value between $\pm2000\,\mbox{GeV}$,  
with the other parameters set at the same values as case~A in Table 1. 
The lines marked by $G$, $C$, $N$, and ``Total" give the complete
gluino, charginolike, neutralinolike, and total ({\it i.e.,} sum of the three) contributions, respectively. Note that the values of the $N$
contributions and those of the $C$ line for 
$\mu_{\!\scriptscriptstyle 0}<-900\,\mbox{GeV}$ are negative. 
}\end{figure}

\eject

\begin{figure}[th]
\vspace*{6.2in}
\includegraphics{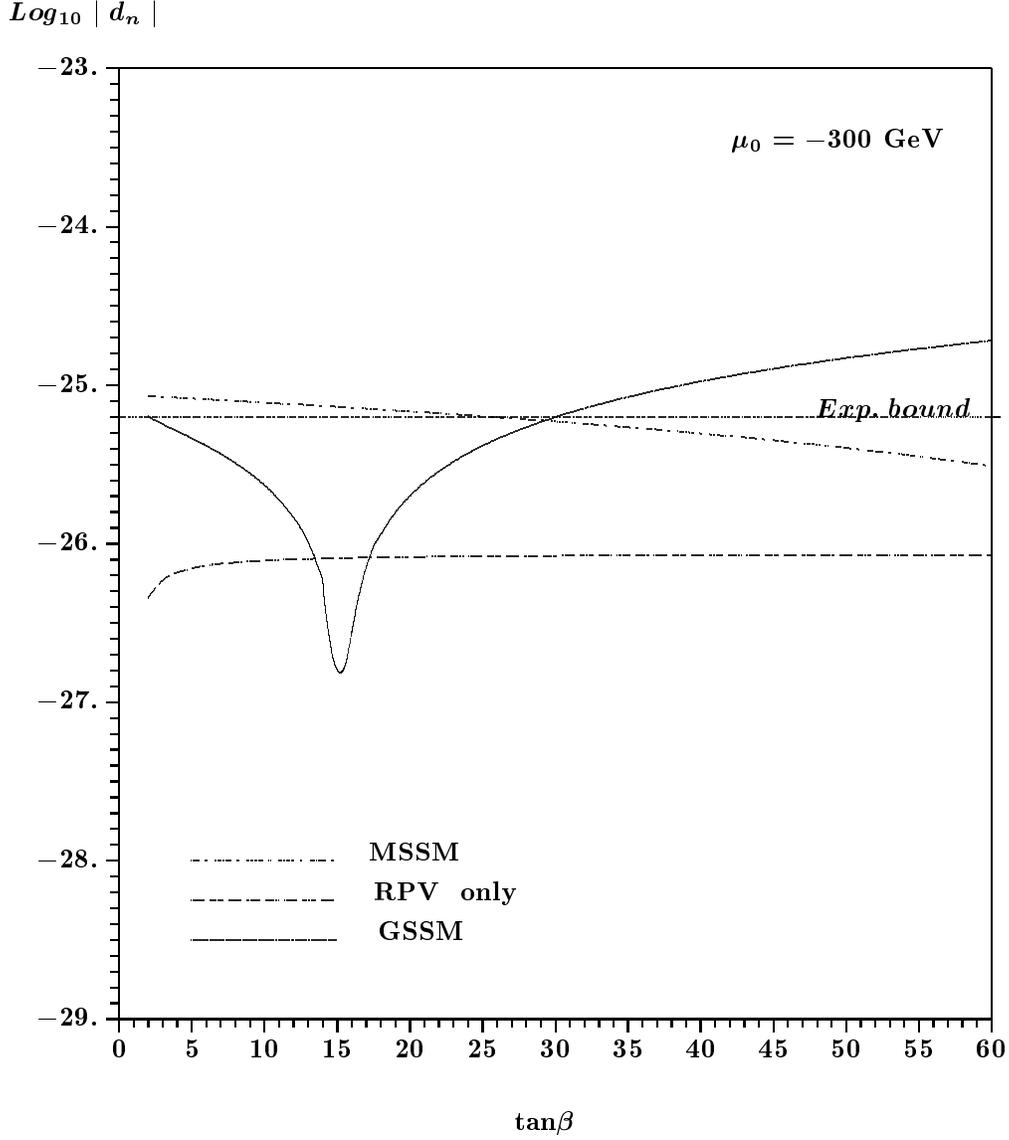}
\vspace*{1in}
\caption{\small
Logarithmic plot of (the magnitude of) the neutron EDM result verses 
$\tan\!\zb$. We show here the MSSM result, our general result with the RPV phase
only, and the generic result  with complex phases of 
both kinds. In particular, the $A$ and $\mu_{\scriptscriptstyle 0}$ 
phases are chosen as $7^o$ and $0.1^o$ respectively, for the MSSM line. They are 
zero for the RPV-only line, with which we have a phase of ${\pi\over 4}$ for 
$\lambda^{\!\prime}_{\scriptscriptstyle 31\!1}$. All the given nonzero values
are used for the three phases for the generic result (from our complete formulae) 
marked by ``GSSM". Again, the other unspecified input parameters are the same as for 
case~A of Table~1.
}\end{figure}
\eject

\begin{figure}[th]
\vspace*{6.5in}
\includegraphics{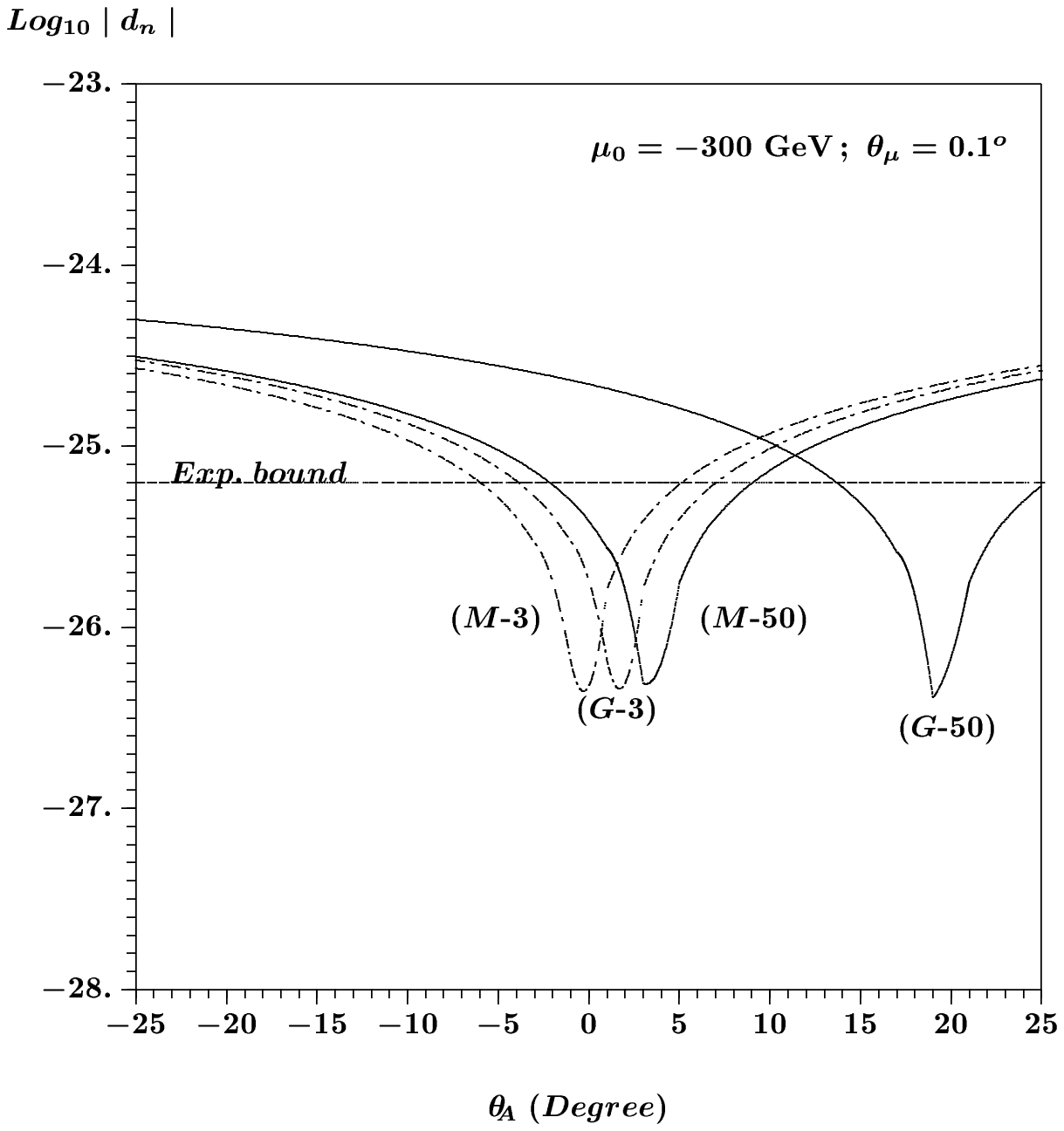}
\vspace*{1in}
\caption{\small
Logarithmic plot of (the magnitude of) the neutron EDM result versus
$\theta_{\!\!\scriptscriptstyle A} $ (the complex phase for the $A$ parameter). 
The four lines shown are characterized by the $\tan\!\zb$ values (~3 or 50~) used 
and whether it is for our GSSM result (again with a phase of ${\pi\over 4}$ for 
$\lambda^{\!\prime}_{\scriptscriptstyle 31\!1}$) --- marked by $G$; or the result
for MSSM --- marked by $M$. Again, the $\lambda^{\!\prime}_{\scriptscriptstyle 31\!1}$
phase is set at ${\pi\over 4}$ for the $G$ lines, and the
$\mu_{\scriptscriptstyle 0}$ phase at $0.1^o$ for all; 
the other unspecified input parameters are the same as for case~A of Table~1. 
}\end{figure}
\eject

\begin{figure}[th]
\vspace*{6.5in}
\includegraphics{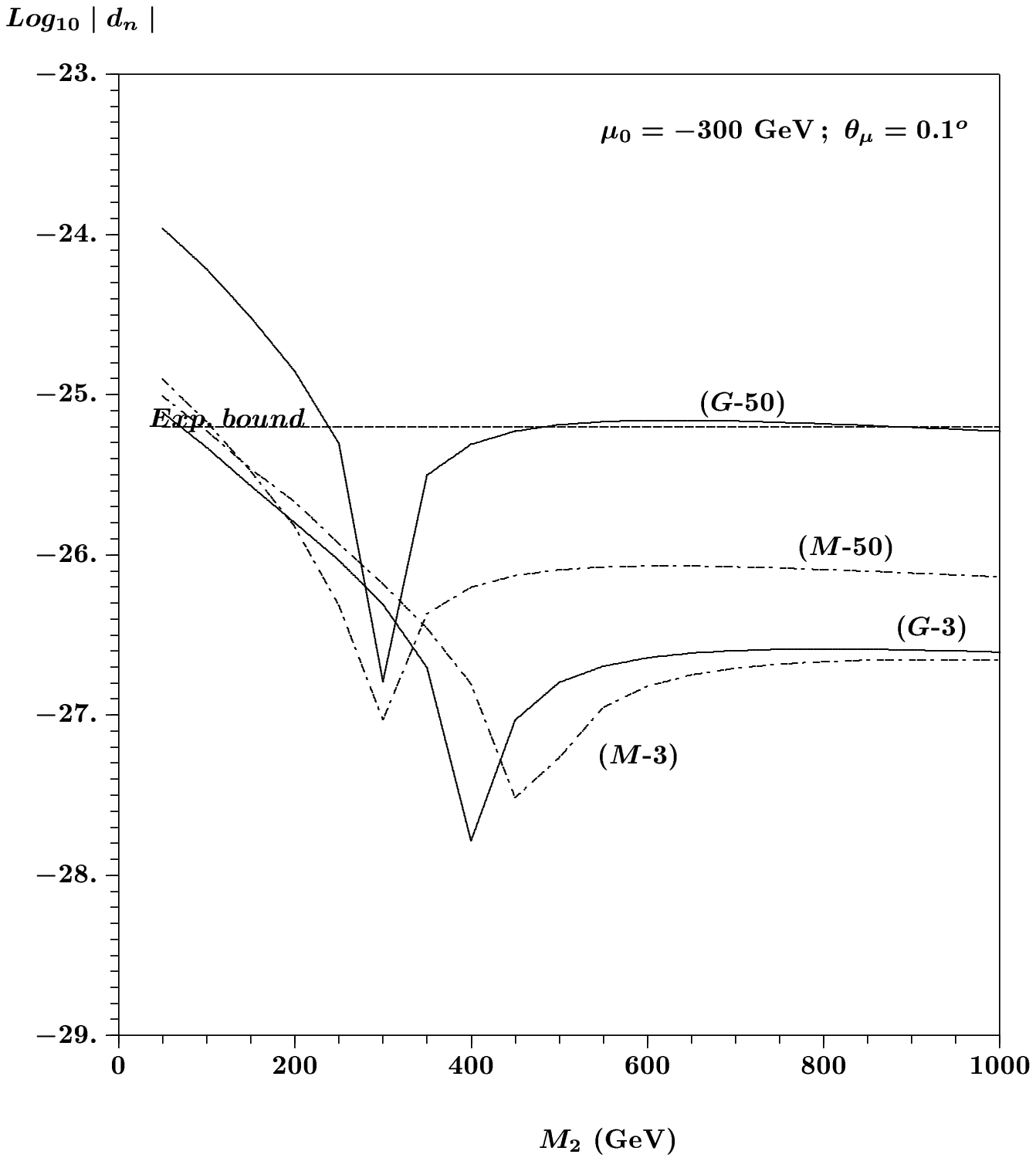}
\vspace*{1in}
\caption{\small
Logarithmic plot of (the magnitude of) the neutron EDM result versus
$M_{\scriptscriptstyle 2}$. The four lines correspond to the four cases of Fig.~12, 
each with $\theta_{\!\!\scriptscriptstyle A} $ set at the dip location, {\it i.e.,} 
$G$-$3$ for GSSM at $\tan\!\zb=3$ with $\theta_{\!\!\scriptscriptstyle A}  = 2^o$, 
$M$-$3$ for MSSM at $\tan\!\zb=3$ with $\theta_{\!\!\scriptscriptstyle A}  = -1^o$, 
$G$-$50$ for GSSM at $\tan\!\zb=50$ with $\theta_{\!\!\scriptscriptstyle A}  = 20^o$, 
$M$-$50$ for MSSM at $\tan\!\zb=50$ with $\theta_{\!\!\scriptscriptstyle A}  = 3^o$. 
All other unspecified input parameters are the same as for Fig.~12. 
}\end{figure}


\newpage
\begin{thebibliography}{99}
\bibitem{susycp}
For a recent review, see
Y. Nir, hep-ph/9911321.
\bibitem{two}
R.M. Godbole, S. Pakvasa, S.D. Rindani, and X. Tata,
Phys. Rev. {\bf D61},  {\it 113003} (2000);
S.A. Abel, A. Dedes, and H.K. Dreiner, JHEP {\bf 05}, 013 (2000);
see also
D. Chang, W.-F. Chang, M. Frank, and W.-Y. Keung,
Phys. Rev. {\bf D62},  {\it 095002} (2000) .
\bibitem{as4} 
Y.-Y. Keum and O.C.W. Kong, 
Phys. Rev. Lett. {\bf 86}, 393 (2001).
\bibitem{cch}
K. Choi, E.J. Chun, and K. Hwang, Phys. Rev. {\bf D63},  {\it 013002} (2000).
\bibitem{as5}
O.C.W. Kong, JHEP {\bf 0009}, 037 (2000).
\bibitem{as8}
For more discussion on the formulation aspect, see
O.C.W. Kong,  {IPAS-HEP-k008},
{\it manuscript in preparation}.
\bibitem{k}
M. Bisset, O.C.W. Kong, C. Macesanu, and L.H. Orr,
Phys. Lett. {\bf B430}, 274 (1998); 
Phys. Rev. {\bf D62},  {\it 035001} (2000).
\bibitem{ok}
O.C.W. Kong, Mod. Phys. Lett. {\bf A14},  903 (1999);
Phys. At. Nucl. {\bf 63}, 1083 (2000).
\bibitem{as1}
K. Cheung and O.C.W. Kong,  
Phys. Rev. {\bf D61},  {\it 113012} (2000).
\bibitem{AL}
A. Abada and M. Losada,  Nucl. Phys. {\bf B585}, 45 (2000).
\bibitem{GH}
Y. Grossman and H.E. Haber,  hep-ph/9906310; see also
 S. Davidson and M. Losada, JHEP {\bf 0005}, {\it 021} (2000).
\bibitem{as7}
K. Cheung and O.C.W. Kong, {IPAS-HEP-k007},
  hep-ph/0101347, {\it submitted to} Phys. Rev. {\bf D}.
\bibitem{cch2}
K. Choi, E.J. Chun, and K. Hwang, Phys. Lett. {\bf B488}, 145 (2000).
\bibitem{as11}
O.C.W. Kong  {\it et al.}, {\it work in progress}.
\bibitem{HKP}
See, for a review, 
X.-G. He, B.H.J. McKellar, and S. Pakvasa,
Int. J. Mod. Phys. {\bf A4}, 5011 (1989).
\bibitem{IN1} 
T. Ibrahim and P. Nath, Phys. Rev. {\bf D57}, 478 (1998);
{\it Erratum --- ibid} {\bf D58}, {\it 019901} (1998);
{\it Erratum --- ibid} {\bf D60}, {\it 079903} (1999);
{\it Erratum --- ibid} {\bf D60}, {\it 119901} (1999).
\bibitem{QCD}
R. Arnowitt, J.L. Lopez, and D.V. Nanopoilos, Phys. Rev. {\bf D42}, 2423 (1990).
\bibitem{exp}
P.G. Harris {\it et al.},
 Phys. Rev. Lett. {\bf 82} 904 (1999).
\bibitem{KiOs}
Y. Kizukuri and N. Oshimo, Phys. Rev. {\bf D46}, 3025 (1992).
\bibitem{IN2} 
T. Ibrahim and P. Nath, Phys. Rev. {\bf D58}, {\it 111301} (1998);
{\it Erratum --- ibid} {\bf D60}, {\it 099902} (1999).
\bibitem{NGKO}
T. Goto, Y.-Y. Keum, T. Nihei, Y. Okada, and Y. Shimizu,
Phys. Lett. {\bf 460B}, 333 (1999).
\bibitem{pd}
L.E. Ib\'a\~nez and G.G. Ross,
Nucl. Phys. {\bf B368}, 3 (1992).
\bibitem{app}
See the appendix of Ref.\cite{as5}.
\bibitem{sK}
Super-Kamiokande Collaboration, Y. Fukuda {\it et al.},
 Phys. Rev. Lett. {\bf 81}, 1562 (1998);
 P. Lipari, hep-ph/9904443;
G.L. Fogli, E. Lisi, A. Marrone, and G. Scioscia,
Phys. Rev. {\bf D59}, {\it 033001} (1999) .
\bibitem{aleph}
R. Barate {\it et al}. (ALEPH Collaboration),
CERN-PPE-97-138, (1997).
\bibitem{lambda}
See, for example,
G. Bhattacharyya, Nucl. Phys. (Proc. Suppl.) {\bf 52A}, 83 (1997);
 V. Bednyakov, A. Faessler, and S. Kovalenko,  hep-ph/9904414.
\bibitem{FO}
T. Falk and K. Olive, Phys. Lett. {\bf B439}, 71 (1998).

\end{thebibliography}
\end{document}